\documentclass[a4paper,UKenglish,cleveref, autoref, thm-restate]{lipics-v2021}

\pdfoutput=1 
\hideLIPIcs  

\usepackage[linesnumbered,ruled,vlined]{algorithm2e}

\usepackage{xcolor,colortbl}

\usepackage{graphicx}
\usepackage{subcaption}

\title{Improved Dominance Filtering for Unions and Minkowski Sums of Pareto Sets} 

\titlerunning{Improved Dominance Filtering for Unions and Minkowski Sums of Pareto Sets} 

\author{Konstantinos Karathanasis}{Department of Computer Engineering and Informatics, University of Patras, Greece \and PIKEI New Technologies, Patras, Greece }{k_karathanasis@ac.upatras.gr}{https://orcid.org/0009-0004-1741-3693}{}

\author{Spyros Kontogiannis}{Department of Computer Engineering and Informatics, University of Patras, Greece \and Computer Technology Institute and Press “Diophantus”, Patras, Greece}{spyridon.kontogiannis@upatras.gr}{https://orcid.org/0000-0002-8559-6418}{}

\author{Christos Zaroliagis}{Department of Computer Engineering and Informatics, University of Patras, Greece \and Computer Technology Institute and Press “Diophantus”, Patras, Greece}{zaro@ceid.upatras.gr}{https://orcid.org/0000-0003-1425-5138}{}

\authorrunning{K.~Karathanasis, S.~Kontogiannis and C.~Zaroliagis} 

\Copyright{Konstantinos Karathanasis, Spyros Kontogiannis and Christos Zaroliagis} 



\category{} 

\relatedversion{} 

\funding{This work was partially supported by the EU I3 Instrument under GA No 101115116 (project AMBITIOUS) and by the University of Patras under GA No 83770 (programme “MEDICUS”).}

\nolinenumbers 

\EventShortTitle{August 2025}

\begin{document}

\maketitle

\begin{abstract}

A key task in multi-objective optimization is to compute the \emph{Pareto subset or frontier} $P$ of
a given $d$-dimensional objective space $F$; that is, a maximal subset $P\subseteq F$ such that every element in $P$ is
\emph{not-dominated} (it is not worse in all criteria) by any element in $F$.
This process, called \textit{dominance-filtering}, often involves handling objective spaces derived from either the \emph{union}
or the \emph{Minkowski sum} of two given partial objective spaces which are Pareto sets themselves, and constitutes a major bottleneck in several multi-objective optimization techniques.
In this work, we introduce three new data structures, ND$^{+}$-trees, QND$^{+}$-trees and TND$^{+}$-trees, which are designed for efficiently indexing non-dominated objective vectors and performing dominance-checks.
We also devise three new algorithms that efficiently filter out dominated objective vectors from the union or the Minkowski sum of two Pareto sets.
An extensive experimental evaluation on both synthetically generated and real-world data sets reveals that our new algorithms outperform state-of-art techniques for dominance-filtering of unions and Minkowski sums of Pareto sets, and scale well w.r.t. the number of $d\ge 3$ criteria and the sets' sizes.

\end{abstract}

\section{Introduction}
\label{section: introduction}

In multiobjective combinatorial optimization (MOCO) problems,
	given an implicit description (e.g., via linear constraints) of a \emph{solution space} $X$
	and the corresponding \emph{objective space} $F$ with $d$-dimensional ($d\ge 2$) objective-value vectors of all elements in $X$, the goal is
	to compute the \emph{Pareto subset or frontier}: a maximal subset of $F$ whose elements are not dominated (are not worse in all criteria) by any other element in $F$.
Many algorithms for MOCO problems, especially when having to work with instances of substantial sizes, rely heavily on the \emph{dominance-filtering} subtask, aiming to efficiently combine (the Pareto frontiers of the objective spaces for) partial solution spaces and filtering out all the dominated objective-value vectors.
In this work we focus on two special cases of dominance-filtering, in which the merged objective space $F$ is created as
	either the \emph{union} $A\cup B$,
	or the \emph{Minkowski sum}\footnote{%
		The \textit{Minkowski sum} $A\oplus B$ contains all the component-wise additions of elements in $A$ and $B$.
If $A = \{(3,5,4), (5,2,1)\}$ and $B = \{(2,1,3), (6,3,2)\}$, then $A \oplus B = \{(5,6,7), (9,8,6), (7,3,4), (11,5,3)\}$.
	}
	$A\oplus B$ of two Pareto sets $A,B$.
These are the two most frequently used variants by solvers of various MOCO problems, e.g.,
	of \emph{decomposition} techniques for multiobjective integer programming  \cite{2019-Schulze-et-al},
	of \emph{Pareto local search} for multiobjective set cover 			\cite{2014-Lust-et-al}, and
	of \emph{dynamic programming} methods for multiobjective shortest paths	(MOSP) \cite{DBLP:journals/cor/PulidoMP15,DBLP:journals/mst/TsaggourisZ09, DBLP:journals/ior/Warburton87},
        multi-objective knapsack \cite{DBLP:journals/ors/EhrgottG00},
		multi-objective vehicle routing 			\cite{DBLP:journals/evi/TanCL23}, or
		multi-objective network design 			\cite{DBLP:journals/eswa/ChenKLK10,DBLP:journals/candie/DehghaniVARS19}.
As manifested in \cite{DBLP:conf/socs/Ulloa0KFS24}, dominance-checking constitutes a major computational burden
of most state-of-the-art algorithms for MOSP problems during the identification of new solutions.
Hence, the development of efficient data structures and algorithms to handle dominance-filtering in unions and Minkowski sums
of Pareto sets is of utmost importance in MOCO problems.

\subparagraph{Related Work and Motivation.}

The literature offers a diverse collection of dominance filtering techniques.
	For $d=2$ objectives, some highly efficient algorithms have been developed \cite{DBLP:conf/esa/Hespe0ST23, DBLP:conf/compgeom/KirkpatrickS85}.
	For the more challenging case of $d\ge 3$ objectives, some general approaches have been explored \cite{DBLP:journals/itor/KerberenesVV23, DBLP:journals/cor/KlamrothLS24}.
	%
	%
	In dynamic settings, where solutions are not known in advance and are revealed gradually, the choice of an indexing data structure plays a crucial role in efficiently updating the Pareto frontier. Several indexing data structures for dominance checking have been proposed in the literature, such as
		\emph{balanced binary search trees} \cite{DBLP:conf/socs/RenZRLC22},
		\emph{ND-trees} \cite{DBLP:journals/tec/JaszkiewiczL18, DBLP:journals/tec/Lang22}, and
		a variant of \emph{k-d trees} \cite{DBLP:journals/cacm/Bentley75, DBLP:journals/comgeo/ChenHT12}.
To the best of our knowledge, the most efficient algorithms
for dominance-filtering of unions and Minkowski sums of Pareto sets for $d\ge 3$ objectives appear in \cite{DBLP:journals/cor/KlamrothLS24}. These methods utilize \emph{space-partitioning ND-trees} \cite{DBLP:journals/tec/JaszkiewiczL18, DBLP:journals/tec/Lang22},
or \emph{divide-and-conquer} strategies.
Despite their effectiveness, these methods suffer from an inherent inefficiency that occurs when the input data emerge from real-world scenarios that typically contain \emph{plateaus} (large collections of objective vectors with identical values in one or more dimensions, e.g., tolls in road networks), and/or are correlated (e.g., distance and time in road networks).
In such cases, ND-trees turn out to be highly unbalanced, which results in significant time bottlenecks for the elementary operations of removing dominated elements from an ND-tree and of re-balancing the tree.
	%

\subparagraph{Our Contribution.}
\label{subsection: contribution}

This work focuses on dominance-filtering techniques for unions and Minkowski sums of Pareto sets for $d\ge 3$ optimization criteria.
Our first contribution are three new data structures for indexing sets of non-dominated elements, which are custom-tailored
to overcome the critical bottlenecks of the algorithms in \cite{DBLP:journals/cor/KlamrothLS24}:
	\noindent (1) \textbf{ND$^{+}$-trees},
		which inherit some desirable features of
    		$k$-$d$ trees \cite{DBLP:journals/cacm/Bentley75} and
    		ND-trees \cite{DBLP:journals/tec/JaszkiewiczL18, DBLP:journals/tec/Lang22}.
	\noindent (2) \textbf{QND$^{+}$-trees},
    		which dynamically adapt partitioning techniques when constructing the indexing tree from a given Pareto set, selecting the most suitable splitting method for each case. This ensures a \emph{provably} balanced tree structure, leading to faster dominance-checks while also achieving dimensionality reduction, whenever this is possible.
	\noindent (3) \textbf{TND$^{+}$-trees}
		which are specially designed for scenarios where large \emph{plateaus} occur that cause severe imbalances, which the TND$^{+}$-trees mitigate while also achieving dimensionality reduction, whenever this is possible.

Our second contribution concerns three new algorithms for dominance-filtering of unions and/or Minkowski sums of two Pareto sets for $d\ge 3$ optimization criteria.
	\noindent (1) \texttt{PlainNDred}
		reduces the problem's dimensionality by lexicographically sorting the elements, and eliminates the need for element removals from the data structure.
	\noindent (2) \texttt{PreND}
		constructs an initial tree from a subset of the Pareto set, thereby reducing the need for frequent re-balancing, and avoids element removals.
	\noindent (3) \texttt{SymND}
		exploits symmetry to compute non-dominated objective vectors, also avoiding element removals.
\texttt{PlainNDred} and \texttt{PreND} are applicable to both the union and the Minkowski sum of two Pareto sets.
They can also be applied to a single objective space, as pure dominance-checks, to extract its Pareto frontier.
\texttt{SymND} is applicable only to the union of two Pareto sets.
All three algorithms are compatible with each of the aforementioned data structures.

Our final contribution is an extensive experimental evaluation to assess the performance of our algorithms and data structures.
We consider all nine combinations of a filtering algorithm among \texttt{PlainNDred}, \texttt{PreND}, and \texttt{SymND} with an indexing data structure from ND$^{+}$-trees, QND$^{+}$-trees, and TND$^{+}$-trees. We compare them with the state-of-the-art algorithms in \cite{DBLP:journals/cor/KlamrothLS24} for $d\ge 3$ criteria.
For our experimental evaluation, we used
	real-world data sets,
	synthetic data sets similar to those in \cite{DBLP:journals/cor/KlamrothLS24}, and
	new synthetic data sets specifically designed to resemble features of real-world instances.
Our experimental results reveal that our algorithms are very efficient and scale well w.r.t.~both
the number of criteria $d$ and the set sizes across all data sets. Notably, they achieve speedups
	up to 5.9$\times$ on real-world data sets and
	up to 13.2$\times$ on synthetic data sets
against the best-performing algorithms from \cite{DBLP:journals/cor/KlamrothLS24}.

\section{Preliminaries}
\label{section: preliminaries}

Let $[n] = \left\{1,2,\ldots,n\right\}$, $\forall n \in \mathbb{Z}^{+}$. In the following, small letters denote scalars,
boldfaced small letters denote vectors, and capital letters denote sets.
For any element or point $\mathbf{p}\in \mathbb{R}^{d}$, let $\mathbf{p}[i]$ denote the value of its $i$-th coordinate, for each $i \in[d]$.
We consider multi-objective minimization problems with $d \geq 2$ objective functions:
\[
	\begin{array}{ll}
	\mathrm{minimize} 	& \mathbf{f}(\mathbf{x}) = \left(~
															\mathbf{f}(\mathbf{x})[1] = f_1(\mathbf{x}),~
															\mathbf{f}(\mathbf{x})[2] = f_2(\mathbf{x}),
															~\ldots~,~
															\mathbf{f}(\mathbf{x})[d]=f_d(\mathbf{x}) ~\right)
	\\
	\mathrm{s.t.} 			& \mathbf{x} \in X
	\end{array}
\]
$X$ is the \emph{solution space}, i.e., the set of feasible solutions for the instance at hand.
\(
	F = \mathbf{f}(X)
	= \left\{\mathbf{p}\in\mathbb{R}^d:\exists\mathbf{x}\in X, \mathbf{p} = \mathbf{f}(\mathbf{x})\right\}
\)
is the corresponding \emph{objective space}, with all $d$-dimensional vectors that appear as \emph{objective-value vectors} for at least one feasible solution from $X$. We refer to these objective vectors simply as \emph{(data) points} and focus on $F\subseteq\mathbb{R}^d$, since all dominance checks are conducted among the points of $F$.

\begin{definition}[Dominance]
	Given two points $\mathbf{p}, \mathbf{p'}$ $\in \mathbb{R}^d$, we say that \emph{$\mathbf{p}$ dominates $\mathbf{p'}$}, denoted as $\mathbf{p} < \mathbf{p'}$, if $\mathbf{p} \neq \mathbf{p'}$ and $\mathbf{p}[i] \leq \mathbf{p'}[i]$, $\forall i \in[d]$.
\end{definition}

\begin{definition}[Pareto frontier/set]
	Given a set $F\subseteq \mathbb{R}^d$, the \emph{Pareto frontier} (a.k.a. the \emph{Skyline}) of $F$ is the maximal subset $P\subseteq F$ of points which are not dominated by any other point in $F$. If $P = F$, then $F$ itself is also called a \emph{Pareto set}.
\end{definition}

\begin{definition}[Minkowski sum]
	Given two sets $A, B \subset \mathbb{R}^d$,
	their \emph{Minkowski sum} is defined as $A \oplus B = \{ \mathbf{a} + \mathbf{b} $ $|$ $  \mathbf{a} \in A, \mathbf{b} \in B \}$.
\end{definition}

\begin{definition}[Pareto union / Pareto sum]
	Given two Pareto sets $A,B\subset\mathbb{R}^d$,
	their \emph{Pareto union} is the Pareto frontier of $A \cup B$, and
	their  \emph{Pareto sum} is the Pareto frontier of $A\oplus B$.
\end{definition}

\begin{definition}[Dominance-filtering]
\label{def:dom-filter}
	Given a set of points $F \subset \mathbb{R}^d$, \emph{dominance-filtering} is the computational problem of filtering out all points in $F$ which are dominated by at least one other point in $F$, so as to construct its Pareto set.
\end{definition}

\section{Algorithmic Background}
\label{sec:algorithmic_background}

A generic approach for dominance-filtering is to process the points of $F=\{\mathbf{p_1},\ldots,\mathbf{p_n}\}$ sequentially,
and keep updating a subset $P$, which will eventually be the Pareto frontier $F$, as follows.
For each new point $\mathbf{p_i}\in F$:
	Compare $\mathbf{p_i}$ (sequentially) with each point $\mathbf{p_j}\in P$ ($j<i$).
	If $\mathbf{p_j} < \mathbf{p_i}$ then reject $\mathbf{p_i}$ (it is dominated by some point in $P$) and proceed with the next point of $F$.
	Otherwise,
		if $\mathbf{p_i} < \mathbf{p_j}$, then remove $\mathbf{p_j}$ from $P$ (it is dominated by $\mathbf{p_i}$);
		if there is no other point in $P$ to compare with, append $\mathbf{p_i}$ to $P$;
		otherwise, proceed with a comparison of $\mathbf{p_i}$ with the next point in $P$.
The efficiency of the data structure used to maintain the current subset $P$ and perform the previously mentioned dominance-checks is critical for the performance of this incremental approach. A well-suited data structure for this task is the ND-tree \cite{DBLP:journals/tec/JaszkiewiczL18}, which we discuss subsequently.

\subsection{ND-trees}

\begin{figure}[ht]
    \centering
    \includegraphics[width=0.8\linewidth]{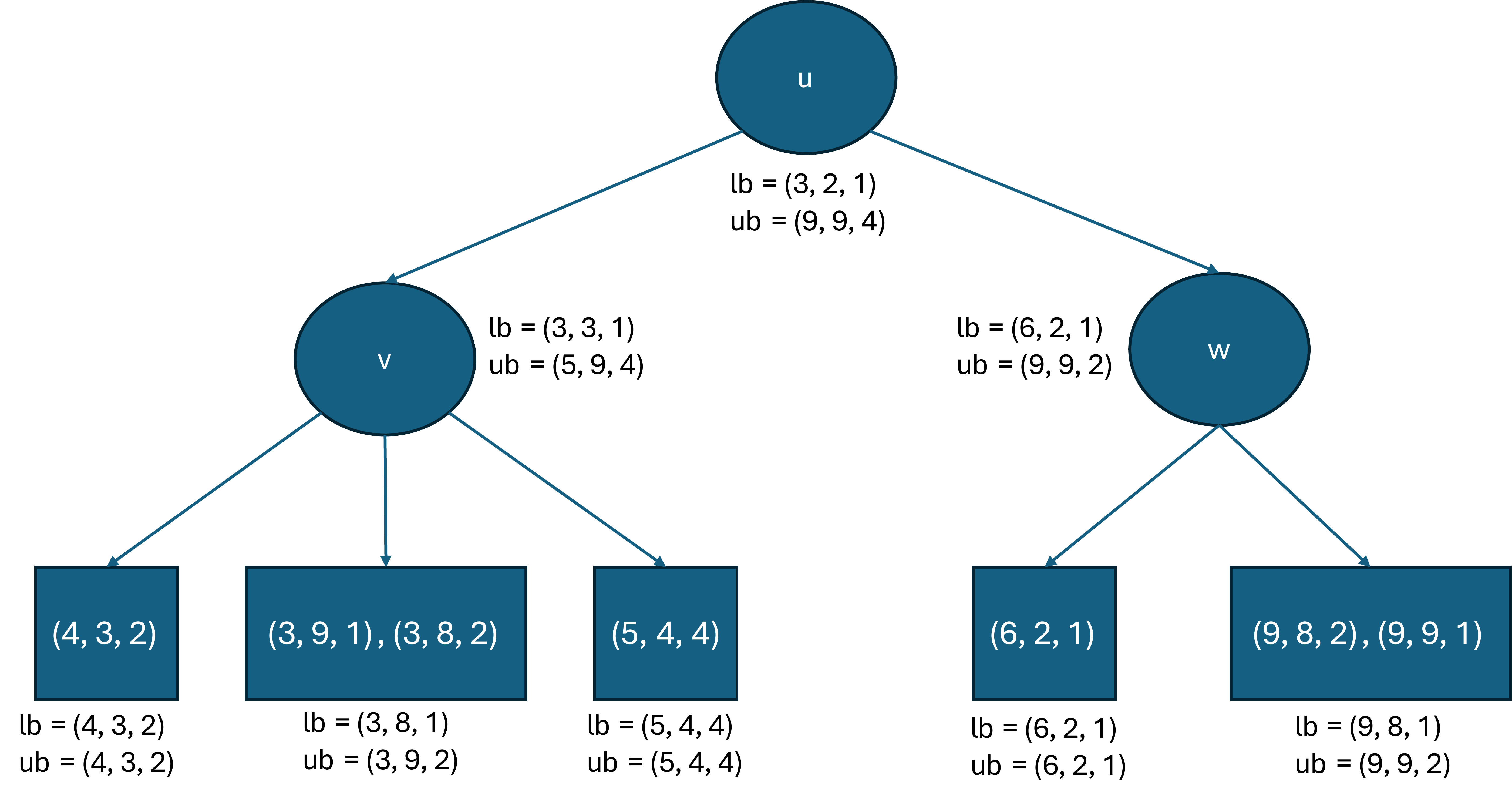}
    \caption{ND-tree containing 3-dimensional points}
    \label{fig:ND-tree}
\end{figure}

An ND-tree is a typical rooted $c$-ary tree $T$, in which a distinct node $r=root(T)$ of degree at most $c$ is the \emph{root node}, all nodes of degree $1$ (except possibly for the root) are its \emph{leaf nodes}, and the remaining nodes of degree from $2$ up to $c+1$ are its \emph{internal nodes}. The ND-trees are \emph{leaf-oriented}, meaning that all data points are stored exclusively in leaf nodes. Each leaf node can store up to $m$ points. The parameters $c$ and $m$ must satisfy the condition $c \leq m + 1$ \cite{DBLP:journals/tec/Lang22}.
Each node $v$ stores a lower-bounding vector $\mathbf{lb_v}$ and an upper-bounding vector $\mathbf{ub_v}$ for all the data points stored in leaves of the subtree $T_v$ of $T$ rooted at $v$. Specifically, for each point $\mathbf{p}$ stored in $T_v$, it holds that $\forall i\in[d], \mathbf{lb_v}[i]\leq \mathbf{p}[i] \leq \mathbf{ub_v}[i]$.
An example of an ND-tree can be found in Figure \ref{fig:ND-tree}.

The lower and upper-bounding vectors are typically used to determine, as early as possible, if a new data point $\mathbf{p}$ is dominated by any data point already in the tree.
For instance,
	if $\exists i\in[d]: \mathbf{p}[i] < \mathbf{lb_v}[i]$, then $\mathbf{p}$ is not dominated by any data point stored in $T_v$, and we do not have to explicitly verify this with all of them (of course, it might still be the case that $\mathbf{p}$ dominates some of these data points).
	If $\mathbf{p} < \mathbf{lb_v}$ then $\mathbf{p}$ dominates all the data points stored in $T_v$.
	Finally, if $\mathbf{p} > \mathbf{ub_v}$, then all the data points stored in leaves of $T_v$ dominate $\mathbf{p}$.
An ND-tree $T$ supports the following operations.
\begin{itemize}

	\item\texttt{NonDomPrune}$(\mathbf{p},T)$:
	This operation effectively utilizes the bounding vectors to perform two tasks,
		a \emph{dominance-check} for a point $\mathbf{p}$ to decide whether it is dominated by any data point in $T$,
		and a \emph{pruning} of $T$ to remove all its data points that are dominated by $\mathbf{p}$.
	If $\mathbf{p}$ is not dominated by any point in $T$, \texttt{NonDomPrune} returns True; otherwise, it returns False.

	\item\texttt{Insert}$(\mathbf{p},v)$:
	This operation inserts a new point $\mathbf{p}$ into a leaf of $T_v$ as follows.
	If $v$ is a non-leaf node,
	then a child node $w$ of $v$ is selected with minimum distance from $\mathbf{p}$. The distance of $\mathbf{p}$ from any node $u$ is the Euclidean distance between $\mathbf{p}$ and $\frac{\mathbf{lb_u}+\mathbf{ub_u}}{2}$ (the center of the bounding box containing all data points stored $T_u$).
	Consequently, $\mathbf{p}$ is recursively requested to be inserted in $T_{w}$.
	For a leaf node $v$,
		if it stores less than $m$ data points,
		then $\mathbf{p}$ is simply appended to its list of stored points;
		otherwise, $v$ is converted into an internal node with $c$ children, and the pending $m+1$ data points are distributed evenly among them.
	Throughout the insertion process, the bounds of all affected nodes are updated accordingly.

	\item\texttt{SPNDBuild$(P)$}:
	This operation, introduced in \cite{DBLP:journals/tec/Lang22}, aims to handle situations in which repeated insertions into an ND-tree might eventually lead to an unbalanced tree structure. It takes as input a Pareto set $P$ (e.g., with all the data points stored in an unbalanced ND-tree) and builds from scratch a \emph{perfectly balanced} ND-tree from it, as no pruning is ever required, in which the bounding areas defined by the upper and lower bounds also \emph{are non-overlapping}.

\end{itemize}

\subsection{ND-Tree based Algorithms for Dominance Filtering}

The following dominance-filtering algorithms, proposed in \cite{DBLP:journals/cor/KlamrothLS24}, are all based on ND-trees and are, to our knowledge, the state-of-the-art techniques for $d\geq 3$ criteria.
\begin{itemize}
	\item\texttt{PlainND}:
This algorithm employs \texttt{NonDomPrune} and \texttt{Insert} to compute either the Pareto union or the Pareto sum of two Pareto sets $A$ and $B$. It begins with an empty ND-tree $T$,  and processes sequentially the points in $F$ (either $A\cup B$, or $A\oplus B$).
	For each point $\mathbf{p} \in F$, it calls \texttt{NonDomPrune}($\mathbf{p},T$).
	If False is returned, $\mathbf{p}$ is discarded.
	Otherwise, it executes \texttt{Insert}($\mathbf{p},T$) to store $\mathbf{p}$ in $T$.
After having processed all points in $F$, the points eventually stored in the leaves of $T$ constitute the Pareto frontier of $F$.

	\item\texttt{PlainSPND}:
This algorithm is similar to \texttt{PlainND}, but it periodically takes the Pareto set $P$ of data items in the current ND-tree, it then destroys the tree, and consequently calls \texttt{SPNDBuild}$(P)$ to create a new, balanced ND-tree. This periodic tree reconstruction can significantly improve the efficiency of intermediate calls to the \texttt{NonDomPrune} and \texttt{Insert} operations, due to limitations in the imbalance of the evolving ND-tree, while the tree reconstruction cost is amortized among consecutive insertion and pruning operations.

	\item\texttt{PruneSPND}:
This algorithm is custom-tailored for computing the Pareto union of two Pareto sets $A$ and $B$. It exploits the fact that points in $A$ may only be dominated by points in $B$, and vice versa. Therefore, for the larger of the two sets (say, $A$) it calls \texttt{SPNDBuild$(A)$} to build a balanced ND-tree $T$. Subsequently, for each point $\mathbf{p}$ in the smaller set (say, $B$), it calls \texttt{NonDomPrune$(\mathbf{p},T)$} to check if $\mathbf{p}$ is dominated and to remove from $T$ all points dominated by $\mathbf{p}$. If $\mathbf{p}$ is dominated, it is removed from $B$.
After having processed all points in $B$, $T$ contains all points of $A$ which are not dominated by any point in $B$, and (eventually) $B$ has only retained those points which are not dominated by any point in $A$. Their union constitutes the Pareto union.
\end{itemize}

\section{New Data Structures for Dominance-Filtering}
\label{sec:new-ds}

We present here our new data structures, ND$^{+}$-trees, QND$^{+}$-trees and TND$^{+}$-trees, designed
to boost the efficiency of dominance-filtering. 

\subsection{\texorpdfstring{ND$^{+}$-trees}{ND+-trees}}

Similar to the ND-trees, the ND$^{+}$-trees are leaf-oriented and each leaf node can store up to $m$ points. Nevertheless, the ND$^{+}$-trees are binary ($c=2$). Additionally, inspired by the k-d trees \cite{DBLP:journals/cacm/Bentley75}, every internal node $v$ is associated with one particular dimension $v.dim\in[d]$ and partitions the set of all the points assigned to its own subtree into two distinct subsets according to their values in this particular dimension $v.dim$. The dimension associated with a node is determined by the modulus of the node's level and the total number of dimensions.
In particular, for $d\geq 2$ dimensions, the root (level 0) node of the tree is associated with dimension $1+(0~\mathbf{mod} ~d) = 1$, its children (level 1 nodes) are associated with dimension $1+(1~\mathbf{mod} ~d)$, and generally, any level-$\ell$ node in the tree is associated with dimension $1+(\ell~\mathbf{mod} ~d)$.
Furthermore, each internal node $v$ stores the median value $v.q$ in the node's associated dimension $v.dim$, chosen from all the points within the subtree rooted at this node. Points in the subtree of $v$ possessing a value less than $v.q$, at dimension $v.dim$, are assigned to $v$'s left subtree, while the remaining points, with a value
greater than or equal to $v.q$, at dimension $v.dim$, are assigned to $v$'s right subtree.
	
\begin{figure}[ht]
	\centering
	\includegraphics[width=0.8\linewidth]{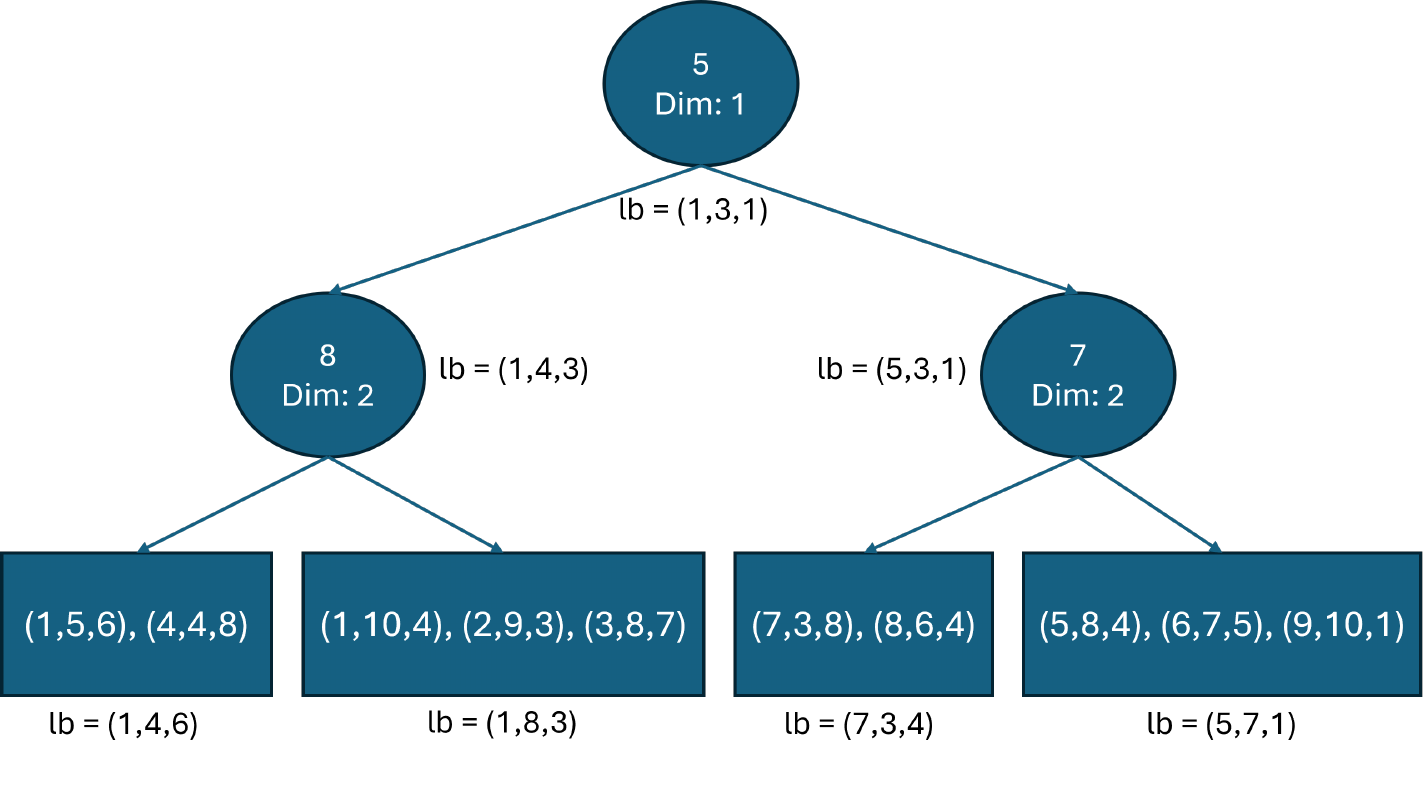}
	\caption{Example of an ND$^+$-tree containing 3-dimensional points with $m = 3$.
}
\label{fig:ND+-tree}
\end{figure}

Contrary to the standard k-d trees, besides the median value in the node's associated dimension, each internal node $v$ of an ND$^{+}$-tree $T$ also maintains the lower-bounding vector ($\mathbf{lb_v}$) of all data points stored in $T_v$, similarly to ND-trees. However, we avoid storing also the upper-bounding vectors, since our experimental evaluation showed that maintaining them is not beneficial for the operations of the ND$^{+}$-trees. An example of an ND$^{+}$-tree is shown in Fig.~\ref{fig:ND+-tree}. We proceed with the description of some elementary operations on ND$^{+}$-trees.

\begin{description}

\item[\texttt{BuildND}$^{+}$($P$, $\ell$, $d$):]
	this method (cf. Algorithm \ref{alg:BuildND}) takes as input a Pareto set $P$ of $n$ points from $\mathbb{R}^d$ and constructs an ND$^{+}$-tree as follows.
	Starting with an empty root $v$ ($\ell=0$), if all points fit within the current node $v$, then $v$ becomes a leaf and stores all these points of $P$.
	Otherwise, we identify and store the median value $v.q$ in the associated dimension $v.dim$ of $v$ (computed using the current level $\ell$ and the number
	of dimensions $d$). The set of points $P$ is then partitioned in two disjoint subsets:
	$L = \left\{\mathbf{p}\in P: \mathbf{p}[v.dim] < v.q\right\}$, and $R = P - L$ (containing the points whose $v.dim$-values are greater than or equal to $v.q$).
	The \texttt{BuildND}$^{+}$ algorithm is then called recursively to construct the left and right subtrees of $v$.

\RestyleAlgo{ruled}
\begin{algorithm}
\small
\SetAlgoLined
\KwIn{Pareto set of Points $P$, the current level $\ell$ and the number of dimensions $d$}
\KwOut{The ND$^{+}$-tree containing all points}
create a new node $v$\;
$v.dim \gets 1 + (\ell\mod d)$\;
\uIf{all points fit into $v$} {
	make $v$ a leaf node and insert all points of $P$ into $v$;
}
\uElse{
	$v.q \gets$ the median value in dimension $v.dim$\;
	Split($P$, $v.q$, $L$, $R$)\;
	$v.left$ $\gets$ BuildND$^{+}$($L$, $\ell+1$, $d$); 	\hfill\tcp{Build  left subtree}
	$v.right$ $\gets$ BuildND$^{+}$($R$, $\ell+1$, $d$); 	\hfill\tcp{Build right subtree}
}
\Return($v$);\
\caption{BuildND$^{+}$($P$, $\ell$, $d$)}
\label{alg:BuildND}
\end{algorithm}

\item[\texttt{ComputeBoundsND}$^{+}$($r$):]
this method (cf.~Algorithm \ref{alg:ComputeBoundsND}) takes as input a node $r$ of a constructed ND$^{+}$-(sub)tree
and computes the lower-bounding vectors for all nodes of $T_r$. In particular, this method is called immediately
after the \texttt{BuildND}$^{+}$ to compute the lower-bounding vectors of the initially constructed ND$^{+}$-tree.
The method works as follows. Each leaf node computes its bounds directly from the $m$ points it stores. Internal
nodes compute their bounds by executing a component-wise minimum operation on their children's lower-bounding vectors.

\RestyleAlgo{ruled}
\begin{algorithm}
\small
\SetAlgoLined
\KwIn{The root $r$ of the tree}
\uIf{$r$ is a leaf node}{
\For{each point $\mathbf{p}$ in $r.points$}{
\For{each dimension $i$}{
$\mathbf{lb_r}[i] \gets min(\mathbf{lb_r}[i], \mathbf{p}[i])$
}
}
}
\uElse{
ComputeBoundsND$^{+}(r.left)$; ComputeBoundsND$^{+}(r.right)$\;
\For{each dimension $i$}{
$\mathbf{lb_r}[i] \gets min(\mathbf{lb_r}[i], \mathbf{lb_{r.left}}[i], \mathbf{lb_{r.right}}[i])$
}
}
\caption{ComputeBoundsND$^{+}(r)$}
\label{alg:ComputeBoundsND}
\end{algorithm}

\item[\texttt{WidenBoundsND}$^{+}(v, \mathbf{p})$:]
this method takes as input a node $v$ and an additional data point $p$ and updates the lower-bounding vector of all nodes in the ND$^{+}$-(sub)tree rooted at $v$, if necessary.
For each dimension $j \in [d]$, if $\mathbf{p}[j] < \mathbf{lb}_v[j]$, then $\mathbf{lb}_v[j] = \mathbf{p}[j]$.


\RestyleAlgo{ruled}
\begin{algorithm}
\small
\SetAlgoLined
\KwIn{Node $v$ and its level $\ell$ in the tree, and the new point $\mathbf{p}$}
\KwOut{The ND$^{+}$-tree containing the new point}
$v.dim \gets$ $1 + (\ell\mod d)$\; 
WidenBoundsND$^{+}$($v$, $\mathbf{p}$)\;
\uIf{$v$ is a leaf node} {
insert $\mathbf{p}$ into $v$\;
\If{$v$ overflows}{
$v.q \gets$ the median value in dimension $v.dim$\;
make $v$ an internal node\;
Split($v.points$, $v.q$, $L$, $R$)\;
$v.left$ $\gets$ BuildND$^{+}$($L$, $\ell+1$, $d$); ComputeBoundsND$^{+}$($v.left$)\;
$v.right$ $\gets$ BuildND$^{+}$($R$, $\ell+1$, $d$); ComputeBoundsND$^{+}$($v.right$)\;
}
}
\lElseIf{$\mathbf{p}[v.dim] < v.q$}{InsertND$^{+}$($v.left$, $\ell+1$, $\mathbf{p}$)}
\lElse{InsertND$^{+}$($v.right$, $\ell+1$, $\mathbf{p}$)}

\caption{InsertND$^{+}$($v$, $\ell$, $\mathbf{p}$)}
\label{alg:InsertND}
\end{algorithm}

\item[\texttt{InsertND}$^{+}(v, \ell,\mathbf{p})$:]
	this method (cf. Algorithm \ref{alg:InsertND}) takes as input a node $v$, its level $\ell$ and a new point $\mathbf{p}$,
and inserts $\mathbf{p}$ into an already constructed ND$^{+}$-(sub)tree $T_v$ as follows.
	Starting from the root, it recursively considers to add $\mathbf{p}$ to the subtree $T_v$ of the current node $v$:
		It first calls \texttt{WidenBoundsND}. Consequently,
		if $v$ is a leaf node with less than $m$ data points, it directly stores $\mathbf{p}$ in $v$'s list. However,
		if $v$ is a leaf node with already $m$ data points in its list, it is converted into an internal node with two children nodes and the $m+1$ data points (including $\mathbf{p}$) are redistributed by \texttt{BuildND}$^{+}$ in $T_v$. Specifically, the median value $v.q$ in $v$'s associated dimension $v.dim = 1+(\ell~\mathbf{mod} ~d)$ is identified, among all previously stored points and the new point $\mathbf{p}$. The left child receives the points with values less than $v.q$, while the right child receives all the points with values greater than or equal to $v.q$. After that, \texttt{ComputeBoundsND}$^{+}$ is executed to calculate the lower-bounding vectors of the two new leaf nodes. Finally,
		if $v$ is an internal node, the value $\mathbf{p}[v.dim]$ is compared with $v.q$ and, based on the result of this comparison, \texttt{InsertND}$^{+}$ is recursively called on the appropriate child of $v$.

\item[\texttt{DominatedND}$^{+}(v, \ell, \mathbf{p})$:]
	this method (cf. Algorithm \ref{alg:DominatedND}) takes as input a node $v$, its level $\ell$ and a new point $\mathbf{p}$,
and decides whether $\mathbf{p}$ is dominated by any other point stored in the ND$^{+}$-(sub)tree $T_v$.
	If $\mathbf{p}[i]<\mathbf{lb_v}[i]$ for some dimension $i\in[d]$, then $\mathbf{p}$ is not dominated by any point in $T_v$. Consequently, there is no need to further examine $T_v$ and the method returns False.
	Similarly, if $\mathbf{p}[v.dim] < v.q$, then $\mathbf{p}$ cannot be dominated by any point in the right subtree of $v$, therefore the method is recursively applied only on $v$'s left child.
	Otherwise, both subtrees of $v$ must be examined recursively for dominance over $\mathbf{p}$.
	When a leaf node is reached and it cannot be determined through the lower bound whether $\mathbf{p}$ is not dominated, a direct comparison of $\mathbf{p}$ with all data points in $v$'s list is necessary.  Given that the maximum number $m$ of points stored in a leaf is sufficiently small, this pairwise comparison remains efficient.

\RestyleAlgo{ruled}
\begin{algorithm}
\small
\SetAlgoLined
\KwIn{The node $v$ and its level $\ell$ in the current tree, and the new point $\mathbf{p}$}
\KwOut{True if the point is dominated, else False}
$v.dim\gets$ $1+(\ell \mod d)$\;
\lIf{$r$ is NULL}{\Return{False}}
\lElseIf{$\exists$ $j\in[1,\dots,d]$ such that $\mathbf{p}[j] < \mathbf{lb_v}[j]$}{\Return{False}}
\uElseIf{$v$ is a leaf node}{
    \lIf{any point in $v$ dominates $\mathbf{p}$}{\Return{True}}
    \lElse{\Return{False}}
}
\lElseIf{$\mathbf{p}[v.dim] < v.q$}{\Return{DominatedND$^{+}$($v.left$, $\ell+1$, $\mathbf{p}$)}}
\lElse{
\Return{DominatedND$^{+}$($v.left$, $\ell+1$, $\mathbf{p}$) $\lor$ DominatedND$^{+}$($v.right$, $\ell+1$, $\mathbf{p}$)}
}
\caption{DominatedND$^{+}$($v$, $\ell$, $\mathbf{p}$)}
\label{alg:DominatedND}
\end{algorithm}
\end{description}
In the remaining part of this section we provide some theoretical guarantees on the complexities of these elementary operations,
when each leaf of the tree stores at most $m$ $d$-dimensional points for (constants) $m,d\in O(1)$.

\begin{theorem}
\label{theorem: BuildND}
Given a Pareto set of $n$ points, \texttt{BuildND}$^{+}$ constructs an ND$^{+}$-tree, with $N=O(n)$ nodes in $O(n \log n)$ time when all splits of a point set produce constant fractions for both parts, and in $O(n^2)$ time otherwise.

\begin{proof}
If all point values are distinct across each dimension, selecting the median (which can be done in linear time using the \texttt{Select} algorithm \cite{DBLP:books/daglib/0023376}) ensures an (almost) even split of the data set at every internal node.
This implies a recurrence of the form $T(n) = 2T(n/2) + O(n)$, which results in time complexity $O(n \log n)$.
For an extreme case where all points have the same value in the current splitting dimension, no meaningful partitioning is possible. However, such a dimension can be ignored, as it provides no new information.
When many (but not all) points in the data set constitute a plateau (i.e., share the same value) around the median value of the splitting dimension, then imbalanced splits may occur, even having $n-o(n)$ points in one part and only $o(n)$ points in the other part.
If this phenomenon appears repeatedly across levels, the resulting tree can have height $O(n)$, yielding the recurrence $T(n) = T(n-o(n)) + O(n)$ which results in time complexity $O(n^2)$.
Recall that since an $N$-node ND$^+$-tree $T$ is a full binary tree (i.e., in which all internal nodes have exactly two children),  it has exactly $L = \frac{N+1}{2}$ leaves. Each leaf must store at least one and at most $m$ of the ($n$ in total) data points of the Pareto set. Therefore, $T$ has $L = \frac{N+1}{2} \in\left\{ \lceil\frac{n}{m}\rceil , n\right\}$ leaves, and $N \in\left\{ 2\lceil\frac{n}{m}\rceil - 1 , 2n - 1  \right\} < 2n$ nodes in total.
\end{proof}
\end{theorem}

It should be noted that the (worst-case) quadratic complexity of \texttt{BuildND}$^{+}$ is rather unlikely for essentially uncorrelated (e.g., chosen uniformly at random) points of the Pareto set. Indeed, this may only happen when the resulting tree resembles the shape of a ``fat'' path, which is rather unlikely to occur in practice. Nevertheless, not so extreme unbalancing may indeed happen, e.g., when large plateaus exist in the coordinate values of the objective vectors.

\begin{theorem}
\label{theorem:NDplus-operations}
Given an ND$^{+}$-tree with $N$ nodes, $n$ points and height $h$, the following bounds hold for its elementary operations:
(i) \texttt{ComputeBoundsND}$^{+}$ takes $O(nmd) = O(n)$ time;
(ii) \texttt{InsertND}$^{+}$ takes $O(hd+m)=O(h)$ time;
(iii) \texttt{DominatedND}$^{+}$ takes $O(dn) = O(n)$ time.

\begin{proof}
(i) Each leaf node of the ND$^{+}$-tree computes its lower-bounding vector by taking the component-wise minimum over the $m$ points it stores,
which takes $O(md)$ time. Internal nodes compute their bounds by taking the component-wise minimum of the bounds of their two children,
which takes $O(d)$ time per node. Since the tree contains $N = O(n)$ nodes in total, as shown in Theorem \ref{theorem: BuildND}, and each node is visited exactly once during a post-order traversal, the total time for all computations is $O(n m d) = O(n)$.

(ii) The insertion of a point $\mathbf{p}$ into an ND$^{+}$-tree requires traversing a path from the root to the appropriate leaf node.
    At each node $v$ along this path, the lower-bounding vector is updated via \texttt{WidenBoundsND}$^{+}$, which takes $O(d)$ time. Therefore, the traversal step requires $O(hd)$ time, where $h$ is the height of the tree. When a leaf node overflows, it is split into two children and its $m+1$ points (including the new point) are redistributed between them. Identifying the median, partitioning these points and computing the lower bounds of the new children takes $O(m)$ time.
    Hence, the overall insertion time is $O(hd + m)$.

(iii) Dominance checks for pairs of $d$-dimensional points requires $O(d)$ time. In the worst case, no pruning will occur and the lower-bounding vectors will provide no conclusive information during the execution of \texttt{DominatedND}$^{+}$, thus the algorithm must examine all $n$ points in the tree.
Consequently, its time complexity is $O(dn)$.
\end{proof}
\end{theorem}

An example of an ND$^{+}$-tree constructed using \texttt{BuildND}$^{+}$ is shown in Figure~\ref{fig:ND+-tree}.

\subsection{\texorpdfstring{QND$^{+}$-trees}{QND+-trees}}

Given a set of $n$ points from $\mathbb{R}^d$ such that all the coordinate values are distinct across every dimension, \texttt{BuildND}$^{+}$ constructs a balanced ND$^{+}$-tree in $O(n\log n)$ time, as the median ensures an even split at each step. 
However, data sets emerging from real-world scenarios rarely adhere to such a strong property. 
Instead, it is common for large subsets of points to possess identical values in certain dimensions, e.g., tolls in road networks.
When multiple points have the same value with the median (i.e., they constitute a large \emph{plateau} around it) of the splitting objective, the resulting partition of the data set may create subsets of (possibly heavily) unequal sizes. 
If this pattern occurs frequently at intermediate nodes of the tree, then the resulting ND$^{+}$-tree  will be heavily unbalanced, leading to quadratic (rather than quasilinear) construction time and linear (instead of logarithmic) time for insertion of new points.
	
To tackle these worst-case performances of the ND$^{+}$-trees, we introduce in this section an alternative data structure, the QND$^{+}$-trees (\emph{Quartile ND$^+$-trees}).
In a nutshell, the QND$^{+}$-trees are also binary, leaf-oriented, and allow up to $m$ data points per leaf. Each level-$\ell$ internal node $v$ is associated with a dimension $v.dim = 1 + (\ell~\mathbf{mod}~d)$ and stores the median value $v.q$, in that dimension, of all data points stored in the subtree $T_v$, along with the corresponding lower-bounding vector ($\mathbf{lb_v}$) for these points.
However, the QND$^{+}$-trees perform a more careful bipartition of the data set to be stored in $T_v$ between $v$'s children. Specifically, when a large plateau of data points (with identical values in dimension $v.dim$ around the median) is detected, it is entirely assigned to the right child (and eventually stored in that subtree), while the remaining data points (outside the plateau) are assigned to the left child of $v$ (and eventually stored in the corresponding subtree). If no plateau is discovered around the median of the splitting dimension, then the data items assigned to $v$ are partitioned between its left and right child depending on their comparison with the value $v.q$.
The advantage of this splitting strategy is that within the right subtree, all points share the same value in dimension $v.dim$. As a result, when performing dominance-checks within this subtree, it is no longer necessary to consider $v.dim$, achieving dimensionality reduction. To indicate whether the right subtree of node $v$ consists entirely of points with identical values, an additional boolean field $v.plateau$ is introduced and set to \texttt{True} in such cases.

	%
In detail, consider a set of $n$ points which are ordered by a particular dimension $k$ (e.g., the set of data points assigned to $v$, ordered by the splitting dimension $k=v.dim$). The \emph{quartiles} $Q_1$, $Q_2$ and $Q_3$ of this set are the $k$-coordinate values of the points residing (roughly) at positions $\frac{n}{4}$ (first quartile), $\frac{n}{2}$ (second quartile, or median) and $\frac{3n}{4}$ (third quartile) in this order, respectively.
We define the following partitioning strategies that are considered by our new dominance-filtering algorithms.
\begin{description}

	\item[\texttt{Median Partitioning (MP)}:]
    	Points with values less than the median are assigned to the left subtree, and the rest to the right subtree.
    	Note that ND$^{+}$-trees always apply this type of partitioning.

	\item[\texttt{Quartile Partitioning (QP)}:]
    	Points with values equal to the median $Q_2$ are assigned to the right subtree, while all others are placed in the left subtree.

\end{description}
By analyzing the quartiles of a multidimensional point set for a particular (e.g., the splitting) dimension, we can efficiently detect plateaus and determine the appropriate partitioning strategy, based on the following cases:
\begin{romanenumerate}

\item $Q_1 = Q_3$:
   At least 50\% of the points constitute a plateau around the median, in this dimension.
   Then \texttt{QP} is preferable, as it effectively reduces the dimensionality for at least half of the data set.

\item $Q_1 = Q_2 < Q_3$:
   At least 25\% but at most 75\% of the points constitute a plateau around the median. However, since \texttt{MP} would split exactly by the value of $Q_2$, if points before $Q_1$ also share the same value as $Q_2$ then we might
   lead to a highly unbalanced structure, even with almost all points stored in the right subtree.
   In contrast, \texttt{QP} ensures a worst-case split of at most 25\%-75\%, maintaining better balance.

\item $Q_1 < Q_2 = Q_3$:
   Again, at least 25\% but no more than 75\% of the points constitute a plateau around the median.
   Since neither partitioning method can create an excessively unbalanced tree, either approach might be used.

\item $Q_1 < Q_2 < Q_3$:
   In this case, neither partitioning method results in a split worse than 75\%-25\%.
\end{romanenumerate}
QND$^+$-trees entail the following elementary operations.
\begin{description}

\item[\texttt{BuildQND$^{+}(P, \ell, d)$}:]
	this method (c.f.~Algorithm \ref{alg:BuildQND}) takes as input a Pareto set $P$ of $n$ points from $\mathbb{R}^d$ and constructs a QND$^{+}$-tree as follows.
	It starts with a new node $v$ (initially level $\ell=0$). If all points fit within $v$, then $v$ becomes a leaf and stores all points of $P$.
	Otherwise, if the data set assigned to $v$ is larger than $m$ points, then $v$ is an internal node and hence a split should occur.
	Then, the method first computes the quartiles $Q_1, Q_2$ for dimension $v.dim = 1+(\ell~\mathbf{mod} ~d)$ and applies either \texttt{QP} (when $Q_1 = Q_2$) or \texttt{MP} (when $Q_1 < Q_2$) to bipartite these data points to the appropriate children. When $Q_1 = Q_2$, this dimension is excluded from subsequent partitioning within the right subtree, as it no longer contributes meaningful separation. The process then recurses on $v$'s children. Otherwise, $v$ is a leaf node and no recursive call is made.
	After completing the tree construction, it calls \texttt{ComputeBoundsND$^{+}$} to compute the lower bounds of all the nodes in the tree.

\begin{algorithm}
	\small
	\SetAlgoLined
	\KwIn{Pareto set of Points $P$, the current level $\ell$ and the number of dimensions $d$}
	\KwOut{The QND$^{+}$-tree containing all points}
	create a new node $v$\;
	$v.dim \gets 1 + (\ell\mod d)$\;
	\lIf{all points fit into $v$}{make $v$ a leaf node and insert all points of $P$ into $v$}
	\uElse{
			get quartiles $Q_1$, $Q_2$ based on dimension $v.dim$\;
			$v.q \gets$ $Q_2$\;
			\uIf{$Q_1 = Q_2$}{
				$L, R \gets$ Quartile Partitioning of $P$\;
				$v.plateau \gets True$\;
			}
			\lElse{$L, R \gets$ Median Partitioning of $P$}
		$v.left$ $\gets$ BuildQND$^{+}$($L$, $\ell+1$, $d$)\;
		$v.right$ $\gets$ BuildQND$^{+}$($R$, $\ell+1$, $d$) ignoring $v.dim$ when $v.plateau = True$;
	}
	\Return($v$);\
	\caption{BuildQND$^{+}$($P$, $\ell$, $d$)}
	\label{alg:BuildQND}
\end{algorithm}

\begin{algorithm}[t]
	\small
	\SetAlgoLined
	\KwIn{Node $v$ and its level $\ell$ in the tree, and the new point $\mathbf{p}$}
	\KwOut{The QND$^{+}$-tree containing the new point}
	$v.dim \gets$ $1 + (\ell\mod d)$; WidenBoundsND$^{+}$($v$, $\mathbf{p}$)\;
	\uIf{$v$ is a leaf node} {
	insert $\mathbf{p}$ into $v$\;
	\uIf{$v$ overflows}{
	get quartiles $Q_1$, $Q_2$ based on dimension $v.dim$; $v.q \gets$ $Q_2$\;
	make $v$ an internal node\;
	\uIf{$Q_1 = Q_2$}{
		$L, R \gets$ Quartile Partitioning of the points of $v$\;
		$v.plateau \gets True$\;
	}
	\lElse{
		$L, R \gets$ Median Partitioning of the points of $v$
	}
	$v.left$ $\gets$ BuildQND$^{+}$($L$, $\ell+1$, $d$); ComputeBoundsND$^{+}$($v.left$)\;	
	$v.right$ $\gets$ BuildQND$^{+}$($R$, $\ell+1$, $d$); ComputeBoundsND$^{+}$($v.right$)\;	
	}
	}
	\uElseIf{$\mathbf{p}[v.dim] < v.q \lor (v.plateau = True \land \mathbf{p}[v.dim] \ne v.q)$}{InsertQND$^{+}$($v.left$, $\ell+1$, $\mathbf{p}$)\;}
	\lElse{InsertQND$^{+}$($v.right$, $\ell+1$, $\mathbf{p}$)}
	
	\caption{InsertQND$^{+}$($v$, $\ell$, $\mathbf{p}$)}
	\label{alg:InsertQND}
\end{algorithm}

\item[\texttt{InsertQND$^{+}(v, \ell,\mathbf{p})$}:]
	this method (c.f.~Algorithm \ref{alg:InsertQND}) takes as input a node $v$, its level $\ell$ and a new point $\mathbf{p}$,
	and inserts $\mathbf{p}$ into the QND$^{+}$-tree $T_v$, similarly to the corresponding operation on ND$^{+}$-trees.
	It begins by updating the lower-bounding vector of the current node $v$ using \texttt{WidenBoundsND}$^{+}$.
	If $v$ is an internal node, the placement of $\mathbf{p}$ depends on whether a plateau is detected.
	If $v.plateau$ is \texttt{True}, then $\mathbf{p}$ is redirected to the right child's subtree when $\mathbf{p}[v.dim] = v.q$, else to the left child's subtree.
	Otherwise (i.e., no plateau is detected), 	$\mathbf{p}$ is redirected to the left subtree when $\mathbf{p}[v.dim] < v.q$
	and to the right subtree when $\mathbf{p}[v.dim] \geq v.q$.
	If $v$ is a leaf node with at most $m-1$ stored data points, then $\mathbf{p}$ is directly stored in its list.
	Otherwise, $v$ is converted into an internal node, and the $m+1$ points (including $\mathbf{p}$) are redistributed between its two children using \texttt{BuildQND}$^{+}$.
	Finally, \texttt{ComputeBoundsND}$^{+}$ is executed to update the corresponding lower bounds of the two new children.

\begin{algorithm}
	\small
	\SetAlgoLined
	\KwIn{Node $v$ and its level $\ell$ in the current tree, the new point $\mathbf{p}$ and the set $D$ (initially empty) of dimensions to be ignored}
	\KwOut{True if the point is dominated, else False}
	\lIf{$v$ is NULL}{\Return{False}}
	\lElseIf{$\exists$ $j\in[d]$ such that $\mathbf{p}[j] < \mathbf{lb_v}[j]$}{\Return{False}}
	\uElseIf{$v$ is a leaf node}{
		\lIf{any point in $v$ dominates $\mathbf{p}$ (ignoring dimensions in $D$)}{\Return{True}}
		\lElse{\Return{False}}
	}
	
	\lElseIf{$\mathbf{p}[v.dim] < v.q$}{\Return{DominatedQND$^{+}$($v.left$, $\ell+1$, $\mathbf{p}$, $D$)}}
	\uElseIf{$v.plateau = True$}{
	$D' \gets D$ $\cup$ $v.dim$\;
	\Return{DominatedQND$^{+}$($v.left$, $\ell+1$, $\mathbf{p}$, $D$) $\lor$ DominatedQND$^{+}$($v.right$, $\ell+1$, $\mathbf{p}$, $D'$)\;}
	}
	\lElse{\Return{DominatedQND$^{+}(v.left,\ell+1, \mathbf{p}, D) \lor$ DominatedQND$^{+}(v.right, \ell+1, \mathbf{p}, D)$}}
	
	\caption{DominatedQND$^{+}$($v$, $\ell$, $\mathbf{p}$, $D$)}
	\label{alg:DominatedQND}
\end{algorithm}

\item[\texttt{DominatedQND$^{+}(v,\ell,\mathbf{p}, D)$}:]
	this method is an adaptation of \texttt{DominatedND$^{+}$} on QND$^+$-trees (c.f.~Algorithm \ref{alg:DominatedQND}).
	It takes as input a node $v$, its level $\ell$, a new point $\mathbf{p}$ and a set $D$ of dimensions to be ignored in the dominance-checks (initially $D=\emptyset$).
	The method starts from the root of the tree and checks whether $\mathbf{p}$ is dominated by any other point in the QND$^{+}$-tree as follows:
	At each internal node $v$, if $\mathbf{p} < \mathbf{lb_v}$ (i.e., $\mathbf{p}$ cannot be dominated by any point in (sub)tree $T_v$), then
	the search terminates. Otherwise, the traversal depends on the value of $v.plateau$.
	When $v.plateau = \texttt{True}$, if $\mathbf{p}[v.dim] < v.q$, then $\mathbf{p}$ is examined only in the left subtree; otherwise, both subtrees are explored.
	However, in the right subtree, all points share the same value $v.q$ in dimension $v.dim$, therefore this dimension can be ignored in further dominance-checks, and hence $v.dim$ is added to the set $D$.
	When $v.plateau = \texttt{False}$, if $\mathbf{p}[v.dim] < v.q$,then only the left subtree is explored; otherwise, both subtrees are explored.
	Upon reaching a leaf node, the algorithm first checks the node’s lower-bounding vector.  If this check is inconclusive, a direct comparison with all points stored in the node is performed to determine dominance.
\end{description}

\begin{theorem}
\label{theorem: BuildQND+}
\emph{Given a Pareto set of $n$ points, BuildQND$^{+}$ constructs a QND$^+$-tree with $N=O(n)$ nodes and height $O(\log n + d) = O(\log n)$ in $O( n (\log n + d) )=O( n \log n )$ time.}

\begin{proof}
At each node $v$, the first step is to compute the quartiles $Q_1$ and $Q_2$, ($Q_3$ is not used) in $v.dim$. This is done in linear time using the \texttt{Select} \cite{DBLP:books/daglib/0023376} algorithm as follows: first, execute \texttt{Select} on the entire data set of points assigned to $v$ to compute the
median value $Q_2 = v.q$; then, apply \texttt{Select} to the lower half of the data set (to compute $Q_1$).
Since the time complexity of \texttt{Select} on $n$ points is $O(n)$ time and because in overall \texttt{Select} operates on partitions of subsets of $n$ and $n/2$, respectively, the total cost remains $O(n)$.
	Even when plateaus do exist, the worst-case split per internal node is $75\%-25\%$, yielding (in worst-case) a recurrence
    			$T(n) = T(0.75n) + T(0.25n) + O(n)$
    	where the linear term is due to the linear-time computations of the quartiles.
    	Nevertheless, because each level in the tree constitutes a full partition of the entire data set, the overall work per level for computing quartiles is $O(n)$.
    	As for the recursion depth, it is dominated by a worst-case root-to-leaf path along which the data set size decreases by a factor of $0.75$.
    	Solving $n \cdot (0.75)^h = 1$ gives the worst-case height of the tree: $h \in O(\log n)$.
	Multiplying by $O(n)$ total work per level, we conclude that $T(n) \in O(n\log n)$.
	In the extreme case where almost all (i.e., $n-o(n)$) points constitute a plateau (this is possible only when $Q_1 = Q_3$),
	partitioning might lead to $T(n) = T(n) + T(o(n)) + O(n) \approx T(n) + O(n)$.
	However, for the giant part of $n-o(n)$ points the splitting dimension can be ignored as it adds no new information, i.e., a dimensionality reduction occurs.
	Clearly, the appearance of such giant parts may only happen at most $d$ times in each root-to-leaf path, resulting in an overall height $h \in O(\log n + d)$ and overall time $T(n) \in O(n(\log n + d))$.
    As shown in Theorem \ref{theorem: BuildND}, the tree contains $N=O(n)$ nodes.
\end{proof}
\end{theorem}

\begin{theorem}
\label{theorem:QNDplus-operations}
\emph{Given a QND$^{+}$-tree with $N$ nodes and height $h$, containing $n$ points, then the following time bounds hold for its elementary operations:
(i) \texttt{InsertQND}$^{+}$ takes $O(hd)=O(h)$ time;
(ii) \texttt{DominatedQND}$^{+}$ takes $O(dn) = O(n)$ time.}

\begin{proof}
The proofs of (i) and (ii) are similar to those of Theorem~\ref{theorem:NDplus-operations} (ii) and (iii).
\end{proof}
\end{theorem}


\subsection{\texorpdfstring{TND$^{+}$-trees}{TND+-trees}}

In this section, we introduce the TND$^{+}$-trees (\emph{Ternary ND$^{+}$-trees}), a data structure designed to exploit at the same time both the tree balance and the dimensionality reduction due to the existence of plateaus of data points.

TND$^{+}$-trees are leaf-oriented, allow up to $m$ points per leaf, and each level-$\ell$ node $v$ is associated with a dimension $v.dim = 1 + (\ell~\mathbf{mod}~d)$.
Additionally, $v$ stores the median value $v.q$ in dimension $v.dim$, computed from all data points (eventually) stored in its subtree $T_v$, along with the lower-bounding vector ($\mathbf{lb_v}$) representing all the minimum values among these points.
The key distinction of the TND$^{+}$-trees is that they are not strictly binary. Each tree node may have up to 3 children.
	The left child receives all points $\mathbf{p}$ with $\mathbf{p}[v.dim] < v.q$, while the right child contains all points $\mathbf{p}$ with $\mathbf{p}[v.dim] > v.q$.
	The middle child, created only in the presence of a plateau around the median, stores all points where $\mathbf{p}[v.dim] = v.q$.
This adaptive structure leads to dimensionality reduction since, within the middle subtree, dimension $v.dim$ is constant and can be
ignored during subsequent dominance-checks.
We will refer to this plateau-based partitioning method as a \texttt{TriPartitioning} (TP).
If no plateau is detected, the standard \texttt{Median Partitioning} (MP) is applied, resulting in only two subtrees (left and right).

The following elementary operations are supported for TND$^{+}$-trees.
\begin{description}
\item[\texttt{BuildTND}$^{+}(P,\ell,d)$:]
this method (c.f.~Algorithm \ref{alg:BuildTND}) takes as input a Pareto set $P$ of $n$ points from $\mathbb{R}^d$ and constructs a TND$^{+}$-tree as follows.
It starts with a new node $v$ (initially $\ell=0$). If all points fit within $v$, then $v$ becomes a lead and stores all points of $P$.
Otherwise, if the data set assigned to $v$ is larger than $m$ points, then it proceeds with the computation of the quartiles $Q_1$, $Q_2$, and $Q_3$
on the dimension $v.dim = 1+(\ell~\mathbf{mod} ~d)$ of the current node, to detect the existence of plateaus.
A middle child is introduced when $Q_1 = Q_2$ or $Q_2 = Q_3$, ensuring that at least 25\% of the data set benefits from dimensionality reduction. Within this middle child, $v.dim$ is excluded from future partitioning steps, as it no longer contributes meaningful separation.
	The possible cases are analyzed below:
	\begin{romanenumerate}
	    \item \textbf{$Q_1 = Q_3$:}
	    In this extreme scenario, at least $50\%$ of the points share the same value $v.q$ in dimension $v.dim$.
	    \item \textbf{$Q_1 = Q_2 < Q_3$:}
	    At most $75\%$ of the points form a plateau with value $v.q$ in dimension $v.dim$.
	    The partitioning results in a worst-case split of $0\%-75\%-25\%$, ensuring that at least 25\% of the points are placed in a different subtree.
	    \item \textbf{$Q_1 < Q_2 = Q_3$:}
	    Similarly, up to 75\% of the points may share the same value, but this time, the worst-case split is $25\%-75\%-0\%$. Again, at least 25\% of the data set avoids being part of the plateau subtree.
	    \item \textbf{$Q_1 < Q_2 < Q_3$:}
	    In this case, no plateau around the median is detected, and \texttt{Median Partitioning} is applied. The worst-case split remains $75\%-25\%$ or $25\%-75\%$, maintaining the logarithmic tree depth.
	\end{romanenumerate}
	After completing the tree construction, it calls \texttt{ComputeBoundsTND$^{+}$}, a slightly modified version of \texttt{ComputeBoundsND$^{+}$} that also takes into consideration the middle child, to compute the lower-bounding vectors of each node in the tree.

\begin{algorithm}
\small
\SetAlgoLined
\KwIn{Pareto set of Points $P$, the current level $\ell$ and the number of dimensions $d$}
\KwOut{The TND$^{+}$-tree containing all points}
create a new node $v$\;
$v.dim \gets 1 + (\ell\mod d)$\;
\lIf{all points fit into $v$} {
	make $v$ a leaf node and insert all points of $P$ into $v$
}
\uElse{
        get quartiles $Q_1$, $Q_2$, $Q_3$ based on dimension $v.dim$;        $v.q \gets$ $Q_2$\;
        \uIf{$Q_1 = Q_2 \lor Q_2 = Q_3$}{
            $L, M, R \gets$ TriPartitioning of $P$\;
            $v.plateau \gets True$\;
        }
        \uElse{
            $L, R \gets$ Median Partitioning of $P$\;
            $M \gets \emptyset$\;
        }
	$v.left$ $\gets$ BuildTND$^{+}$($L$, $\ell+1$, $d$); 	\hfill\tcp{Build  left subtree}
    $v.middle$ $\gets$ BuildTND$^{+}$($M$, $\ell+1$, $d$) ignoring $v.dim$; 	\hfill\tcp{Build  middle subtree}
	$v.right$ $\gets$ BuildTND$^{+}$($R$, $\ell+1$, $d$); 	\hfill\tcp{Build right subtree}
}
\Return($v$);\
\caption{BuildTND$^{+}$($P$, $\ell$, $d$)}
\label{alg:BuildTND}
\end{algorithm}

\item[\texttt{InsertTND$^{+}(v,\ell,\mathbf{p})$}:]
this method (cf.~Algorithm \ref{alg:InsertTND}) takes as input a node $v$, its level $\ell$ and a new point $\mathbf{p}$,
and inserts $\mathbf{p}$ to a TND$^+$-(sub)tree $T_v$ rooted at $v$, resembling \texttt{InsertQND$^{+}$}.
	It first updates $v$'s lower-bounding vector using \texttt{WidenBoundsND$^{+}$}, ensuring that the bounding information remains valid.
	If $v$ is an internal node,
	then a recursive call of the method is executed to insert $\mathbf{p}$ to the appropriate child of $v$, taking also into account whether $v$ possesses a middle child.
	If $v$ is a leaf node that stores less than $m$ points,
	then it simply stores $\mathbf{p}$ at $v$'s list of points,
	otherwise $v$ is converted into an internal node and the $m+1$ now pending points (including $\mathbf{p}$) are redistributed among  its (either two or three, depending on the presence of a plateau) children, using \texttt{BuildTND$^{+}$}.
	Upon completion of the insertion process, \texttt{ComputeBoundsTND$^{+}$} is executed to compute the lower bounds of the newly created children.

\begin{algorithm}[t]
	\small
	\SetAlgoLined
	\KwIn{Node $v$ and its level $\ell$ in the tree, and the new point $\mathbf{p}$}
	\KwOut{The TND$^{+}$-tree containing the new point}
	$v.dim \gets$ $1 + (\ell\mod d)$\; WidenBoundsND$^{+}$($v$, $\mathbf{p}$)\;
	\uIf{$v$ is a leaf node} {
	insert $\mathbf{p}$ into $v$\;
	\uIf{$v$ overflows}{
	get quartiles $Q_1$, $Q_2$, $Q_3$ based on dimension $v.dim$; $v.q \gets$ $Q_2$\;
	make $v$ an internal node\;
	\lIf{$Q_1 = Q_2 \lor Q_2 = Q_3$}{
		$L, M, R \gets$ TriPartitioning of the points of $v$
	}
	\uElse{
		$L, R \gets$ Median Partitioning of the points of $v$\;
		$M \gets \emptyset$;
	}
	$v.left$ $\gets$ BuildTND$^{+}$($L$, $\ell+1$, $d$); ComputeBoundsTND$^{+}$($v.left$)\;
	$v.middle$ $\gets$ BuildTND$^{+}$($M$, $\ell+1$, $d$); 	ComputeBoundsTND$^{+}$($v.middle$)\;
	$v.right$ $\gets$ BuildTND$^{+}$($R$, $\ell+1$, $d$); 	ComputeBoundsTND$^{+}$($v.right$)\;
	}
	}
	\lElseIf{$\mathbf{p}[v.dim] < v.q$}{InsertTND$^{+}$($v.left$, $\ell+1$, $\mathbf{p}$)}
	\lElseIf{($\mathbf{p}[v.dim] = v.q \land v.middle$ is not NULL)}{InsertTND$^{+}$($v.middle$, $\ell+1$, $\mathbf{p}$)}
	\lElse{InsertTND$^{+}$($v.right$, $\ell+1$, $\mathbf{p}$)}
	\caption{InsertTND$^{+}$($v$, $\ell$, $\mathbf{p}$)}
	\label{alg:InsertTND}
\end{algorithm}


\begin{algorithm}
	\small
	\SetAlgoLined
	\KwIn{Node $v$ and its level $\ell$ in the current tree, the new point $\mathbf{p}$ and the set $D$ (initially empty) of dimensions to be ignored}
	\KwOut{True if the point is dominated, else False}
	\lIf{$v$ is NULL}{\Return{False}}
	\lElseIf{$\exists$ $j\in[1,\dots,d]$ such that $\mathbf{p}[j] < \mathbf{lb_r}[j]$}{\Return{False}}
	\uIf{$v$ is a leaf node}{
		\lIf{any point in $v$ dominates $\mathbf{p}$ (ignoring dimensions in $D$)}{\Return{True}}
		\lElse{\Return{False}}
	}
	\lElseIf{$\mathbf{p}[v.dim] < v.q$}{\Return{DominatedTND$^{+}$($v.left$, $\ell+1$, $\mathbf{p}$, $D$)}}
		\lElseIf{DominatedTND$^{+}$($v.left$, $\ell+1$, $\mathbf{p}$, $D$)}{\Return{True}}
		\uElseIf{$v.mid$ is not NULL}{
			$D' \gets D$ $\cup$ $v.dim$\;
			\Return{DominatedTND$^{+}(v.middle, \ell+1, \mathbf{p}, D') \lor$ DominatedTND$^{+}(v.right, \ell+1, \mathbf{p}, D)$\;}
		}
		\lElse{\Return{DominatedTND$^{+}(v.right, \ell+1, \mathbf{p}, D)$}}
	\caption{DominatedTND$^{+}$($v$, $\ell$, $\mathbf{p}$, $D$)}
	\label{alg:DominatedTND}
\end{algorithm}

\item[\texttt{DominatedTND$^{+}(v,\ell,\mathbf{p},D)$}:]
this method (c.f.~Algorithm \ref{alg:DominatedTND}) takes as input a node $v$, its level $\ell$, a new point $\mathbf{p}$
and a set $D$ of dimensions to be ignored in the dominance-checks (initially $D=\emptyset$), and determines whether $\mathbf{p}$
is dominated by any other point in the TND$^{+}$-(sub)tree $T_v$ rooted at $v$.
If $\mathbf{p}$ has a smaller value in any dimension than $\mathbf{lb_v}$, it cannot be dominated by any point in $v$'s subtree, and the search terminates.
If $\mathbf{p}[v.dim] < v.q$, only the left subtree of $T_v$ is explored, since the points in the other subtree(s) have greater value than $v.q$ on dimension $v.dim$.
Otherwise, all (including the middle, if it exists) subtrees of $v$ in $T_v$ must be examined, with appropriate recursive calls of the method.
In the middle subtree, all points share the value $v.q$ in dimension $v.dim$, allowing this dimension to be ignored within that subtree, therefore $v.dim$ is added to $D$.
If $v$ is a leaf node, the method first checks the $\mathbf{lb_v}$.
If this check is inconclusive w.r.t. dominance on/by $\mathbf{p}$, a direct comparison with all points stored in $v$'s list is conducted to check dominance on/by $\mathbf{p}$.

\end{description}

\begin{theorem}
\label{theorem: BuildTND}
\emph{Given a Pareto set of $n$ points, \texttt{BuildTND}$^{+}$  constructs an TND$^{+}$-tree with $N=O(n)$ nodes and height $O(\log n + d) = O(\log n)$ in $O( n ( \log n + d ) ) = O( n \log n )$ time.}

\begin{proof}
	At each node $v$, the first step is to compute the quartiles $Q_1$, $Q_2$, and $Q_3$ in dimension $v.dim$.
	As shown in Theorem \ref{theorem: BuildQND+}, this is done in linear time.
	When plateaus exist, the worst-case scenario occurs when the data set is split into subsets of proportions $75\%$, $25\%$ and $0\%$, yielding the recurrence
		$T(n) = T(0.75n) + T(0.25n) + O(n)$
	which again leads to root-to-leaf paths of length at most $h\in O(\log n)$ and, thus, $T(n) = O(n \log n)$, as shown in Theorem \ref{theorem: BuildQND+}.
	In the extreme case where $n-o(n)$ points of $v$ share the same value in dimension $v.dim$ (this may only occur when $Q_1 = Q_2 = Q_3$),
	(TP) results in an almost $0\%-100\%-0\%$ split, leading to $T(n) = T(n-o(n)) + 2T(o(n)) + O(n) \approx T(n) + O(n)$.
	However, as shown in Theorem \ref{theorem: BuildQND+}, this may happen at most $d$ times along a root-to-leaf path, since each time the dimensionality of the giant part is reduced by one. Therefore, again, the time complexity of the method is $O(n (\log n + d))$. Similar to Theorem \ref{theorem: BuildND}, the tree contains $N=O(n)$ nodes.
\end{proof}
\end{theorem}

\begin{theorem}
\label{theorem:TNDplus-operations}
\emph{Given a TND$^{+}$-tree with $n$ nodes and height $h$, then the following time bounds hold for its elementary operations:
(i) \texttt{InsertTND}$^{+}$ takes $O(hd)=O(h)$ time;
(ii) \texttt{DominatedTND}$^{+}$ takes $O(dn) = O(n)$ time.}

\begin{proof}
The proofs of (i) and (ii) are similar to those of Theorem~\ref{theorem:NDplus-operations} (ii) and (iii).
\end{proof}
\end{theorem}

\subsection{Comparison of the New Data Structures}

To illustrate the differences between the three data structures, and to demonstrate how QND$^{+}$- and TND$^{+}$-trees produce more balanced structures than ND$^{+}$-trees in the presence of plateaus, consider building ND$^{+}$-, QND$^{+}$-, and TND$^{+}$-trees from the point set
$S = \{(1,10,2),\allowbreak\ (2,9,6),\allowbreak\ (2,8,7),\allowbreak\ (2,12,0),\allowbreak\ (2,7,8),\allowbreak\ (2,11,1),\allowbreak\ (4,7,4),\allowbreak\ (5,7,3),\allowbreak\ (6,7,2),\allowbreak\ (7,6,1),\allowbreak\ (8,6,0)\}$.
Assume that each leaf node can store up to $m = 4$ points.

\begin{itemize}

	\begin{figure}[ht]
		\centering
		\includegraphics[width=0.87\linewidth]{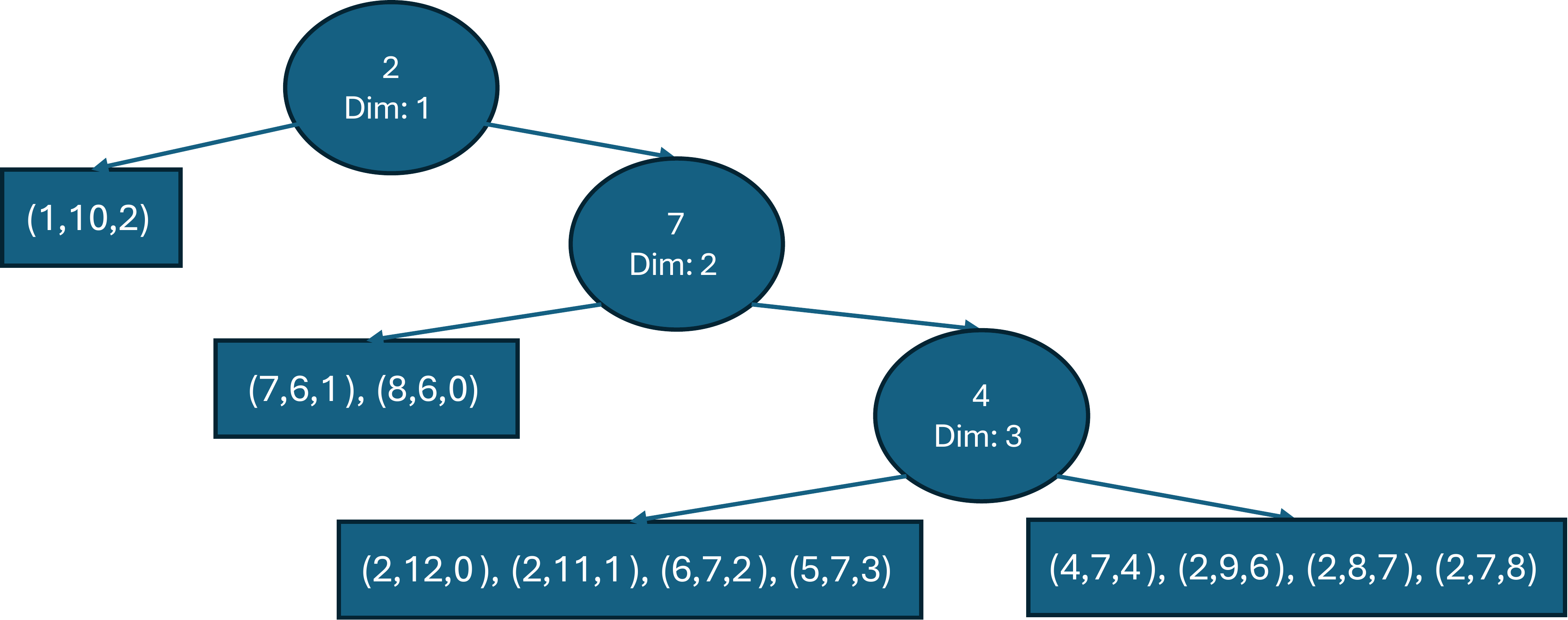}
		\caption{ND$^+$-tree containing 3-dimensional points with $m=4$.}
		\label{fig:nd_plateau}
	\end{figure}

	\item{\textbf{ND$^{+}$-tree:}} Median Partitioning (MP) is first applied in the first dimension, yielding the median value 2. This assigns $(1,10,2)$ to the left subtree and all remaining points to the right, giving $L = \{(1,10,2)\}$ and $R = \{(7,6,1),\allowbreak\ (8,6,0),\allowbreak\ (2,7,8),\allowbreak\ (4,7,4), \allowbreak\ (5,7,3),\allowbreak\ (6,7,2),\allowbreak\ (2,8,7),\allowbreak\ (2,9,6),\allowbreak\ (2,11,1), \allowbreak\ (2,12,0)\}$. MP is then applied to $R$ in the second dimension, with median 7, producing $RL = \{(7,6,1), \allowbreak\ (8,6,0)\}$ and $RR = \{(2,12,0), \allowbreak\ (2,11,1), \allowbreak\ (6,7,2), \allowbreak\ (5,7,3), \allowbreak\ (4,7,4), \allowbreak\ (2,9,6), \allowbreak\ (2,8,7), \allowbreak\ (2,7,8)\}$. Since $RR$ exceeds the leaf size $m$, it is split once more using MP in the third dimension, where the median is 4, giving $RRL = \{(2,12,0), \allowbreak\ (2,11,1), \allowbreak\ (6,7,2), \allowbreak\ (5,7,3)\}$ and $RRR = \{(4,7,4), \allowbreak\ (2,9,6), \allowbreak\ (2,8,7), \allowbreak\ (2,7,8)\}$. All resulting subtrees contain at most $m$ points, completing the tree (see Figure~\ref{fig:nd_plateau}).
	
	\begin{figure}[ht]
		\centering
		\includegraphics[width=0.88\linewidth]{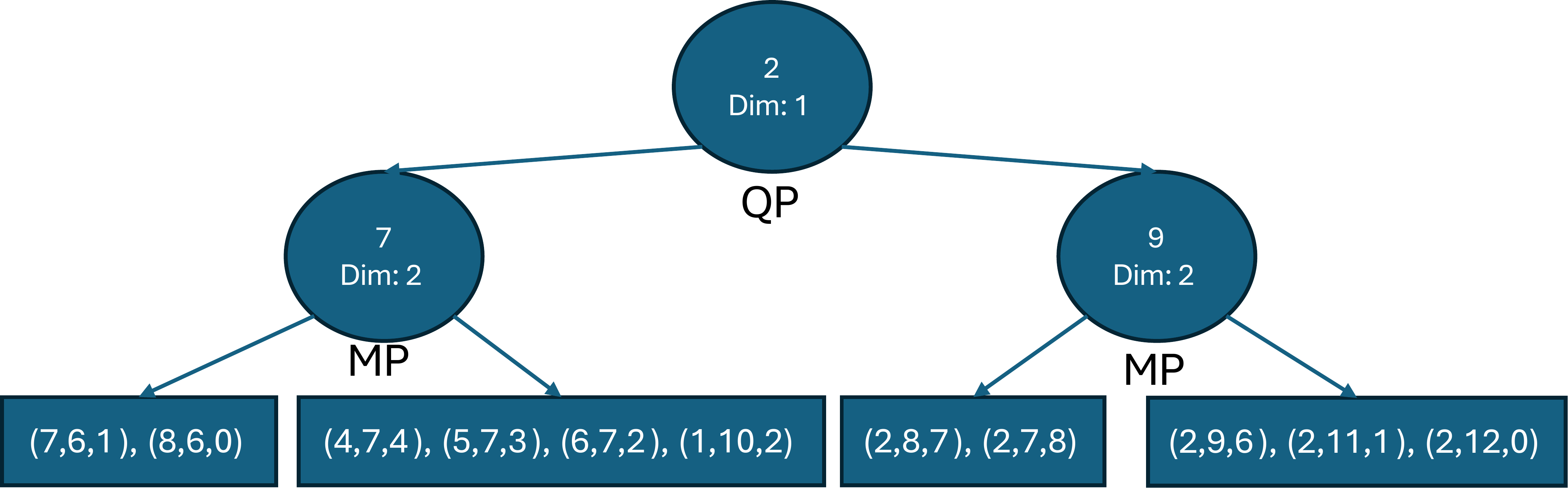}
		\caption{QND$^+$-tree containing 3-dimensional points with $m=4$.}
		\label{fig:qnd_plateau}
	\end{figure}
	
	\item{\textbf{QND$^{+}$-tree:}} Since $Q_1 = Q_2 = 2$ in the first dimension, Quartile Partitioning (QP) is used, yielding $L = \{(7,6,1), \allowbreak\ (8,6,0), \allowbreak\ (4,7,4), \allowbreak\ (5,7,3), \allowbreak\ (6,7,2), \allowbreak\ (1,10,2)\}$ and $R = \{(2,7,8), \allowbreak\ (2,8,7), \allowbreak\ (2,9,6), \allowbreak\ (2,11,1), \allowbreak\ (2,12,0)\}$. In $L$, $Q_1 = 6 \neq 7 = Q_2$ in the second dimension, so MP is applied, splitting it into $LL = \{(7,6,1), \allowbreak\ (8,6,0)\}$ and $LR = \{(4,7,4), \allowbreak\ (5,7,3), \allowbreak\ (6,7,2), \allowbreak\ (1,10,2)\}$, both of which satisfy the leaf size constraint. In $R$, $Q_1 = 8 \neq 9 = Q_2$ in the second dimension, so MP is again applied, giving $RL = \{(2,8,7), \allowbreak\ (2,7,8)\}$ and $RR = \{(2,9,6), \allowbreak\ (2,11,1), \allowbreak\ (2,12,0)\}$, each also within the allowed limit. The tree is thus complete (see Figure~\ref{fig:qnd_plateau}).
	
	\begin{figure}[ht]
		\centering
		\includegraphics[width=1\linewidth]{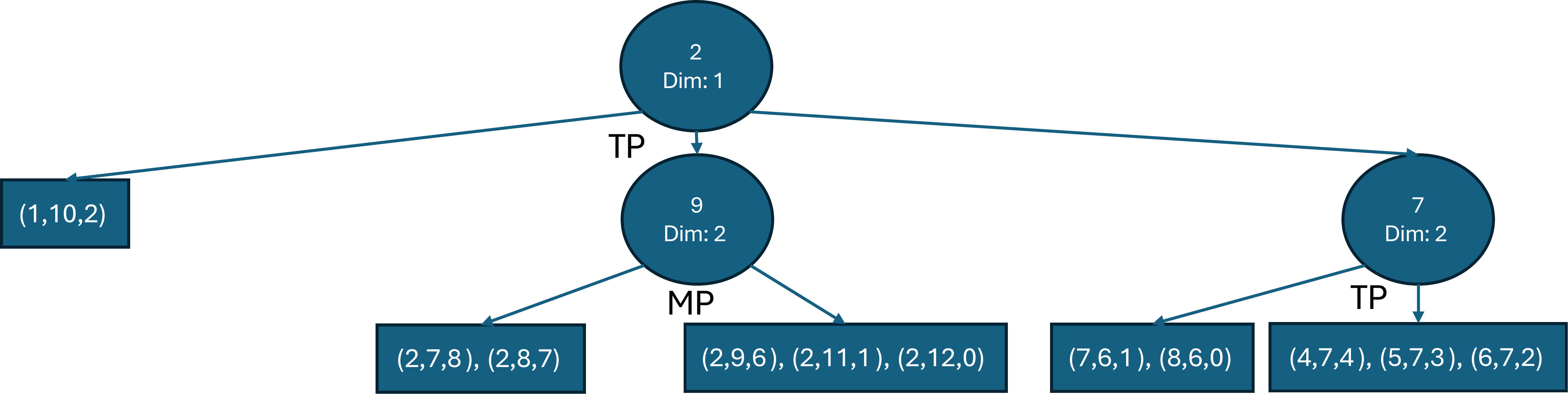}
		\caption{TND$^+$-tree containing 3-dimensional points with $m=4$.}
		\label{fig:tnd_plateau}
	\end{figure}
	
	\item{\textbf{TND$^{+}$-tree:}} $Q_1 = Q_2 = 2$ in the first dimension, so TriPartitioning (TP) is applied, producing $L = \{(1,10,2)\}$, $M = \{(2,7,8), \allowbreak\ (2,8,7), \allowbreak\ (2,9,6), \allowbreak\ (2,11,1), \allowbreak\ (2,12,0)\}$, and $R = \{(7,6,1), \allowbreak\ (8,6,0), \allowbreak\ (4,7,4), \allowbreak\ (5,7,3), \allowbreak\ (6,7,2)\}$. $L$ needs no further processing. In $M$, no plateau is present in the second dimension, so MP is used with median 9, resulting in $ML = \{(2,7,8), \allowbreak\ (2,8,7)\}$ and $MR = \{(2,9,6), \allowbreak\ (2,11,1), \allowbreak\ (2,12,0)\}$. In $R$, a plateau is found: $Q_2=7=Q_3$, so TP is applied again, yielding $RL = \{(7,6,1), \allowbreak\ (8,6,0)\}$ and $RM = \{(4,7,4), \allowbreak\ (5,7,3), \allowbreak\ (6,7,2)\}$. All resulting subtrees respect the leaf size constraint, completing the tree (see Figure~\ref{fig:tnd_plateau}).	
\end{itemize}


\section{New Algorithms for Pareto Unions and Sums}

We now present our new algorithms for computing Pareto unions or Pareto sums of Pareto sets. These algorithms are designed to work
with any of the three data structures of Section~\ref{sec:new-ds}. For simplicity, all algorithms are presented w.r.t.~ND$^{+}$-trees.
In our experimental evaluation
(cf.~Section~\ref{section: experimental evaluation}) we consider all possible combinations of algorithms and data structures.

\subsection{PlainNDred}

As noted in \cite{DBLP:journals/cor/KlamrothLS24}, the main computational burden in \texttt{PlainND}, \texttt{PlainSPND} and \texttt{PruneSPND} is the execution of \texttt{NonDomPrune} operations (cf.~Section~\ref{sec:algorithmic_background}).
The most demanding task is the removal of all dominated points from the tree, as new points are inserted to it.
The main idea behind the \texttt{reduced PlainND} algorithm (\texttt{PlainNDred} in short, cf.~Algorithm \ref{alg:PlainNDred}) is to avoid this costly \emph{pruning} task of the evolving tree, by ensuring that any point that is inserted to the tree is actually a member of the required Pareto frontier of $F$.
To achieve this, \texttt{PlainNDred} first lexicographically sorts $F$, in quasilinear time.
Then, to efficiently manage \emph{dominance-checks}, it processes the data points in that order and uses one of the new indexing structures  (ND$^{+}$-trees in \texttt{PlainNDred}, QND$^{+}$-trees in \texttt{PlainQNDred}, and TND$^{+}$-trees in \texttt{PlainTNDred}) to store only the non-dominated ones of $F$ so far.
In particular, for each point $\mathbf{p}$ in the lexicographic order, the algorithm must only check if it is dominated by any point in the tree, since $\mathbf{p}$ cannot dominate any of the preceding points in that order, as shown next.
\begin{lemma}
	\label{lemma: lex-order non-dominance}
	Let
		$S = \left( \mathbf{p_1}, \mathbf{p_2},\ldots,\mathbf{p_n}\right)$
	be a lexicographic order of a set $F \subset \mathbb{R}^d$ of $n$ points.
	Then the following non-dominance property holds:
	$\forall 1\leq i < j\leq n$, $\mathbf{p_j}$ cannot dominate $\mathbf{p_i}$.

	\begin{proof}
		Due to the lexicographic order, $\forall 1\leq i < j \leq n$,
		$\mathbf{p_i}$ and $\mathbf{p_j}$ may have a prefix of coordinates with the same values,
		but for the first coordinate in which they differ it certainly holds that $\mathbf{p_j}$'s value is larger than that of $\mathbf{p_i}$.
		Therefore, $\mathbf{p_j}$ cannot dominate $\mathbf{p_i}$.
	\end{proof}
\end{lemma}

If $\mathbf{p}$ is not dominated by any point already in the tree, the algorithm inserts it.
	Moreover, \emph{dominance-checks} can safely ignore dimension  $1$, since $\mathbf{p}[1] \geq \mathbf{q}[1]$ for any point preceding $\mathbf{p}$ in the lexicographic order. Therefore, \texttt{PlainNDred} needs only check the remaining $d-1$ dimensions.
	As a result, the algorithm builds a tree considering only the last $d-1$ dimensions of the points in $F$. We denote as $\mathbf{p_{red}}$ the projection of $\mathbf{p}$ on the last $d-1$ dimensions.
	%

\RestyleAlgo{ruled}
\begin{algorithm}
\small
\SetAlgoLined
\KwIn{Set $F$ of points and the number of dimensions $d$}
\KwOut{The non-dominated points of $F$}
Lexicographically sort $F$\;
$T\gets$ empty ND$^{+}$-tree of $d-1$ dimensions\;
\For{each point $\mathbf{p} \in U$ according to the lexicographic order}{
  \lIf{$\lnot$DominatedND$^{+}$($T.root$, $0$, $\mathbf{p_{red}}$)}{
     InsertND$^{+}$($T.root$, $0$, $\mathbf{p_{red}}$)}
}
\Return{all points $\mathbf{p} \in U$ whose $\mathbf{p_{red}}$ projections are in $T$}\;
\caption{PlainNDred($F$, $d$)}
\label{alg:PlainNDred}
\end{algorithm}

\begin{theorem}
	\label{theorem:PlainNDred}
	For an $n$-point set $F$ that is either the Minkowski sum or the union of two Pareto sets, the time complexity of \texttt{PlainNDred} algorithm is $O(n^2(d-1)) = O(n^2)$.

	\begin{proof}
		The points of $F$ are first lexicographically sorted, in time $O(n \log n)$.
		Then, for each point, we perform a dominance-check against the previously processed points that belong to the tree.
		Each pairwise dominance-check takes time $O(d-1)$ since only the last $d-1$ dimensions matter.
		Even if all points in $F$ are non-dominated and no pruning occurs, each point is compared to all previously processed points.
		Therefore, the algorithm makes at most $(d-1)\frac{n(n-1)}{2} \in O((d-1) n^2)$ comparisons, for all dominance-checks.
	\end{proof}
\end{theorem}

\subsection{PreND}

The consecutive insertions into the tree by \texttt{PlainNDred} are likely to gradually unbalance it, diminishing the efficiency of subsequent dominance-checks and insertions. To address this challenge, one option would be to periodically rebuild the tree, as is done by \texttt{PlainSPND} in \cite{DBLP:journals/cor/KlamrothLS24}. However, building the tree from scratch is not a trivial task.
To avoid that, we could leverage once more the lexicographic order of the point set $F=(\mathbf{p_1},\ldots,\mathbf{p_n})$ to compute first an initial subset $P$ of the Pareto frontier, which is then used to construct a \emph{balanced} ND$^{+}$-tree that will be large enough so that the subsequent insertions of the remaining non-dominated points, again examined in lexicographic order, will not be able to unbalance it severely. This is exactly the main idea of the \texttt{presorted ND} algorithm (\texttt{PreND} in short, cf.~Algorithm \ref{alg:PreND}).

To compute this subset of the Pareto frontier, we deploy the \texttt{ParetoSubset} algorithm (cf. Algorithm \ref{alg:ParetoSubset}). It starts with the initialization of a vector $\mathbf{y}$ with infinite values in all dimensions, and then makes a single pass over the data points, in lexicographic order, using $\mathbf{y}$ to keep track of the smallest values seen in each dimension, up to the current point $\mathbf{p_i}$.
For the next point in order, $\mathbf{p_{i+1}}$,
	if there exists a dimension $j\in[d]$ such that $\mathbf{p_{i+1}}[j] < \mathbf{y}[j]$,
	then $\mathbf{p_{i+1}}$ is not dominated by any preceding point. Moreover, due to the lexicographic order, $\mathbf{p_{i+1}}$ cannot be dominated by any subsequent point $\mathbf{p_{k}} : k\geq i+2$ (cf. Lemma~\ref{lemma: lex-order non-dominance}).
	Therefore, $\mathbf{p_{i+1}}$ is certainly a non-dominated point in $F$ and is appended to $P$, and $\mathbf{y}$ is updated to always keep the smallest value seen so far, per dimension.
	Otherwise, when $\mathbf{y} \leq \mathbf{p_{i+1}}$, then $\mathbf{p_{i+1}}$ is appended to another subset $Q$, for further examination.
	Note that, since all points are processed in lexicographic order, $Q$ remains lexicographically sorted.
	Observe also that, for $d=2$, \texttt{ParetoSubset} already computes the entire Pareto frontier of $F$.

\RestyleAlgo{ruled}
\begin{algorithm}
\small
\SetAlgoLined
\KwIn{Set $F$ of points and the number of dimensions $d$}
\KwOut{The Pareto subset $P$ and the rest of the points $Q$}
Lexicographically sort $F$\;
$\mathbf{y}[1 \dots d] \gets (+\infty, \dots, +\infty)$;
$P \gets \emptyset$;
$Q \gets \emptyset$\;
\For{each point $\mathbf{p} \in U$}{
add\_in\_P $\gets$ False\;
  \For{$k=2,\dots,d$}{
    \uIf{$\mathbf{p}[k] < \mathbf{y}[k]$}{
    $\mathbf{y}[k]\gets \mathbf{p}[k]$\;
    add\_in\_P $\gets$ True\;
    }
  }
  \lIf{add\_in\_P}{$P$.append($\mathbf{p}$)}
  \lElse{$Q$.append($\mathbf{p}$)}
}
\Return{$P$, $Q$}\;
\caption{ParetoSubset($F$, $d$)}
\label{alg:ParetoSubset}
\end{algorithm}

\RestyleAlgo{ruled}
\begin{algorithm}
\small
\SetAlgoLined
\KwIn{Set $F$ of points and the number of dimensions $d$}
\KwOut{The non-dominated points of $F$}
$P$, $Q$ = ParetoSubset($F$, $d$)\;
$T \gets$ BuildND$^{+}$($P$, 0, $d$)\;
\For{each point $\mathbf{q} \in Q$}{
  \uIf{$\lnot$DominatedND$^{+}$($T.root$, $0$, $\mathbf{q}$)}{
    InsertND$^{+}$($T.root$, $0$, $\mathbf{q}$)\;
    $P$.append($\mathbf{q}$)\;
  }
}
\Return{$P$}\;
\caption{PreND($F$, $d$)}
\label{alg:PreND}
\end{algorithm}

The \texttt{PreND} algorithm
	initially calls \texttt{ParetoSubset} to get the sets $P$ and $Q$.
	It then calls \texttt{BuildND}$^{+}$ to construct an initial ND$^{+}$-tree with the points of $P$.
	Subsequently, for each point $\mathbf{q}\in Q$,
		it calls \texttt{DominatedND$^{+}$} to determine if $\mathbf{q}$ is dominated by any point of the tree.
		If it is dominated, then it is discarded.
		Otherwise, it is appended to $P$ and inserted to the ND$^{+}$-tree by calling \texttt{InsertND$^{+}$}.
	After having processed all points in $Q$, $P$ is the required Pareto frontier of $F$.
Note that \texttt{PreND} can be used for constructing the Pareto union or the Pareto sum of two Pareto sets, but also for identifying the Pareto frontier of a single set.

\begin{theorem}
	\label{theorem:PreND}
	For an $n$-point set $F$ that is either the Minkowski sum or the union of two Pareto sets, the time complexity of
	\texttt{PreND} is $O(n^2)$.
	\begin{proof}
		\texttt{ParetoSubset} first lexicographically sorts the set of points, in $O(n \log n)$ time.
		Then, it iterates through all $n$ points and for each point determines in $O(d)$ time whether it belongs to $P$ or $Q$.
		Thus, \texttt{ParetoSubset} runs in $O(n \log n + n d) = O(n\log n)$ time.
		Next, an ND$^{+}$-tree is constructed from the points in $P$. In the worst case, this step takes $O(n^2)$ time ($O(n\log n)$ for QND$^{+}$ and TND$^{+}$-trees), as previously described in the context of building ND$^{+}$-trees.
		After constructing the tree, for each point in $Q$, we perform a dominance-check against the points already in the tree. In the worst case, where no pruning occurs and every point is non-dominated, each dominance-check involves comparing the point with all previously processed points. Since the pairwise dominance-checks are executed in time $O(d)$, and we perform this check for at most $\frac{n(n-1)}{2}$ pairs of points, the total time complexity for the dominance-checks is $O(d n^2)$.
		Hence, the overall time complexity of the \texttt{PreND} algorithm is $O(d n^2) = O(n^2)$.
	\end{proof}
\end{theorem}

\subsection{SymND}

Especially for the union of two Pareto sets $A,B$, recall that points of one set may be dominated only by points of the other set. Thus, applying some sort of symmetric dominance-filtering could be extremely efficient. This is exactly what the \texttt{symmetric ND} algorithm (\texttt{SymND} in short) does (cf. Algorithm~\ref{alg:SymND}).
	It constructs first an ND$^{+}$-tree using the points of $A$ and then executes \texttt{DominatedND$^{+}$} for each point in $B$, to remove from $B$ all those points which are dominated by a point in $A$.
	Then, it constructs another ND$^{+}$-tree, using only the remaining points in $B$, and
	executes \texttt{DominatedND$^{+}$} for each point in $A$, to remove those points which are dominated by a point in $B$.
	In the end, both surviving subsets of $A$ and $B$ contain only non-dominated points, and their union constitutes the Pareto union of $A$ and $B$.

\RestyleAlgo{ruled}
\begin{algorithm}
\small
\SetAlgoLined
\KwIn{Pareto sets $A, B$ and the number of dimensions $d$}
\KwOut{The non-dominated points of $A \cup B$}
$T \gets$ BuildND$^{+}$($A$, 0, $d$)\;
\For{each point $\mathbf{p} \in B$}{
  \lIf{DominatedND$^{+}$($T.root$, $0$, $\mathbf{p}$)}{remove $\mathbf{p}$ from $B$}
  }
$T \gets$ BuildND$^{+}$($B$, 0, $d$)\;
\For{each point $\mathbf{p} \in A$}{
  \lIf{DominatedND$^{+}$($T.root$, $0$, $\mathbf{p}$)}{remove $\mathbf{p}$ from $A$}
}
\Return{$A \cup B$} \hfill\tcp{Now $A$ and $B$ contain only non-dominated points}
\caption{SymND($A$, $B$, $d$)}
\label{alg:SymND}
\end{algorithm}

\begin{theorem}
	\label{theorem:SymND}
	Given two Pareto sets $A$ and $B$, the time complexity of \texttt{SymND} to compute their Pareto union is $O(|A|\cdot|B|)$.
	\begin{proof}
		Let $|A|=n_1 \leq |B| = n_2$. Assume also that the size of their Pareto union (to be computed) is $k$.
		First, an ND$^{+}$-tree is constructed using the points of the smaller set $A$, which takes time
			$O(d n_1^2)$ in the worst case for ND$^+$-trees, and
			$O(n_1\log n_1)$ for QND$^{+}$ and TND$^{+}$-trees.
		Then, for each point in set $B$, the \texttt{DominatedND$^{+}$} method is applied.
		Since there are $n_2$ points in set $B$, and each \texttt{DominatedND$^{+}$} operation takes time $O(d n_1)$ in the worst case (when each point of $B$ is compared to all points in the tree), the total complexity of this step is $O(d n_1 n_2)$.
	
		Consequently, we construct a second ND$^{+}$-tree using the remaining (at most $\hat{n}_2 = \min\{k,n_2\}$) points from set $B$. This tree construction requires in worst case time
			$O(d \hat{n}_2^2)$ time for ND$^+$-trees and
			$O(\hat{n}_2\log \hat{n}_2)$ for QND$^{+}$ and TND$^{+}$-trees.
		Next, the \texttt{DominatedND$^{+}$} method is applied for each point in set $A$, removing any dominated points from $A$. As there are $n_1$ points in set $A$, the total complexity of this second phase is $O(d n_1 \hat{n}_2)$.
		Therefore, the overall time complexity of the \texttt{SymND} algorithm is
			$O( d n_1^2 + d n_1 n_2 +  d \hat{n}_2^2 + d n_1 \hat{n}_2 ) = O( d n_1 (n_2 + \min\{k,n_2\}) )$ with ND$^+$-trees and
			$O( d n_1\log n_1 + d n_1 n_2 +  d \hat{n}_2\log \hat{n}_2 + d n_1 \hat{n}_2 ) = O( d n_1 n_2 )$ with QND$^+$-trees and TND$^+$-trees.
	\end{proof}
\end{theorem}

\section{Experimental Evaluation}
\label{section: experimental evaluation}

In our experimental evaluation, we implemented all nine combinations of our proposed indexing data structures and dominance-filtering algorithms. We distinguish each combination with an appropriate naming as follows:
	For each algorithm, its short name is used to indicate an implementation with ND$^{+}$-trees, and
	variants with QND$^{+}$-trees and TND$^{+}$-trees are indicated by the appearance in the short name of the substrings ``QND'' and ``TND'', respectively.
For example,
	\texttt{PlainNDred} indicates the implementation of \texttt{reduced PlainND} with ND$^{+}$-trees,
	\texttt{PreQND} indicates the implementation of \texttt{presorted ND} with QND$^{+}$-trees, and
	\texttt{SymTND} indicates the implementation of \texttt{symmetric ND} with TND$^{+}$-trees.

In addition, we implemented nine algorithms of \cite{DBLP:journals/cor/KlamrothLS24}, which constitute, to the best of our knowledge, the state-of-the-art dominance-filtering algorithms for MOCO problems with $d \geq 3$ dimensions.
Apart from the algorithms \texttt{PlainSPND} and \texttt{PruneSPND} which were discussed in Section~\ref{sec:algorithmic_background},
several more algorithms were provided in \cite{DBLP:journals/cor/KlamrothLS24}:
	\texttt{NonDomDC} explores divide-and-conquer strategies that partition the initial set of solutions into smaller subsets to reduce unnecessary comparisons;
	\texttt{FilterX2} and \texttt{FilterSym} are bidirectional filters that are also based on divide-and-conquer techniques;
	\texttt{BatchedSPND} that utilizes SPND-trees;
	\texttt{LimMem} provides a memory-efficient alternative for scenarios where memory availability is limited;
	finally, \texttt{Doubling(Filter)} and \texttt{Doubling(Tree)} are adaptations of \texttt{FilterSym} and \texttt{PruneSPND}, respectively, which are custom-tailored for Minkowski sums.
All implemented algorithms are listed in Table \ref{tab:table1}, where their applicability on the specific dominance-filtering variant (union and/or Minkowski sum) is also mentioned.
All experiments were conducted on a single core of a 2.2 GHz AMD EPYC 7552 48-Core processor with 256 GB of RAM.
All algorithms were implemented in C++ (g++ v.11.4.0 with -O3 optimization flag).
The reported running times in the result tables provided in Section~\ref{subsec: runtime tables}, are averages over 5 independent runs.

\begin{table}[ht]
    \caption{Algorithms that were implemented and tested in our experimental evaluation.}
    \centering
    	\begin{small}
	\begin{tabular}{|>{	\cellcolor{gray!50!white}}r|>{\cellcolor{gray!25!white}}c|>{\cellcolor{gray!50!white}}r|>{\cellcolor{gray!25!white}}c|}
	    \hline
	    \multicolumn{2}{|c|}{\bf\cellcolor{gray!50!white}Reference: \cite{DBLP:journals/cor/KlamrothLS24}}
	    &
	    \multicolumn{2}{c|}{\bf\cellcolor{gray!50!white}Reference: this work}
	    \\
	    \hline
	    \rowcolor{gray!25!white}
	    \textbf{Algorithm} & \textbf{Usage} 				
	    &
	    \textbf{Algorithm} & \textbf{Usage}
	    \\
	    \hline
	    FilterX2 			& Union 				
	    &
	    SymND 			& Union 		
	    \\
	    FilterSym 		& Union 				
	    &
	    SymQND 		& Union 		
	    \\
	    BatchedSPND 	& Minkowski sum 				
	    &
	    SymTND			& Union 		
	    \\
	    Doubling(Filter) 	& Minkowski sum 				
	    &
	    PreND 			& Union \& Minkowski sum 		
	    \\
	    Doubling(Tree) 	& Minkowski sum 				
	    &
	    PreQND			& Union \& Minkowski sum 		
	    \\
	    LimMem 		& Minkowski sum 				
	    &
	    PreTND 			& Union \& Minkowski sum 		
	    \\
	    NonDomDC 		& Union \& Minkowski sum 	
	    &
	    PlainNDred 		& Union \& Minkowski sum 		
	    \\
	    PlainSPND 		& Union \& Minkowski sum 	
	    &
	    PlainQNDred		& Union \& Minkowski sum
	    \\
	    PruneSPND 		& Union  	
	    &
	    PlainTNDred		& Union \& Minkowski sum 		
	    \\
	    \hline
	\end{tabular}
	\end{small}
    \label{tab:table1}
\end{table}

\subsection{Data Sets}
\label{sec:data sets}
To evaluate the performance of our algorithms, we used both real-world and synthetic data sets.
Note that real-world data sets with three or more objectives ($d \geq 3$) are very rare.
Since our main goal is to test scalability with dimensionality, we have also used two families of synthetic data sets with up to $d=10$ objectives:
	the randomly constructed data sets of \cite{DBLP:journals/cor/KlamrothLS24}, and
	some new, carefully generated synthetic data sets that resemble some crucial features of real-world instances for MOCO problems.
Below, we provide an overview of all these data sets. 
Table~\ref{tab:data-sets} summarizes all data sets used in our experiments along with the corresponding dimensionalities evaluated.
\begin{description}

	\item[RW: Real-world data sets.]
	We used the New York City road network from the 9th DIMACS Implementation Challenge \cite{demetrescu2009shortest}, which contains 264{,}346 nodes and 733{,}846 edges.
	Each edge is associated with two costs: travel time and distance.
	To extend these instances to higher dimensions, we followed established augmentation techniques.
	For three-dimensional instances, we adopted the approach from \cite{DBLP:conf/socs/RenZRLC22}, where a third objective—related to hazardous material transportation \cite{ERKUT2007539}—is introduced.
	For five-dimensional instances, we followed the setup from \cite{ijcai2023p850}, where the fourth objective is a random integer between 1 and 100, and the fifth is a random integer between 1 and the number of edges in the graph.
	To obtain Pareto sets, we applied the publicly available \texttt{Enhanced Multi-objective A$^*$} algorithm (\texttt{EMOA$^*$} in short) \cite{DBLP:conf/socs/RenZRLC22} to various randomly selected source-target pairs.

	In particular, for the Minkowski sum experiment,
			set $A$ contains objective vectors for Pareto-optimal paths from a source node $s$ to an intermediate node $v_1$, computed using \texttt{EMOA$^*$}.
			Set $B$ contains objective vectors for Pareto-optimal paths from $v_1$ to the target node $t$.
			The Pareto sum $S_1$ of $A$ and $B$ then represents non-dominated $s$–$t$ paths that pass through $v_1$.
	Repeating the same process with a different intermediate node $v_2$ yields a second set $S_2$ representing non-dominated $s$–$t$ paths that go through $v_2$.
	Then, for the union experiment, we use the Pareto sets $S_1$ and $S_2$, resulting in a set of non-dominated paths from $s$ to $t$ that pass through $v_1$, $v_2$, or both.
	It is important to choose $v_1$ and $v_2$ such that the resulting sets $S_1$ and $S_2$ are not entirely dominated by one another, ensuring that their union is meaningful. To this end, we selected appropriate intermediate nodes through a trial-and-error procedure.

	We also constructed two additional datasets derived from the real-world instances for the five-dimensional case. In the first variant (RWP), we introduced plateaus in the last two objectives, while in the second variant (RWC), we applied correlation to those same objectives.
	\item[URS: Uniform-Random Synthetic Pareto sets.]
	These were generated by sampling points uniformly in $d$-dimensional space, following the method described in \cite{DBLP:journals/cor/KlamrothLS24}.
	To enforce non-dominance within each set, we project them into the unit sphere by converting each point $\mathbf{p}$ to $\mathbf{p} / ||\mathbf{p}||_{2}$, thus ensuring that the set of points is a Pareto set.
	\item[URSP: Uniform-Random Synthetic Pareto sets with Plateaus.]
	To model more realistic data sets with repeated or flat objective values, we introduced plateaus into the synthetic Pareto sets.
	After generating a set as above, we randomly selected half of the dimensions and assigned identical values to one-fifth of the points in those dimensions.
	This modification breaks the strict non-dominance condition, so we applied a dominance-filtering algorithm to restore the Pareto property.
	To reach the desired number of non-dominated points, we began with a significantly larger initial set and repeated the entire generation process—creating a new random set, introducing plateaus, and applying filtering—until the resulting Pareto set matched the target size.
	\item[URSC: Uniforn-Random Synthetic Pareto sets with Correlations.]
	To simulate objective interdependencies (often encountered in real-world data, e.g., road networks), we introduced correlations among three randomly selected dimensions. Specifically, after generating an initial random set, we modified two of the objectives based on a third: one was made directly proportional to the selected dimension, and another inversely proportional.
	A small amount of noise was added to avoid strict linearity and introduce slight variation.
	Since this transformation can introduce dominated points, we applied a dominance-filtering algorithm to restore Pareto optimality.
	As with the plateau case, we began with a larger initial set and repeated the process—generation, correlation, and filtering—until we obtained a Pareto set of the desired size.
	\item[URSPC: Uniform-Random Synthetic Pareto sets with Plateaus and Correlations.]
	To increase the structural complexity of the data sets, we combined both correlation and plateau features in a single Pareto set.
	First, we selected three random dimensions to introduce correlations, modifying two of them to be directly and inversely related to the third, with small perturbations added for variability.
	Then, we randomly selected one-third of the dimensions and introduced plateaus by assigning identical values to one-fifth of the points within those dimensions.
	Notably, these two modifications were applied independently, allowing a dimension to simultaneously participate in both correlation and plateau transformations.
	As before, a dominance-filtering step was used to ensure Pareto-optimality, and the process—generation, transformation, and filtering—was repeated until the final set reached the desired number of non-dominated points.
\end{description}

\begin{table}[ht]
	\small
    \centering
    \caption{Data sets and dimensionality used in our experimental evaluation}
    \label{tab:data-sets}
    \begin{tabular}{|>{\cellcolor{gray!50!white}}r|>{\cellcolor{gray!30!white}}l|}
        \hline
        \textbf{Data Set} & \textbf{Dimensions} \\
        \hline
		RW    & 3, 5 \\
		RWP   & 5 \\
		RWC   & 5 \\
        URS   & 4, 6, 8, 10 \\
        URSP  & 4, 6, 8, 10 \\
        URSC  & 4, 6, 8, 10 \\
		URSPC & 5, 6, 8, 10 \\
        \hline
    \end{tabular}
\end{table}

\subsection{Real-world data sets}

Table~\ref{tab:graph_sum} reports the running times of all algorithms for computing the Pareto sum of two Pareto sets $A$ and $B$ in 3 and 5 dimensions.
These sets consist of shortest paths extracted from the New York City road network, as described in Section~\ref{sec:data sets}.
In each instance, the sets $A$ and $B$ contain the same number of points, with $n \in \{100,\allowbreak  200,\allowbreak  300,\allowbreak  400,\allowbreak  500,\allowbreak  600,\allowbreak  700,\allowbreak  800,\allowbreak  900,\allowbreak  1000\}$.
The sizes shown in the table correspond to the total number of input points after forming the Minkowski sum, i.e., $n^2$.
Across all instances, our algorithms consistently outperform those of \cite{DBLP:journals/cor/KlamrothLS24}.
\texttt{PlainQNDred} is the fastest in most 3-dimensional cases, while \texttt{PreND} performs best in the 5-dimensional setting.
In particular, \texttt{PlainQNDred} achieves speedups ranging from 1.1$\times$ to 3.6$\times$, and \texttt{PreND} from 1.1$\times$ to 3.8$\times$, compared to the best-performing baseline from \cite{DBLP:journals/cor/KlamrothLS24}.

Table~\ref{tab:graph_union} presents the running times for computing the Pareto union of two sets, $S_1$ and $S_2$, in 3 and 5 dimensions.
As described in Section~\ref{sec:data sets}, each set is derived from a Minkowski sum of two Pareto sets, followed by non-dominance filtering.
This process significantly reduces the size of the resulting sets, which is why our union instances are naturally smaller compared to those in the Minkowski sum experiment.
In each instance, both sets contain the same number of points, with $n \in \{10000,\allowbreak  20000,\allowbreak  30000,\allowbreak  40000,\allowbreak  50000,\allowbreak  60000,\allowbreak  70000,\allowbreak  80000,\allowbreak  90000,\allowbreak  100000\}$.
The sizes shown in the table correspond to the total number of input points after forming the union, i.e., $2n$.
Again, all our algorithms outperform the methods from \cite{DBLP:journals/cor/KlamrothLS24}.
The best performers, \texttt{PreQND} and \texttt{PlainQNDred}, achieve speedups of 2.1$\times$–5.9$\times$ and 2.2$\times$–5.8$\times$, respectively, over the fastest baseline in each case.



For the 5-dimensional setting, we conducted two additional experiments. In the first, we introduced plateaus in the fourth and fifth objectives—those that, following \cite{ijcai2023p850}, are originally assigned random integer values between 1 and 100, and between 1 and the number of edges in the graph, respectively.
In the second experiment, we introduced correlation between the same two objectives, making the 5th objective inversely proportional to the 4th.
Table~\ref{tab:graph_sum_plateau_correlation} reports the running times for computing the Pareto sum in these two modified settings. The input sizes are identical to those used in Table~\ref{tab:graph_sum}. In the plateau scenario, \texttt{PreND} achieves the best performance, with speedups ranging from 2.9$\times$ to 4.4$\times$ over the fastest baseline. In the correlation scenario, \texttt{PlainQNDred} performs best, achieving speedups between 1.4$\times$ and 2.6$\times$.
Table~\ref{tab:graph_union_plateau_correlation} presents the corresponding running times for computing the Pareto union (the sizes are identical to that of Table~\ref{tab:graph_union}).
In this case, \texttt{PreQND} is the top performer in both scenarios, achieving speedups of 2.0$\times$–4.9$\times$ in the plateau case and 1.7$\times$–4.5$\times$ in the correlation case, compared to the best algorithm from \cite{DBLP:journals/cor/KlamrothLS24}.

\subsection{Uniform-Random Pareto sets}

Table~\ref{tab:union_random} presents the running times (in seconds) of all algorithms for computing the Pareto union of two equally sized Uniform-Random Pareto sets in 4, 6, 8, and 10 dimensions. Each set contains $n \in \{25000,\allowbreak  50000,\allowbreak  100000,\allowbreak  250000,\allowbreak  500000\}$ points, resulting in a total input size of $2n$ after forming the union.
In all instances, our algorithms consistently outperform those of \cite{DBLP:journals/cor/KlamrothLS24}. The variants of \texttt{PlainNDred} are the best performers in most cases, though all our methods exhibit similar performance.
Among them, \texttt{PlainQNDred} stands out, achieving the highest speedups—ranging from 1.9$\times$ to 6.8$\times$ compared to the best baseline.

Table~\ref{tab:sum_random} shows the running times (in seconds) for all algorithms on the same dimensional settings (4, 6, 8, and 10), this time for computing the Pareto sum of two equally sized Uniform-Random Pareto sets, with $n \in \{225,\allowbreak  325,\allowbreak  450,\allowbreak  708,\allowbreak  1000\}$.
Once again, the \texttt{PlainNDred} and its variants are the fastest, with \texttt{PlainQNDred} achieving the best results—showing speedups ranging from 2.9$\times$ up to 6.9$\times$ compared to the best-performing baseline from \cite{DBLP:journals/cor/KlamrothLS24}.

\begin{figure}[htbp]
    \centering
    \begin{subfigure}{0.49\textwidth}
        \includegraphics[width=\linewidth]{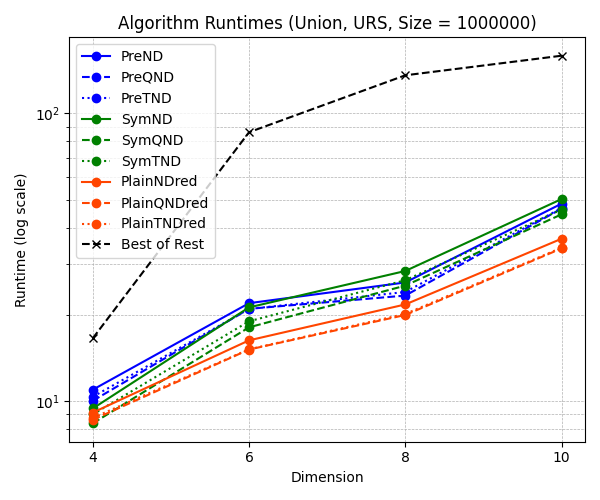}
    \end{subfigure}
    \hfill
    \begin{subfigure}{0.49\textwidth}
        \includegraphics[width=\linewidth]{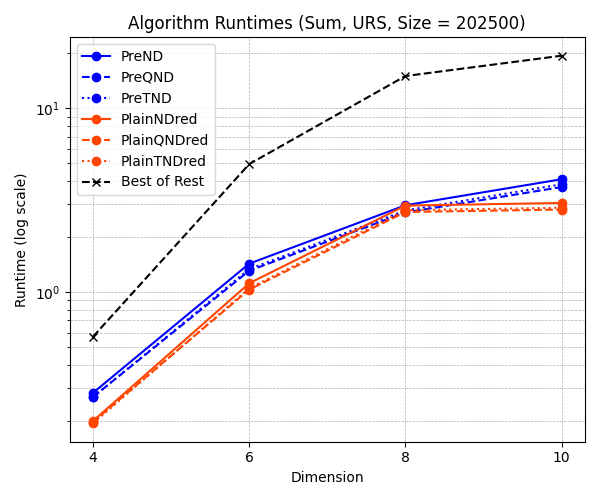}
    \end{subfigure}
    \caption{Runtime of all proposed algorithms across dimensions 4, 6, 8, 10 for the Pareto union (left) and sum (right) operations on URS, evaluated at the largest tested input size in each setting. For each dimension, the baseline corresponds to the best runtime achieved by any method from~\cite{DBLP:journals/cor/KlamrothLS24}.}
	\label{fig:urs}
\end{figure}

Figure~\ref{fig:urs} presents the runtime of our algorithms compared against the best runtime achieved by any baseline algorithm from~\cite{DBLP:journals/cor/KlamrothLS24} for the largest tested input size in each setting.

\subsection{Uniform-Random Pareto Sets with Plateaus}

Tables~\ref{tab:union_plateau} and \ref{tab:sum_plateau} show the running times (in seconds) for all algorithms on 4, 6, 8, and 10 dimensions, for the Pareto union and Pareto sum of two equally sized Uniform-Random Pareto sets with plateaus. For the Pareto union, each set contains $n \in \{5000,\allowbreak 10000,\allowbreak  25000,\allowbreak  50000,\allowbreak  100000\}$ points, and for the Pareto sum $n \in \{100, 150, 225, 325, 450\}$ points.

\begin{figure}[htbp]
    \centering
    \begin{subfigure}{0.49\textwidth}
        \includegraphics[width=\linewidth]{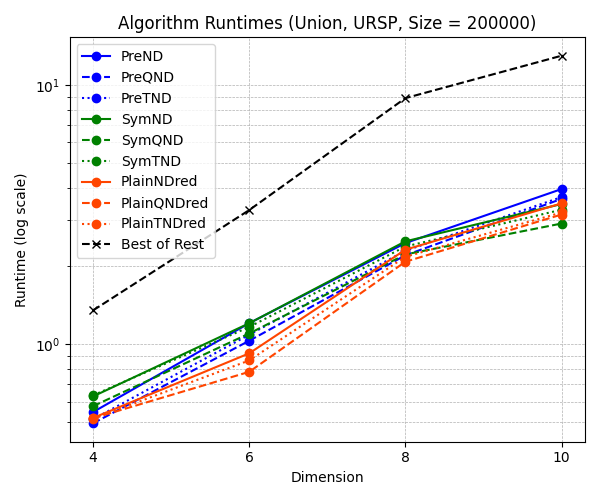}
    \end{subfigure}
    \hfill
    \begin{subfigure}{0.49\textwidth}
        \includegraphics[width=\linewidth]{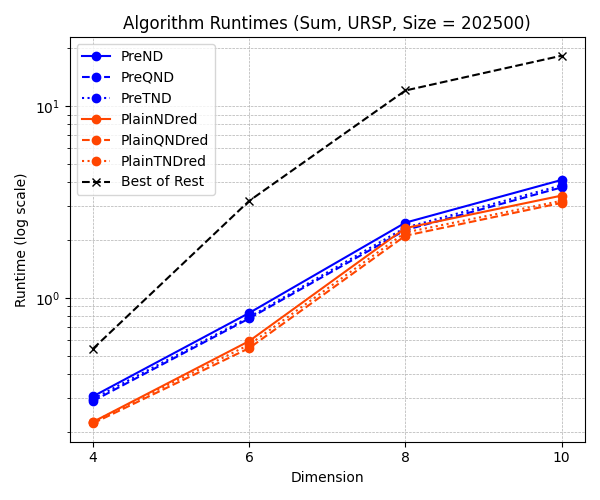}
    \end{subfigure}
    \caption{Runtime of all proposed algorithms across dimensions 4, 6, 8, 10 for the Pareto union (left) and sum (right) operations on URSP, evaluated at the largest tested input size in each setting. For each dimension, the baseline corresponds to the best runtime achieved by any method from~\cite{DBLP:journals/cor/KlamrothLS24}.}
    \label{fig:ursp}
\end{figure}

The algorithms based on QND$^{+}$-trees outperform others. For the union, \texttt{PreQND} achieves speedups ranging from 1.8$\times$ to 4$\times$, \texttt{SymQND} achieves speedups of 1.4$\times$ to 4.4$\times$, and \texttt{PlainQNDred} shows speedups from 1.9$\times$ to 4.3$\times$ compared to the best-performing baseline from \cite{DBLP:journals/cor/KlamrothLS24}. For the sum, \texttt{PreQND} delivers speedups of 1.8$\times$ to 5.3$\times$, while \texttt{PlainQNDred} offers the highest speedups, ranging from 2.4$\times$ to 9.2$\times$.
Notably, in this scenario where plateaus are present in the data set, algorithms utilizing QND$^{+}$-trees and TND$^{+}$-trees consistently outperform those that use ND$^{+}$-trees.
Figure~\ref{fig:ursp} presents runtime plots at the largest tested size for the URSP data set. As before, we compare all proposed algorithms against the best baseline runtime per dimension from~\cite{DBLP:journals/cor/KlamrothLS24}.

\subsection{Uniform-Random Pareto Sets with Correlations}

Table~\ref{tab:union_correlation} and Table~\ref{tab:sum_correlation} present the running times (in seconds) for all algorithms on 4, 6, 8, and 10 dimensions, for the Pareto union and Pareto sum of two equally sized Uniform-Random Pareto sets with correlations. The sizes in these tables are consistent with those in Tables~\ref{tab:union_plateau} and~\ref{tab:sum_plateau}, respectively.
For the union, \texttt{SymQND} consistently outperforms other algorithms, achieving speedups ranging from 1.6$\times$ to 3.9$\times$ compared to \texttt{PruneSPND}, the fastest algorithm of \cite{DBLP:journals/cor/KlamrothLS24} for these data sets. \texttt{PruneSPND} is faster than \texttt{PreND}, \texttt{PlainNDred}, and their QND$^{+}$ and TND$^{+}$ variants in some instances (as shown in the Avg column of Table~\ref{tab:union_urs_speedup_summary}, \texttt{PreND} and \texttt{PlainNDred} achieve identical or better average performance than \texttt{PruneSPND} in most cases). However, \texttt{PruneSPND} never surpasses \texttt{SymND} and its variants in performance.
For the Pareto sum, \texttt{PlainQNDred} (in most cases) and \texttt{PlainTNDred} are the fastest, achieving speedups ranging from 1.7$\times$ to 7.5$\times$ for the former, and from 1.7$\times$ to 7.2$\times$ for the latter.

\begin{figure}[htbp]
    \centering
    \begin{subfigure}{0.49\textwidth}
        \includegraphics[width=\linewidth]{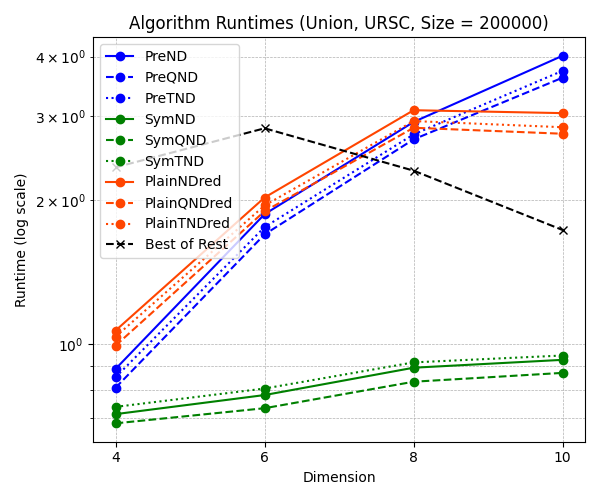}
    \end{subfigure}
    \hfill
    \begin{subfigure}{0.49\textwidth}
        \includegraphics[width=\linewidth]{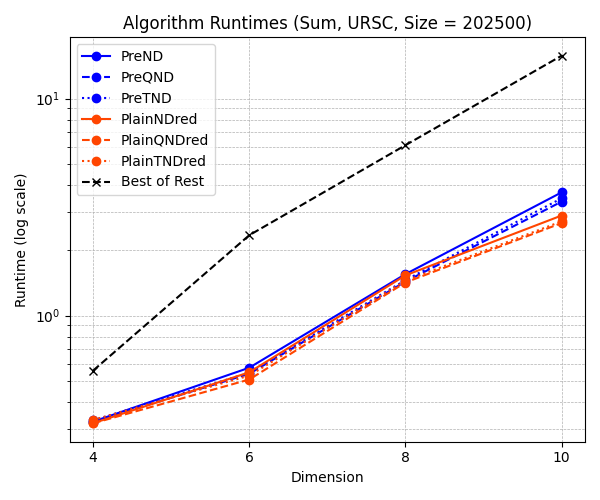}
    \end{subfigure}
    \caption{Runtime of all proposed algorithms across dimensions 4, 6, 8, 10 for the Pareto union (left) and sum (right) operations on URSC, evaluated at the largest tested input size in each setting. For each dimension, the baseline corresponds to the best runtime achieved by any method from~\cite{DBLP:journals/cor/KlamrothLS24}.}
    \label{fig:ursc}
\end{figure}

Figure~\ref{fig:ursc} shows the runtime of our algorithms compared against the best runtime achieved by any baseline algorithm from~\cite{DBLP:journals/cor/KlamrothLS24} for the largest tested input size in each setting.

\subsection{Uniform-Random Pareto sets with Plateaus and Correlations}

Table~\ref{tab:union_plateau_correlation} and Table~\ref{tab:sum_plateau_correlation} show the running times (in seconds) for all algorithms on 5, 6, 8, and 10 dimensions, for the Pareto union and Pareto sum of two equally sized Uniform-Random Pareto sets with both plateaus and correlations. The sizes in these tables align with those in Tables~\ref{tab:union_plateau} and~\ref{tab:sum_plateau}, respectively.
For the union, similar to the previous experiment, \texttt{SymQND} consistently leads in performance, delivering speedups from 1.0$\times$ to 3.7$\times$ over \texttt{PruneSPND}, the fastest algorithm from \cite{DBLP:journals/cor/KlamrothLS24} for these data sets. Although \texttt{PruneSPND} outperforms \texttt{PreND}, \texttt{PlainNDred}, and their QND$^{+}$ and TND$^{+}$ variants in some cases, it never surpasses \texttt{SymND} and its variants (however, as shown in the Avg column of Table~\ref{tab:union_urs_speedup_summary}, \texttt{PreND}, \texttt{PlainNDred} and their variants achieve identical or better average performance than \texttt{PruneSPND}).

\begin{figure}[htbp]
    \centering
    \begin{subfigure}{0.49\textwidth}
        \includegraphics[width=\linewidth]{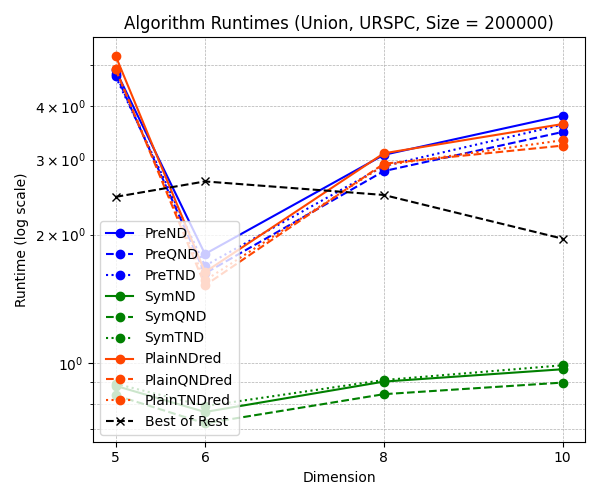}
    \end{subfigure}
    \hfill
    \begin{subfigure}{0.49\textwidth}
        \includegraphics[width=\linewidth]{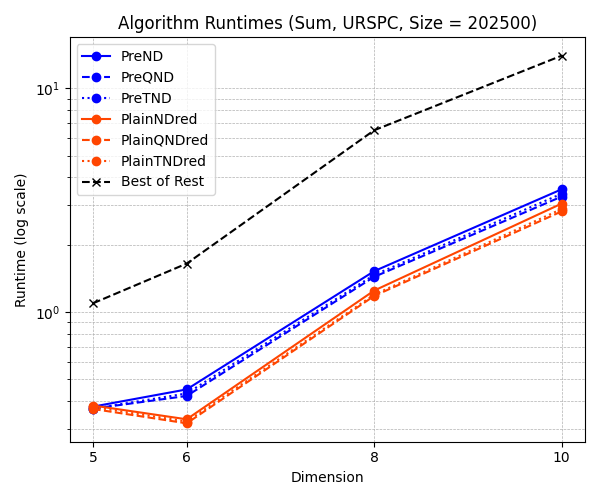}
    \end{subfigure}
    \caption{Runtime of all proposed algorithms across dimensions 5, 6, 8, 10 for the Pareto union (left) and sum (right) operations on URSPC, evaluated at the largest tested input size in each setting. For each dimension, the baseline corresponds to the best runtime achieved by any method from~\cite{DBLP:journals/cor/KlamrothLS24}.}
    \label{fig:urspc}
\end{figure}

For the Pareto sum, \texttt{PlainQNDred} emerges as the fastest, with speedups ranging from 3$\times$ to 13.2$\times$ compared to the best-performing baseline.
Figure~\ref{fig:urspc} shows the runtime of our algorithms compared against the best runtime achieved by any baseline algorithm from~\cite{DBLP:journals/cor/KlamrothLS24} for the largest tested input size in each setting.

\subsection{Overview of Experimental Results}

\subparagraph{Comparison with the state-of-the-art.}
Table~\ref{tab:sum_speedup_summary} summarizes the results of Tables~\ref{tab:graph_sum}, \ref{tab:graph_sum_plateau_correlation}, \ref{tab:sum_random}, \ref{tab:sum_plateau}, \ref{tab:sum_correlation}, and \ref{tab:sum_plateau_correlation}, while Table~\ref{tab:union_speedup_summary} summarizes those of Tables~\ref{tab:graph_union}, \ref{tab:graph_union_plateau_correlation}, \ref{tab:union_random}, \ref{tab:union_plateau}, \ref{tab:union_correlation}, and \ref{tab:union_plateau_correlation}, all of which are included in Section \ref{subsec: runtime tables}.
We compute the speedup factor per (algorithm, data set) pair, as follows:
	For each (dimensionality, input size) pair,
		we identify the fastest algorithm of \cite{DBLP:journals/cor/KlamrothLS24}.
		We then compute the ratio of its runtime to that of our own algorithm.
	In the tables we report the minimum and maximum speedups observed across all (dimensionality, input size) pairs.

\begin{table}[ht]
\centering
\setlength{\tabcolsep}{3pt}
\renewcommand{\arraystretch}{1.2}
\caption{Min and Max speedups for each algorithm for the \emph{Pareto Sum} operation.
}
\begin{small}
\begin{tabular}{|r|cc|cc|cc|cc|cc|cc|cc|}
\hline
\rowcolor{gray!50!white}
\textbf{Algorithm}
& \multicolumn{2}{c|}{\textbf{RW}}
& \multicolumn{2}{c|}{\textbf{RWP}}
& \multicolumn{2}{c|}{\textbf{RWC}}
& \multicolumn{2}{c|}{\textbf{URS}}
& \multicolumn{2}{c|}{\textbf{URSP}}
& \multicolumn{2}{c|}{\textbf{URSC}}
& \multicolumn{2}{c|}{\textbf{URSPC}}
\\
\rowcolor{gray!25!white}
& Min & Max & Min & Max & Min & Max & Min & Max & Min & Max & Min & Max & Min & Max \\
\hline
PreND       & 1.1 & 3.8 & 2.9 & 4.4 & 1.4 & 2.6 & 1.8 & 5.2 & 1.8 & 4.9 & 1.7 & 4.7 & 2.9 & 7.7 \\
PreQND      & 1.0 & 3.7 & 2.8 & 4.3 & 1.4 & 2.5 & 1.7 & 5.6 & 1.8 & 5.3 & 1.7 & 5.2 & 2.9 & 8.2 \\
PreTND      & 1.0 & 3.7 & 2.8 & 4.2 & 1.4 & 2.5 & 1.7 & 5.5 & 1.8 & 5.2 & 1.7 & 5.0 & 2.9 & 7.8 \\
PlainNDred  & 1.1 & 3.7 & 2.6 & 4.5 & 1.4 & 2.5 & 2.9 & 10.8 & 2.4 & 7.0 & 1.7 & 6.2 & 2.9 & 12.2 \\
PlainQNDred & 1.1 & 3.6 & 2.6 & 4.3 & 1.4 & 2.6 & 2.9 & 11.8 & 2.4 & 9.2 & 1.7 & 7.5 & 3.0 & 13.2 \\
PlainTNDred & 1.1 & 3.6 & 2.5 & 4.4 & 1.4 & 2.6 & 3.0 & 8.6  & 2.4 & 8.8 & 1.7 & 7.2 & 2.9 & 12.7 \\
\hline
\end{tabular}
\end{small}
\label{tab:sum_speedup_summary}
\end{table}

\begin{table}[ht]
\centering
\setlength{\tabcolsep}{3pt}
\renewcommand{\arraystretch}{1.2}
\caption{Min and Max speedups for each algorithm for the \emph{Pareto Union} operation.
}
\begin{small}
\begin{tabular}{|r|cc|cc|cc|cc|cc|cc|cc|}
\hline
\rowcolor{gray!50!white}
\textbf{Algorithm}
& \multicolumn{2}{c|}{\textbf{RW}}
& \multicolumn{2}{c|}{\textbf{RWP}}
& \multicolumn{2}{c|}{\textbf{RWC}}
& \multicolumn{2}{c|}{\textbf{URS}}
& \multicolumn{2}{c|}{\textbf{URSP}}
& \multicolumn{2}{c|}{\textbf{URSC}}
& \multicolumn{2}{c|}{\textbf{URSPC}}
\\
\rowcolor{gray!25!white}
& Min & Max & Min & Max & Min & Max & Min & Max & Min & Max & Min & Max & Min & Max \\
\hline
PreND			& 1.9 & 5.5 & 2.0 & 4.2 & 1.7 & 4.1 & 1.5 & 5.2 & 1.7 & 3.6 & 0.4 & 2.6 & 0.5 & 1.5 \\
PreQND		& 2.1 & 5.9 & 2.0 & 4.9 & 1.7 & 4.5 & 1.6 & 5.8 & 1.8 & 4.0 & 0.5 & 2.9 & 0.5 & 1.7 \\
PreTND			& 2.0 & 5.7 & 1.7 & 4.6 & 1.6 & 4.3 & 1.6 & 5.7 & 1.8 & 3.9 & 0.5 & 2.8 & 0.5 & 1.6 \\
SymND			& 1.4 & 3.0 & 1.4 & 3.0 & 1.1 & 2.9 & 1.7 & 4.8 & 1.3 & 3.7 & 1.4 & 3.6 & 1.0 & 3.5 \\
SymQND		& 1.5 & 3.2 & 1.4 & 3.4 & 1.1 & 3.1 & 1.7 & 5.4 & 1.4 & 4.4 & 1.6 & 3.9 & 1.0 & 3.7 \\
SymTND		& 1.4 & 3.1 & 1.3 & 3.1 & 1.1 & 2.9 & 1.6 & 5.2 & 1.3 & 3.9 & 1.4 & 3.5 & 1.0 & 3.4 \\
PlainNDred		& 2.1 & 5.3 & 1.7 & 2.6 & 1.7 & 3.1 & 1.7 & 6.3 & 1.8 & 3.8 & 0.5 & 2.2 & 0.5 & 1.6 \\
PlainQNDred	& 2.2 & 5.8 & 2.3 & 3.0 & 2.0 & 3.6 & 1.9 & 6.8 & 1.9 & 4.3 & 0.6 & 2.4 & 0.5 & 1.6 \\
PlainTNDred	& 2.2 & 5.6 & 2.0 & 2.7 & 2.0 & 3.3 & 1.8 & 6.7 & 1.8 & 4.1 & 0.6 & 2.3 & 0.5 & 1.7 \\
\hline
\end{tabular}
\end{small}
\label{tab:union_speedup_summary}
\end{table}

For Pareto sums (Table \ref{tab:sum_speedup_summary}),
	\texttt{PreND}, \texttt{PlainNDred} and their variants exhibit nearly identical performance, consistently outperforming all other algorithms on RW/RWC/RWP, with speedups reaching up to 4.5$\times$. For the synthetic data sets, all our algorithms significantly outperform the algorithms of \cite{DBLP:journals/cor/KlamrothLS24}, with \texttt{PlainNDred} and its variants achieving the largest speedups, up to 11.8$\times$ for URS, 9.2$\times$ for URSP, 7.5$\times$ for URSC, and 13.2$\times$ for URSPC.

For Pareto unions (Table \ref{tab:union_speedup_summary}),
	the variants of \texttt{PreND} achieve the greatest speedup on RW/RWP/RWC.
	On URS/URSP, all our algorithms show similar speedup ranges.
	On URSC/URSPC, the variants of \texttt{SymND} emerge as the top performers.
    Note that, although the variants of \texttt{PreND} and \texttt{PlainNDred} seem to occasionally be slower
    than \texttt{PruneSPND} on URSC and URSPC, they are faster or identical in average in most cases (cf.~Table \ref{tab:union_urs_speedup_summary}).
	\texttt{SymND} and it variants outperform \texttt{PruneSPND} for all data sets, with speedups of up to $3.9\times$.

Regarding the three data structures,
	for the Pareto union of two Pareto sets, QND$^{+}$-trees and TND$^{+}$-trees outperform ND$^{+}$-trees across almost all data sets. However,
	for the Pareto sum, ND$^{+}$-trees generally perform better than QND$^{+}$-trees and TND$^{+}$-trees on the real-world data sets.
	In contrast, for the synthetic data sets, QND$^{+}$-trees are typically the most efficient, followed by TND$^{+}$-trees, with ND$^{+}$-trees trailing behind.

\subparagraph{Exploring the impact of tree height on algorithmic performances.}

The balance of a tree is crucial for the performance of our algorithms. When the tree is well-balanced, pruning mechanisms at each level can reduce more efficiently the search space and limit the number of candidate points that may dominate a new one. To evaluate this in practice, we conducted an experiment using a URSP set $A$ consisting of $n = 100,000$ points, with a plateau of size $n/2$ around the median in a random dimension. We constructed an ND$^{+}$, a QND$^{+}$ and a TND$^{+}$ tree using their \texttt{Build} methods on $A$. For each tree, we computed the average height, the balance indicator $BI = \max(\text{height}) - \min(\text{height})$, and the number of dominance-checks required per point from a second set $B$ when queried against the respective trees.

\begin{table}[ht]
\centering
\setlength{\tabcolsep}{3pt}
\renewcommand{\arraystretch}{1.2}
\caption{Avg. tree height, balance indicator (BI) and \#dominance-checks for $100K$ points.}
\begin{small}
\begin{tabular}{|r|ccc|ccc|ccc|ccc|}
\hline
\rowcolor{gray!50!white}
\textbf{Tree} & \multicolumn{3}{c|}{\textbf{4 Dimensions}} & \multicolumn{3}{c|}{\textbf{6 Dimensions}} & \multicolumn{3}{c|}{\textbf{8 Dimensions}} & \multicolumn{3}{c|}{\textbf{10 Dimensions}}
\\
\rowcolor{gray!25!white}
 & Height & BI & Checks & Height & BI & Checks & Height & BI & Checks & Height & BI & Checks \\
\hline
ND$^{+}$   & 16 & 10 & 56  & 15 & 7 & 243  & 14 & 4 & 511  & 14 & 2 & 537 \\
QND$^{+}$  & 13 & 0  & 68  & 13 & 0 & 230  & 13 & 0 & 431  & 13 & 0 & 423 \\
TND$^{+}$  & 13 & 1  & 38  & 13 & 1 & 137  & 13 & 1 & 371  & 13 & 1 & 428 \\
\hline
\end{tabular}
\end{small}
\label{tab:tree_height_domchecks}
\end{table}

The results, summarized in Table~\ref{tab:tree_height_domchecks}, show that QND$^{+}$ and TND$^{+}$ trees are consistently more balanced than ND$^{+}$ trees and also exhibit shorter heights. Although ND$^{+}$ trees are only slightly taller on average, their significantly higher balance indicator values reveal a more skewed structure. Consequently—except in the case of 4 dimensions—the ND$^{+}$ trees required more dominance-checks than the other two variants.
It is important to note that, as highlighted in our complexity analysis, the worst-case scenario may require a point to be compared against all nodes in the tree. However, the experimental results indicate that in practice, the number of dominance-checks is substantially lower. This highlights the practical efficiency of our tree structures and the effectiveness of their lower-bounding mechanisms.

\section{Conclusions and Future Work}

We introduced three new data structures and three efficient algorithms for computing the Pareto unions and Pareto sums of Pareto sets, for which we provided a theoretical analysis for their worst-case performances and conducted a thorough experimental evaluation against state-of-art techniques (for $d\geq 3$) from \cite{DBLP:journals/cor/KlamrothLS24}, on several real-world and synthetically generated data sets.
In all instances and dominance-filtering scenarios all of our algorithms consistently outperformed each algorithm in \cite{DBLP:journals/cor/KlamrothLS24}
(except for the case of Pareto unions on URSC and URSPC data sets, in which only \texttt{SymND} and its variants outperformed the best
algorithm in \cite{DBLP:journals/cor/KlamrothLS24}).
Future work will focus on enhancing the performance of \texttt{PreND} by developing an alternative method to \texttt{ParetoSubset}, so as to precompute larger subsets of the Pareto frontier without significantly increasing computational costs.



\bibliography{references}

\appendix
\clearpage

\section{Runtime Tables}
\label{subsec: runtime tables}

This section provides the detailed running times for all nine combinations of our proposed data structures (ND$^{+}$-trees, QND$^{+}$-trees, TND$^{+}$-trees)
and dominance-filtering algorithms (\texttt{PlainNDred}, \texttt{PreND}, and \texttt{SymND}), as well as the running times of the algorithms from \cite{DBLP:journals/cor/KlamrothLS24}, across all datasets (RW, RWP, RWC, URS, URSP, URSC, and URSPC). Input sizes range from 10K to 1M points,
and the number of objectives (dimensions) $d$ varies from 3 to 10.

Tables~\ref{tab:graph_sum}, \ref{tab:graph_sum_plateau_correlation}, \ref{tab:sum_random}, \ref{tab:sum_plateau}, \ref{tab:sum_correlation}, and \ref{tab:sum_plateau_correlation}, present the timings for the Pareto sum of two Pareto sets, while Tables~\ref{tab:graph_union}, \ref{tab:graph_union_plateau_correlation}, \ref{tab:union_random}, \ref{tab:union_plateau}, \ref{tab:union_correlation}, and \ref{tab:union_plateau_correlation} present the timings for the Pareto union of two Pareto sets.
Finally,  speedup statistics—reporting the minimum, maximum, and average speedups for each algorithm, as described in Section~\ref{section: experimental evaluation}—are presented in Tables~\ref{tab:sum_rw_complete_speedup_summary} and \ref{tab:sum_urs_speedup_summary} for the Pareto sum experiments, and in Tables~\ref{tab:union_rw_complete_speedup_summary} and \ref{tab:union_urs_speedup_summary} for the Pareto union experiments. These tables provide an overview of how each of our methods performs relative to the best baseline algorithm from \cite{DBLP:journals/cor/KlamrothLS24} across all relevant datasets.

\begin{table}[ht!]
	\centering
	\setlength{\tabcolsep}{4pt}
	\caption{Running times (in seconds) for all algorithms on 3 and 5  dimensions (objectives) for computing the Pareto sum of two equally sized Pareto sets $A$ and $B$, each containing multi-criteria shortest path solutions from the New York City road network (RW).}
	\begin{small}
\begin{tabular}{|r|rrrrrrrrrr|}
	\hline
	\rowcolor{gray!50!white}
	\textbf{Algorithm} & \multicolumn{10}{c|}{\textbf{Sum Size ($\mathbf{d=3}$, RW)}} \\ 
	\rowcolor{gray!25!white}
	& \textbf{10k} & \textbf{40k} & \textbf{90k} & \textbf{160k} & \textbf{250k} & \textbf{360k} & \textbf{490k} & \textbf{640k} & \textbf{810k} & \textbf{1M} \\ \hline
	NonDomDC        & 0.010 & 0.035 & 0.079 & 0.157 & 0.251 & 0.399 & 0.575 & 0.851 & 1.135 & 1.465 \\
	PlainSPND         & 0.023 & 0.097 & 0.239 & 0.409 & 0.655 & 0.988 & 1.372 & 1.825 & 2.301 & 2.710 \\
	BatchedSPND     & 0.026 & 0.147 & 0.449 & 0.604 & 1.174 & 1.777 & 2.603 & 3.051 & 4.133 & 5.985 \\
	Doubling(Tree) & 0.036 & 0.163 & 0.385 & 0.749 & 1.230 & 1.843 & 2.614 & 3.499 & 4.369 & 5.409 \\
	Doubling(Filter) & 0.024 & 0.099 & 0.233 & 0.457 & 0.736 & 1.125 & 1.605 & 2.173 & 2.755 & 3.373 \\
	LimMem          & 0.013 & 0.050 & 0.113 & 0.209 & 0.334 & 0.514 & 0.745 & 1.061 & 1.384 & 1.740 \\ \hline
	\textbf{PreND} & 0.006 & 0.024 & 0.061 & 0.137 & 0.206 & 0.367 & 0.500 & 0.811 & 0.956 & 1.111 \\
	\textbf{PreQND} & 0.006 & 0.024 & 0.060 & 0.138 & 0.205 & 0.361 & 0.503 & 0.813 & 0.961 & 1.132 \\
	\textbf{PreTND} & 0.006 & 0.024 & 0.061 & 0.139 & 0.209 & 0.368 & 0.510 & 0.826 & 0.979 & 1.144 \\
	\textbf{PlainNDred} & \textbf{0.005} & \textbf{0.022} & \textbf{0.057} & \textbf{0.129} & 0.201 & 0.349 & \textbf{0.478} & 0.753 & 0.926 & 1.114 \\
	\textbf{PlainQNDred} & \textbf{0.005} & \textbf{0.022} & \textbf{0.057} & \textbf{0.129} & \textbf{0.200} & \textbf{0.344} & 0.480 & \textbf{0.742} & \textbf{0.906} & \textbf{1.104} \\
	\textbf{PlainTNDred} & \textbf{0.005} & \textbf{0.022} & 0.058 & 0.139 & 0.204 & 0.355 & 0.484 & 0.763 & 0.939 & 1.139 \\
	\hline
	\end{tabular}
	\end{small}
	
	\vspace{1em}
	
	\begin{small}
	\begin{tabular}{|r|rrrrrrrrrr|}
	\hline
	\rowcolor{gray!50!white}
	\textbf{Algorithm} & \multicolumn{10}{c|}{\textbf{Sum Size ($\mathbf{d=5}$, RW)}} \\
	\rowcolor{gray!25!white}
	& \textbf{10k} & \textbf{40k} & \textbf{90k} & \textbf{160k} & \textbf{250k} & \textbf{360k} & \textbf{490k} & \textbf{640k} & \textbf{810k} & \textbf{1M} \\ \hline
	NonDomDC        & 0.019 & 0.070 & 0.167 & 0.350 & 0.588 & 1.015 & 1.582 & 2.492 & 3.727 & 4.995 \\
	PlainSPND        & 0.024 & 0.086 & 0.239 & 0.431 & 0.609 & 0.982 & 1.230 & 1.652 & 2.052 & 2.430 \\
	BatchedSPND     & 0.022 & 0.109 & 0.329 & 0.496 & 1.051 & 1.424 & 1.684 & 2.419 & 3.079 & 3.972 \\
	Doubling(Tree) & 0.035 & 0.152 & 0.366 & 0.692 & 1.093 & 1.693 & 2.231 & 3.082 & 3.845 & 4.808 \\
	Doubling(Filter) & 0.026 & 0.107 & 0.244 & 0.453 & 0.721 & 1.093 & 1.486 & 1.953 & 2.755 & 3.373 \\
	LimMem          & 0.024 & 0.089 & 0.206 & 0.412 & 0.702 & 1.141 & 1.789 & 2.685 & 3.984 & 5.379 \\ \hline
	\textbf{PreND} & 0.006 & 0.023 & \textbf{0.057} & \textbf{0.128} & \textbf{0.158} & 0.316 & \textbf{0.325} & \textbf{0.660} & 0.693 & \textbf{0.683} \\
	\textbf{PreQND}          & 0.006 & 0.024 & 0.059 & 0.130 & 0.176 & 0.314 & 0.334 & 0.666 & 0.693 & 0.699 \\
	\textbf{PreTND}          & 0.006 & 0.024 & 0.059 & 0.130 & 0.177 & \textbf{0.312} & 0.329 & 0.674 & \textbf{0.692} & 0.697 \\
	\textbf{PlainNDred}       & \textbf{0.005} & \textbf{0.022} & 0.059 & 0.131 & 0.177 & 0.326 & 0.383 & 0.692 & 0.757 & 0.818 \\
	\textbf{PlainQNDred}      & \textbf{0.005} & \textbf{0.022} & 0.060 & 0.131 & 0.176 & 0.340 & 0.371 & 0.772 & 0.772 & 0.816 \\
	\textbf{PlainTNDred}      & \textbf{0.005} & \textbf{0.022} & 0.062 & 0.139 & 0.189 & 0.346 & 0.396 & 0.803 & 0.803 & 0.871 \\
	\hline
	\end{tabular}
	\end{small}
	\label{tab:graph_sum}
\end{table}

\begin{table}
	\centering
	\setlength{\tabcolsep}{4pt}
	\caption{Running times (in seconds) for all algorithms on 3 and 5 dimensions (objectives) for computing the Pareto union of two equally sized Pareto sets $S_1$ and $S_2$, each containing multi-criteria shortest path solutions on the New York City road network (RW).}
	\begin{small}
	\begin{tabular}{|r|rrrrrrrrrr|}
	\hline
	\rowcolor{gray!50!white}
	\textbf{Algorithm} & \multicolumn{10}{c|}{\textbf{Union Size ($\mathbf{d=3}$, RW)}} \\
	\rowcolor{gray!25!white}
	& \textbf{20k} & \textbf{40k} & \textbf{60k} & \textbf{80k} & \textbf{100k} & \textbf{120k} & \textbf{140k} & \textbf{160k} & \textbf{180k} & \textbf{200k} \\ \hline
	NonDomDC        & 0.105 & 0.318 & 0.609 & 1.062 & 1.464 & 1.857 & 2.277 & 2.672 & 3.211 & 3.478 \\
	FilterX2        & 0.089 & 0.249 & 0.482 & 0.850 & 1.236 & 1.736 & 2.250 & 3.303 & 4.585 & 5.761 \\
	FilterSym       & 0.082 & 0.266 & 0.513 & 0.751 & 1.078 & 1.518 & 1.882 & 2.747 & 3.685 & 4.689 \\
	PlainSPND        & 0.391 & 1.013 & 1.749 & 2.616 & 3.480 & 4.511 & 5.696 & 7.393 & 8.974 & 10.51 \\
	PruneSPND       & 0.060 & 0.130 & 0.242 & 0.349 & 0.449 & 0.569 & 0.770 & 0.989 & 1.309 & 1.464 \\ \hline
	\textbf{PreND }          & 0.030 & 0.068 & 0.097 & 0.142 & 0.182 & 0.231 & 0.335 & 0.444 & 0.540 & 0.629 \\
	\textbf{PreQND}          & 0.027 & 0.061 & \textbf{0.087} & \textbf{0.128} & \textbf{0.166} & \textbf{0.213} & \textbf{0.305} & 0.406 & 0.505 & 0.590 \\
	\textbf{PreTND}          & 0.029 & 0.065 & 0.094 & 0.137 & 0.178 & 0.221 & 0.318 & 0.432 & 0.525 & 0.615 \\
	\textbf{SymND }          & 0.042 & 0.094 & 0.136 & 0.204 & 0.253 & 0.317 & 0.470 & 0.605 & 0.738 & 0.819 \\
	\textbf{SymQND }         & 0.039 & 0.087 & 0.124 & 0.190 & 0.235 & 0.287 & 0.436 & 0.574 & 0.682 & 0.757 \\
	\textbf{SymTND }         & 0.043 & 0.096 & 0.140 & 0.208 & 0.258 & 0.322 & 0.476 & 0.612 & 0.732 & 0.815 \\
	\textbf{PlainNDred}       & 0.027 & 0.063 & 0.104 & 0.150 & 0.200 & 0.263 & 0.332 & 0.399 & 0.480 & 0.551 \\
	\textbf{PlainQNDred}      & \textbf{0.026} & \textbf{0.059} & 0.098 & 0.141 & 0.189 & 0.250 & 0.311 & \textbf{0.377} & \textbf{0.449} & \textbf{0.523} \\
	\textbf{PlainTNDred}      & 0.027 & 0.062 & 0.103 & 0.145 & 0.194 & 0.262 & 0.323 & 0.390 & 0.464 & 0.536 \\
	\hline
	\end{tabular}
	\end{small}
	\vspace{1em}
	
	\begin{small}
	\begin{tabular}{|r|rrrrrrrrrr|}
	\hline
	\rowcolor{gray!50!white}
	\textbf{Algorithm} & \multicolumn{10}{c|}{\textbf{Union Size ($\mathbf{d=5}$, RW)}} \\
	\rowcolor{gray!25!white}
	& \textbf{20k} & \textbf{40k} & \textbf{60k} & \textbf{80k} & \textbf{100k} & \textbf{120k} & \textbf{140k} & \textbf{160k} & \textbf{180k} & \textbf{200k} \\ \hline
	NonDomDC        & 0.262 & 0.776 & 1.483 & 2.318 & 3.292 & 4.292 & 5.245 & 6.264 & 7.158 & 8.147 \\
	FilterX2        & 0.176 & 0.499 & 0.920 & 1.552 & 2.699 & 3.712 & 4.854 & 7.091 & 9.690 & 12.44 \\
	FilterSym       & 0.144 & 0.456 & 0.896 & 1.457 & 2.161 & 3.000 & 3.865 & 5.003 & 6.061 & 8.669 \\
	PlainSPND        & 0.508 & 1.366 & 2.409 & 3.663 & 5.018 & 6.703 & 8.079 & 10.11 & 11.884& 14.18 \\
	PruneSPND       & 0.110 & 0.280 & 0.512 & 0.766 & 1.053 & 1.483 & 1.773 & 2.500 & 2.977 & 3.866 \\ \hline
	\textbf{PreND}           & 0.036 & 0.083 & 0.129 & 0.203 & 0.265 & 0.331 & 0.447 & 0.551 & 0.616 & 0.705 \\
	\textbf{PreQND }         & 0.033 & 0.075 & 0.119 & 0.183 & \textbf{0.242} & \textbf{0.307} & 0.418 & 0.518 & 0.577 & \textbf{0.660} \\
	\textbf{PreTND}          & 0.034 & 0.078 & 0.124 & 0.192 & 0.256 & 0.325 & 0.433 & 0.537 & 0.599 & 0.681 \\
	\textbf{SymND}           & 0.050 & 0.120 & 0.195 & 0.306 & 0.421 & 0.549 & 0.688 & 0.903 & 1.058 & 1.278 \\
	\textbf{SymQND}          & 0.048 & 0.113 & 0.177 & 0.287 & 0.399 & 0.506 & 0.647 & 0.844 & 0.990 & 1.195 \\
	\textbf{SymTND }         & 0.051 & 0.121 & 0.191 & 0.302 & 0.413 & 0.531 & 0.672 & 0.877 & 1.029 & 1.247 \\
	\textbf{PlainNDred}       & 0.032 & 0.075 & 0.126 & 0.193 & 0.265 & 0.355 & 0.426 & 0.520 & 0.598 & 0.723 \\
	\textbf{PlainQNDred}      & \textbf{0.031} & \textbf{0.071} & \textbf{0.117} & \textbf{0.182} & 0.243 & 0.320 & \textbf{0.393} & \textbf{0.481} & \textbf{0.544} & 0.662 \\
	\textbf{PlainTNDred}      & 0.032 & 0.074 & 0.122 & 0.186 & 0.254 & 0.333 & 0.410 & 0.504 & 0.579 & 0.686 \\
	\hline
	\end{tabular}
	\end{small}
	\label{tab:graph_union}
\end{table}


\begin{table}[ht]
\centering
\setlength{\tabcolsep}{3.5pt}
\caption{Running times (in seconds) for all algorithms when computing the Pareto sum of two equally sized Pareto sets $A$ and $B$, each containing multi-criteria shortest path solutions from the New York City road network, modified to induce plateaus (RWP) and correlation (RWC) at the 4th and 5th criteria.}
\begin{small}
\begin{tabular}{|r|rrrrrrrrrr|}
\hline
\rowcolor{gray!50!white}
\textbf{Algorithm} & \multicolumn{10}{c|}{\textbf{Sum Size ($\mathbf{d=5}$, RWP)}} \\
\rowcolor{gray!25!white}
& \textbf{10k} & \textbf{40k} & \textbf{90k} & \textbf{160k} & \textbf{250k} & \textbf{360k} & \textbf{490k} & \textbf{640k} & \textbf{810k} & \textbf{1M} \\ \hline
NonDomDC & 0.022 & 0.086 & 0.202 & 0.401 & 0.651 & 1.073 & 1.612 & 2.478 & 3.492 & 4.821 \\
PlainND & 0.026 & 0.095 & 0.228 & 0.437 & 0.671 & 1.009 & 1.313 & 1.892 & 2.415 & 2.985 \\
BatchedSPND & 0.023 & 0.122 & 0.264 & 0.634 & 0.865 & 1.420 & 1.754 & 2.921 & 3.604 & 5.000 \\
Doubling(Tree) & 0.035 & 0.162 & 0.374 & 0.699 & 1.117 & 1.719 & 2.320 & 3.223 & 4.119 & 5.046 \\
Doubling(Filter) & 0.027 & 0.112 & 0.267 & 0.504 & 0.801 & 1.206 & 1.639 & 2.198 & 2.756 & 3.463 \\
LimMem & 0.026 & 0.107 & 0.244 & 0.470 & 0.766 & 1.209 & 1.841 & 2.689 & 3.791 & 5.257 \\ \hline
\textbf{PreND} & \textbf{0.005} & 0.024 & \textbf{0.057} & 0.132 & \textbf{0.154} & \textbf{0.309} & \textbf{0.326} & \textbf{0.651} & 0.743 & \textbf{0.676} \\
\textbf{PreQND} & \textbf{0.005} & 0.025 & 0.059 & 0.134 & 0.160 & 0.318 & 0.332 & 0.685 & \textbf{0.730} & 0.698 \\
\textbf{PreTND} & \textbf{0.005} & 0.025 & 0.059 & \textbf{0.131} & 0.158 & 0.315 & 0.333 & 0.669 & 0.766 & 0.711 \\
\textbf{PlainNDred} & \textbf{0.005} & \textbf{0.023} & 0.061 & 0.135 & 0.176 & 0.332 & 0.384 & 0.721 & 0.836 & 0.855 \\
\textbf{PlainQNDred} & \textbf{0.005} & \textbf{0.023} & 0.062 & 0.139 & 0.183 & 0.337 & 0.388 & 0.725 & 0.818 & 0.860 \\
\textbf{PlainTNDred} & \textbf{0.005} & \textbf{0.023} & 0.063 & 0.140 & 0.182 & 0.350 & 0.402 & 0.746 & 0.845 & 0.890 \\
\hline
\end{tabular}
\end{small}
\vspace{1em}

\begin{small}
\begin{tabular}{|r|rrrrrrrrrr|}
\hline
\rowcolor{gray!50!white}
\textbf{Algorithm} & \multicolumn{10}{c|}{\textbf{Sum Size ($\mathbf{d=5}$, RWC)}} \\
\rowcolor{gray!25!white}
& \textbf{10k} & \textbf{40k} & \textbf{90k} & \textbf{160k} & \textbf{250k} & \textbf{360k} & \textbf{490k} & \textbf{640k} & \textbf{810k} & \textbf{1M} \\ \hline
NonDomDC & 0.018 & 0.055 & 0.126 & 0.250 & 0.423 & 0.691 & 1.081 & 1.617 & 2.352 & 3.381 \\
PlainND & 0.060 & 0.274 & 0.651 & 1.179 & 1.798 & 2.639 & 3.668 & 4.560 & 5.544 & 6.740 \\
BatchedSPND & 0.058 & 0.328 & 0.880 & 1.819 & 2.803 & 4.089 & 6.296 & 9.188 & 10.04 & 13.03 \\
Doubling(Tree) & 0.064 & 0.303 & 0.743 & 1.369 & 2.174 & 3.218 & 4.447 & 5.915 & 7.509 & 9.350 \\
Doubling(Filter) & 0.061 & 0.250 & 0.592 & 1.070 & 1.678 & 2.443 & 3.321 & 4.282 & 5.404 & 6.905 \\
LimMem & 0.025 & 0.086 & 0.182 & 0.329 & 0.531 & 0.853 & 1.270 & 1.846 & 2.621 & 3.701 \\ \hline
\textbf{PreND} & 0.008 & 0.035 & 0.088 & 0.181 & 0.282 & 0.438 & 0.599 & 0.828 & 1.048 & 1.313 \\
\textbf{PreQND} & 0.008 & 0.034 & 0.089 & 0.179 & 0.285 & 0.426 & \textbf{0.589} & 0.830 & 1.052 & 1.337 \\
\textbf{PreTND} & 0.008 & 0.035 & 0.087 & 0.184 & 0.282 & 0.437 & 0.604 & 0.830 & 1.055 & 1.335 \\
\textbf{PlainNDred} & \textbf{0.007} & 0.033 & 0.088 & 0.181 & 0.298 & 0.439 & 0.629 & 0.852 & 1.073 & 1.358 \\
\textbf{PlainQNDred} & \textbf{0.007} &\textbf{ 0.031} &\textbf{ 0.086} & \textbf{0.172} & \textbf{0.278} & \textbf{0.420} & 0.596 & \textbf{0.797} & \textbf{1.027} & \textbf{1.297} \\
\textbf{PlainTNDred} & \textbf{0.007} & 0.032 & \textbf{0.086} & 0.175 & 0.294 & 0.428 & 0.602 & 0.820 & 1.039 & 1.314 \\
\hline
\end{tabular}
\end{small}
\label{tab:graph_sum_plateau_correlation}
\end{table}


\begin{table}[ht]
\centering
\setlength{\tabcolsep}{4pt}
\caption{Running times (in seconds) for all algorithms when computing the Pareto sum of two equally sized Pareto sets $A$ and $B$, each containing multi-criteria shortest path solutions from the New York City road network, modified to induce plateaus (RWP) and correlation (RWC) at the 4th and 5th criteria.}
\begin{small}
\begin{tabular}{|r|rrrrrrrrrr|}
\hline
\rowcolor{gray!50!white}
\textbf{Algorithm} & \multicolumn{10}{c|}{\textbf{Union Size ($\mathbf{d=5}$, RWP)}} \\
\rowcolor{gray!25!white}
& \textbf{20k} & \textbf{40k} & \textbf{60k} & \textbf{80k} & \textbf{100k} & \textbf{120k} & \textbf{140k} & \textbf{160k} & \textbf{180k} & \textbf{200k} \\ \hline
NonDomDC & 0.878 & 0.775 & 1.367 & 2.071 & 2.844 & 3.660 & 4.495 & 5.377 & 6.310 & 7.250 \\
FilterX2 & 0.669 & 0.622 & 1.081 & 1.667 & 2.285 & 2.992 & 3.719 & 4.524 & 5.359 & 6.306 \\
FilterSym & 0.603 & 0.709 & 1.368 & 2.219 & 3.215 & 4.083 & 5.113 & 6.220 & 7.468 & 8.635 \\
PlainSPND & 0.957 & 1.791 & 3.194 & 4.954 & 6.845 & 8.903 & 11.02 & 13.55 & 16.33 & 19.16 \\
PruneSPND & 0.195 & 0.347 & 0.609 & 0.910 & 1.233 & 1.702 & 2.091 & 2.958 & 3.603 & 4.533 \\ \hline
\textbf{PreND} & 0.096 & 0.113 & 0.186 & 0.297 & 0.405 & 0.514 & 0.698 & 0.805 & 0.894 & 1.084 \\
\textbf{PreQND} & 0.096 & \textbf{0.101} & \textbf{0.161} & \textbf{0.260} & \textbf{0.344} & \textbf{0.441} & \textbf{0.607} & \textbf{0.715} & \textbf{0.813} & \textbf{0.930} \\
\textbf{PreTND} & 0.112 & 0.109 & 0.180 & 0.285 & 0.379 & 0.480 & 0.659 & 0.763 & 0.839 & 0.986 \\
\textbf{SymND} & 0.143 & 0.151 & 0.229 & 0.361 & 0.482 & 0.625 & 0.774 & 0.976 & 1.182 & 1.498 \\
\textbf{SymQND} & 0.144 & 0.142 & 0.210 & 0.335 & 0.434 & 0.549 & 0.707 & 0.888 & 1.064 & 1.316 \\
\textbf{SymTND} & 0.146 & 0.150 & 0.226 & 0.358 & 0.464 & 0.592 & 0.758 & 0.954 & 1.158 & 1.445 \\
\textbf{PlainNDred} & 0.093 & 0.132 & 0.255 & 0.424 & 0.642 & 0.909 & 1.201 & 1.571 & 1.938 & 2.510 \\
\textbf{PlainQNDred} & \textbf{0.087} & 0.117 & 0.215 & 0.339 & 0.517 & 0.674 & 0.918 & 1.205 & 1.502 & 1.868 \\
\textbf{PlainTNDred} & 0.094 & 0.129 & 0.240 & 0.399 & 0.576 & 0.832 & 1.071 & 1.385 & 1.700 & 2.153 \\
\hline
\end{tabular}
\end{small}

\vspace{1em}

\begin{small}
\begin{tabular}{|r|rrrrrrrrrr|}
\hline
\rowcolor{gray!50!white}
\textbf{Algorithm} & \multicolumn{10}{c|}{\textbf{Union Size ($\mathbf{d=5}$, RWC)}} \\
\rowcolor{gray!25!white}
& \textbf{20k} & \textbf{40k} & \textbf{60k} & \textbf{80k} & \textbf{100k} & \textbf{120k} & \textbf{140k} & \textbf{160k} & \textbf{180k} & \textbf{200k} \\ \hline
NonDomDC & 0.699 & 1.225 & 0.816 & 1.200 & 1.596 & 1.966 & 2.376 & 2.753 & 3.102 & 3.517 \\
FilterX2 & 0.585 & 1.059 & 0.703 & 1.041 & 1.386 & 1.774 & 2.177 & 2.633 & 3.116 & 3.639 \\
FilterSym & 0.494 & 1.294 & 0.820 & 1.028 & 1.352 & 1.736 & 2.156 & 2.577 & 2.946 & 3.388 \\
PlainSPND & 0.980 & 2.335 & 3.050 & 4.618 & 6.311 & 8.246 & 10.15 & 12.55 & 14.72 & 17.68 \\
PruneSPND & 0.169 & 0.375 & 0.492 & 0.725 & 0.962 & 1.352 & 1.609 & 2.236 & 2.598 & 3.368 \\ \hline
\textbf{PreND} & 0.098 & 0.172 & 0.154 & 0.243 & 0.303 & 0.377 & 0.523 & 0.613 & 0.691 & 0.816 \\
\textbf{PreQND} & 0.099 & 0.173 & \textbf{0.136} & \textbf{0.211} & \textbf{0.269} & \textbf{0.335} & \textbf{0.465} & \textbf{0.552} & \textbf{0.612} & \textbf{0.741} \\
\textbf{PreTND} & 0.105 & 0.177 & 0.154 & 0.225 & 0.293 & 0.354 & 0.493 & 0.601 & 0.667 & 0.783 \\
\textbf{SymND} & 0.155 & 0.247 & 0.197 & 0.301 & 0.383 & 0.537 & 0.675 & 0.829 & 0.930 & 1.172 \\
\textbf{SymQND} & 0.158 & 0.244 & 0.185 & 0.288 & 0.359 & 0.486 & 0.621 & 0.795 & 0.874 & 1.092 \\
\textbf{SymTND} & 0.157 & 0.254 & 0.197 & 0.303 & 0.382 & 0.519 & 0.668 & 0.822 & 0.928 & 1.156 \\
\textbf{PlainNDred} & 0.097 & 0.163 & 0.178 & 0.281 & 0.371 & 0.495 & 0.622 & 0.769 & 0.894 & 1.085 \\
\textbf{PlainQNDred} & 0.086 & \textbf{0.160} & 0.153 & 0.240 & 0.312 & 0.430 & 0.534 & 0.653 & 0.761 & 0.941 \\
\textbf{PlainTNDred} & \textbf{0.085} & 0.162 & 0.165 & 0.261 & 0.354 & 0.481 & 0.584 & 0.720 & 0.848 & 1.024 \\
\hline
\end{tabular}
\end{small}
\label{tab:graph_union_plateau_correlation}
\end{table}


\begin{table}[ht]
\centering
\setlength{\tabcolsep}{3pt}
\renewcommand{\arraystretch}{1.2}
\caption{Min, Avg, and Max speedups for each algorithm for the \emph{Pareto Sum} operation for RW, RWP and RWC data sets (same with Table \ref{tab:sum_speedup_summary}, but also including average speedup).}
\begin{small}
\begin{tabular}{|r|ccc|ccc|ccc|}
\hline
\rowcolor{gray!50!white}
\textbf{Algorithm}
& \multicolumn{3}{c|}{\textbf{RW}}
& \multicolumn{3}{c|}{\textbf{RWP}}
& \multicolumn{3}{c|}{\textbf{RWC}} \\
\rowcolor{gray!25!white}
& Min & Avg & Max & Min & Avg & Max & Min & Avg & Max \\
\hline
PreND       & 1.1 & 2.2 & 3.8 & 2.9 & 3.6 & 4.4 & 1.4 & 1.8 & 2.6    \\
PreQND      & 1.0 & 2.2 & 3.7 & 2.8 & 3.6 & 4.3 & 1.4 & 1.8 & 2.5    \\
PreTND      & 1.0 & 2.2 & 3.7 & 2.8 & 3.6 & 4.2 & 1.4 & 1.8 & 2.5    \\
PlainNDred  & 1.1 & 2.2 & 3.7 & 2.6 & 3.4 & 4.5 & 1.4 & 1.8 & 2.5    \\
PlainQNDred & 1.1 & 2.2 & 3.6 & 2.6 & 3.3 & 4.3 & 1.4 & 1.9 & 2.6    \\
PlainTNDred & 1.1 & 2.1 & 3.6 & 2.5 & 3.3 & 4.4 & 1.4 & 1.9 & 2.6    \\
\hline
\end{tabular}
\end{small}
\label{tab:sum_rw_complete_speedup_summary}
\end{table}

\begin{table}[ht]
\centering
\setlength{\tabcolsep}{3pt}
\renewcommand{\arraystretch}{1.2}
\caption{Min, Avg, and Max speedups for each algorithm for the \emph{Pareto Union} operation for RW, RWP and RWC data sets (same with Table \ref{tab:union_speedup_summary}, but also including average speedup).}
\begin{small}
\begin{tabular}{|r|ccc|ccc|ccc|}
\hline
\rowcolor{gray!50!white}
\textbf{Algorithm}
& \multicolumn{3}{c|}{\textbf{RW}}
& \multicolumn{3}{c|}{\textbf{RWP}}
& \multicolumn{3}{c|}{\textbf{RWC}} \\
\rowcolor{gray!25!white}
& Min & Avg & Max & Min & Avg & Max & Min & Avg & Max \\
\hline
PreND       & 1.9 & 3.2 & 5.5 & 2.0 & 3.3 & 4.2 & 1.7 & 3.1 & 4.1 \\
PreQND      & 2.1 & 3.5 & 5.9 & 2.0 & 3.7 & 4.9 & 1.7 & 3.5 & 4.5 \\
PreTND      & 2.0 & 3.4 & 5.7 & 1.7 & 3.4 & 4.6 & 1.6 & 3.2 & 4.3 \\
SymND       & 1.4 & 2.1 & 3.0 & 1.4 & 2.6 & 3.0 & 1.1 & 2.3 & 2.9 \\
SymQND      & 1.5 & 2.3 & 3.2 & 1.4 & 2.9 & 3.4 & 1.1 & 2.5 & 3.1 \\
SymTND      & 1.4 & 2.2 & 3.1 & 1.3 & 2.7 & 3.1 & 1.1 & 2.3 & 2.9 \\
PlainNDred  & 2.1 & 3.3 & 5.3 & 1.7 & 2.0 & 2.6 & 1.7 & 2.6 & 3.1 \\
PlainQNDred & 2.2 & 3.6 & 5.8 & 2.3 & 2.5 & 3.0 & 2.0 & 3.0 & 3.6 \\
PlainTNDred & 2.1 & 3.4 & 5.6 & 2.0 & 2.2 & 2.7 & 2.0 & 2.8 & 3.3 \\
\hline
\end{tabular}
\end{small}
\label{tab:union_rw_complete_speedup_summary}
\end{table}


\begin{table}[ht]
\centering
\setlength{\tabcolsep}{4pt}
\caption{Running times (in seconds) for all algorithms on 4, 6, 8, and 10 dimensions for the Pareto union of two equally sized Uniform-Random Pareto sets (URS).}
\begin{small}
\begin{tabular}{|r|rrrrr|rrrrr|}
\hline
\rowcolor{gray!50!white}
\textbf{Algorithm} & \multicolumn{5}{c|}{\textbf{Union Size ($\mathbf{d=4}$, URS)}}
& \multicolumn{5}{c|}{\textbf{Union Size ($\mathbf{d=6}$, URS)}} \\
\rowcolor{gray!25!white}
& \textbf{50k} & \textbf{100k} & \textbf{200k} & \textbf{500k} & \textbf{1M}
& \textbf{50k} & \textbf{100k} & \textbf{200k} & \textbf{500k} & \textbf{1M} \\ \hline
NonDomDC        & 0.645 & 1.748 & 4.981 & 27.97 & 98.61 & 1.622 & 4.336 & 11.47 & 42.12 & 110.1 \\
FilterX2        & 0.540 & 1.406 & 3.747 & 17.77 & 57.61 & 1.222 & 3.313 & 8.893 & 33.04 & 86.05 \\
FilterSym       & 1.277  & 4.997 & 19.46 & 123.6 & 579.3 & 1.967 & 7.087 & 26.26 & 104.5 & 300.8 \\
PlainSPND       & 1.043  & 2.768 & 7.510 & 29.97 & 94.12 & 1.817 & 4.927 & 13.69 & 54.56 & 166.1 \\
PruneSPND       & 0.308 & 0.883 & 2.359 & 7.239 & 16.55 & 0.6720 & 1.964 & 5.679 & 27.90 & 100.9 \\
\hline
\textbf{PreND}  & 0.139 & 0.385 & 1.082 & 3.969 & 10.97 & 0.235 & 0.613 & 1.705 & 6.542 & 21.92 \\
\textbf{PreQND} & 0.121 & \textbf{0.336} & 0.979 & 3.670 & 10.04 & \textbf{0.207} & 0.556 & 1.575 & 5.978 & 20.95 \\
\textbf{PreTND} & 0.129 & 0.353 & 1.001 & 3.802 & 10.37 & 0.217 & 0.576 & 1.590 & 6.026 & 20.93 \\
\textbf{SymND}  & 0.148 & 0.385 & 1.039 & 4.196 & 9.441 & 0.237 & 0.588 & 1.523 & 6.019 & 21.19 \\
\textbf{SymQND} & 0.134 & 0.351 & \textbf{0.956} & 4.174 & \textbf{8.383} & 0.209 & \textbf{0.536} & \textbf{1.373} & \textbf{5.350} & 18.07 \\
\textbf{SymTND} & 0.151 & 0.383 & 1.028 & 4.512 & 9.005 & 0.228 & 0.573 & 1.470 & 5.741 & 18.96 \\
\textbf{PlainNDred} & 0.120 & 0.390 & 1.392 & \textbf{3.527} & 9.136 & 0.242 & 0.639 & 1.632 & 5.746 & 16.28 \\
\textbf{PlainQNDred} & \textbf{0.111} & 0.370 & 1.266 & 3.568 & 8.624 & 0.236 & 0.607 & 1.515 & 5.361 & \textbf{15.13} \\
\textbf{PlainTNDred} & 0.119 & 0.383 & 1.340 & 3.646 & 8.757 & 0.248 & 0.625 & 1.585 & 5.505 & 15.16 \\
\hline
\end{tabular}
\end{small}

\vspace{1.2em}

\begin{small}
\begin{tabular}{|r|rrrrr|rrrrr|}
\hline
\rowcolor{gray!50!white}
\textbf{Algorithm} & \multicolumn{5}{c|}{\textbf{Union Size ($\mathbf{d=8}$, URS)}}
& \multicolumn{5}{c|}{\textbf{Union Size ($\mathbf{d=10}$, URS)}} \\
\rowcolor{gray!25!white}
& \textbf{50k} & \textbf{100k} & \textbf{200k} & \textbf{500k} & \textbf{1M}
& \textbf{50k} & \textbf{100k} & \textbf{200k} & \textbf{500k} & \textbf{1M} \\ \hline
NonDomDC        & 2.034 & 5.848 & 16.86 & 66.72 & 187.7 & 2.247 & 6.480 & 18.90 & 78.67 & 225.6 \\
FilterX2        & 1.447 & 4.189 & 12.12 & 47.85 & 135.7 & 1.576 & 4.547 & 13.34 & 55.37 & 158.7 \\
FilterSym       & 2.244 & 8.133 & 30.20 & 182.5 & 433.2 & 2.391 & 8.664 & 32.47 & 188.3 & 452.0 \\
PlainSPND       & 2.349 & 7.206 & 21.58 & 93.04 & 352.3 & 3.056 & 8.778 & 27.00 & 105.7 & 388.6 \\
PruneSPND       & 0.942 & 2.861 & 13.05 & 64.30 & 199.7 & 1.143 & 3.595 & 12.60 & 82.06 & 276.5 \\
\hline
\textbf{PreND}  & 0.329 & 0.992 & 2.595 & 11.01 & 25.87 & 0.399 & 1.156 & 3.372 & 12.81 & 46.71 \\
\textbf{PreQND} & 0.280 & 0.898 & 2.343 & 10.02 & 23.30 & 0.353 & 1.025 & 3.083 & 13.16 & 46.91 \\
\textbf{PreTND} & 0.297 & 0.930 & 2.426 & 10.29 & 23.98 & 0.377 & 1.076 & 3.174 & 12.72 & 50.41 \\
\textbf{SymND}  & 0.307 & 0.933 & 2.406 & 11.39 & 26.15 & 0.373 & 1.046 & 3.133 & 13.02 & 47.76 \\
\textbf{SymQND} & 0.277 & 0.850 & 2.313 & 10.41 & 24.04 & 0.347 & 1.010 & 3.035 & 12.60 & 46.09 \\
\textbf{SymTND} & 0.319 & 0.962 & 2.469 & 11.12 & 25.49 & 0.391 & 1.123 & 3.343 & 13.26 & 47.18 \\
\textbf{PlainNDred} & 0.268 & 0.871 & 2.273 & 9.319 & 23.11 & 0.336 & 0.992 & 2.942 & 11.10 & 43.38 \\
\textbf{PlainQNDred} & \textbf{0.240} & \textbf{0.809} & \textbf{2.182} & \textbf{8.993} & \textbf{22.11} & \textbf{0.316} & \textbf{0.947} & \textbf{2.827} & \textbf{10.87} & \textbf{41.63} \\
\textbf{PlainTNDred} & 0.263 & 0.862 & 2.267 & 9.230 & 23.23 & 0.323 & 0.974 & 2.890 & 11.30 & 42.11 \\
\hline
\end{tabular}
\end{small}
\label{tab:union_random}
\end{table}

\clearpage

\begin{table}[ht]
\centering
\setlength{\tabcolsep}{3.7pt}
\caption{Running times (in seconds) for all algorithms on 4, 6, 8, and 10 dimensions for the Pareto sum of two equally sized Uniform-Random Pareto sets (URS).}
\begin{small}
\begin{tabular}{|r|rrrrr|rrrrr|}
\hline
\rowcolor{gray!50!white}
\textbf{Algorithm}
& \multicolumn{5}{c|}{\textbf{Sum Size ($\mathbf{d=4}$, URS)}}
& \multicolumn{5}{c|}{\textbf{Sum Size ($\mathbf{d=6}$, URS)}} \\
\rowcolor{gray!25!white}
& \textbf{50k} & \textbf{105k} & \textbf{202k} & \textbf{501k} & \textbf{1M}
& \textbf{50k} & \textbf{105k} & \textbf{202k} & \textbf{501k} & \textbf{1M} \\ \hline

NonDomDC         & 0.804 & 0.675 & 0.571 & 1.330 & 2.740 & 0.988 & 2.396 & 4.962 & 11.69 & 27.94 \\
PlainSPND        & 0.578 & 0.486 & 0.724 & 1.472 & 3.078 & 1.409 & 3.453 & 7.362 & 26.61 & 67.41 \\
BatchedSPND      & 0.278 & 0.407 & 0.690 & 1.683 & 4.105 & 1.642 & 4.347 & 9.807 & 26.45 & 64.08 \\
Doubling(Tree)   & 1.218 & 1.091 & 1.603 & 3.644 & 7.707 & 1.707 & 4.408 & 9.660 & 31.65 & 83.53 \\
Doubling(Filter) & 2.131 & 1.673 & 1.519 & 3.662 & 7.614 & 2.049 & 5.154 & 11.17 & 31.86 & 86.29 \\
LimMem           & 1.036 & 0.822 & 0.686 & 1.610 & 3.288 & 1.669 & 3.931 & 7.865 & 16.50 & 39.67 \\ \hline
\textbf{PreND}         & 0.135 & 0.229 & 0.283 & 0.630 & 1.181 & 0.229 & 0.568 & 1.426 & 2.921 & 7.096 \\
\textbf{PreQND}        & 0.139 & 0.229 & 0.268 & 0.611 & 1.134 & 0.211 & 0.528 & 1.305 & 2.731 & 6.588 \\
\textbf{PreTND}        & 0.142 & 0.233 & 0.268 & 0.591 & 1.148 & 0.209 & 0.540 & 1.334 & 2.774 & 6.701 \\
\textbf{PlainNDred}    & 0.071 & 0.128 & 0.199 & 0.383 & 0.753 & 0.199 & 0.502 & 1.116 & 2.100 & 5.164 \\
\textbf{PlainQNDred}   & \textbf{0.056} & \textbf{0.115} & 0.196 & \textbf{0.375} & \textbf{0.743} & \textbf{0.187} & \textbf{0.477} & \textbf{1.032} & \textbf{1.955} & \textbf{4.799} \\
\textbf{PlainTNDred}   & 0.060 & 0.120 & \textbf{0.193} & 0.378 & 0.747 & 0.194 & 0.479 & 1.049 & 2.010 & 4.900 \\ \hline
\end{tabular}
\end{small}

\vspace{1.2em}

\begin{small}
\begin{tabular}{|r|rrrrr|rrrrr|}
\hline
\rowcolor{gray!50!white}
\textbf{Algorithm}
& \multicolumn{5}{c|}{\textbf{Sum Size ($\mathbf{d=8}$, URS)}}
& \multicolumn{5}{c|}{\textbf{Sum Size ($\mathbf{d=10}$, URS)}} \\
\rowcolor{gray!25!white}
& \textbf{50k} & \textbf{105k} & \textbf{202k} & \textbf{501k} & \textbf{1M}
& \textbf{50k} & \textbf{105k} & \textbf{202k} & \textbf{501k} & \textbf{1M} \\ \hline

NonDomDC         & 1.984 & 5.867 & 14.94 & 51.62 & 135.0 & 2.308 & 7.149 & 19.28 & 77.08 & 219.9 \\
PlainSPND        & 2.461 & 7.742 & 21.07 & 54.78 & 153.2 & 3.030 & 9.228 & 26.22 & 83.43 & 253.7 \\
BatchedSPND      & 4.029 & 14.76 & 47.02 & 105.1 & 290.6 & 5.455 & 20.40 & 83.16 & 229.4 & 647.5 \\
Doubling(Tree)   & 2.715 & 8.339 & 21.61 & 52.82 & 140.5 & 3.195 & 9.692 & 26.79 & 74.29 & 212.3\\
Doubling(Filter) & 3.085 & 9.663 & 23.96 & 101.2 & 191.8 & 3.656 & 10.86 & 29.84 & 141.5 & 431.5 \\
LimMem           & 3.907 & 11.40 & 28.70 & 97.29 & 242.8 & 4.615 & 14.35 & 38.46 & 153.2 & 434.7 \\ \hline
\textbf{PreND}         & 0.425 & 1.184 & 2.963 & 10.88 & 28.92 & 0.450 & 1.392 & 4.106 & 16.82 & 59.74 \\
\textbf{PreQND}        & \textbf{0.370} & 1.060 & \textbf{2.707} & 10.00 & 26.594 & 0.413 & 1.281 & 3.723 & 15.48 & 55.31 \\
\textbf{PreTND}        & 0.399 & 1.107 & 2.776 & 10.32 & 27.05 & 0.425 & 1.307 & 3.854 & 15.74 & 55.88 \\
\textbf{PlainNDred}    & 0.404 & 1.103 & 2.951 & 8.769 & 22.57 & 0.444 & 1.379 & 3.051 & 15.76 & 50.41 \\
\textbf{PlainQNDred}   & 0.379 & \textbf{1.030} & 2.723 & \textbf{8.087} & \textbf{20.41} & \textbf{0.412} & \textbf{1.277} & \textbf{2.808} & \textbf{14.71} & \textbf{46.51} \\
\textbf{PlainTNDred}   & 0.395 & 1.063 & 2.787 & 8.215 & 20.625 & 0.437 & 1.304 & 2.873 & 14.91 & 46.86 \\ \hline

\end{tabular}
\end{small}
\label{tab:sum_random}
\end{table}


\begin{table}[ht]
\centering
\setlength{\tabcolsep}{4pt}
\caption{Running times (in seconds) for all algorithms on 4, 6, 8, and 10 dimensions for the Pareto union of two equally sized Uniform-Random Pareto sets with plateaus (URSP).}
\begin{small}
\begin{tabular}{|r|rrrrr|rrrrr|}
\hline
\rowcolor{gray!50!white}
\textbf{Algorithm}
& \multicolumn{5}{c|}{\textbf{Union Size ($\mathbf{d=4}$, URSP)}}
& \multicolumn{5}{c|}{\textbf{Union Size ($\mathbf{d=6}$, URSP)}} \\
\rowcolor{gray!25!white}
& \textbf{10k} & \textbf{20k} & \textbf{50k} & \textbf{100k} & \textbf{200k}
& \textbf{10k} & \textbf{20k} & \textbf{50k} & \textbf{100k} & \textbf{200k} \\ \hline

NonDomDC        & 0.053 & 0.141 & 0.368 & 1.128 & 2.958 & 0.133 & 0.332 & 1.025 & 2.681 & 6.539 \\
FilterX2        & 0.032 & 0.099 & 0.183 & 0.851 & 1.723 & 0.097 & 0.228 & 0.713 & 1.929 & 4.223 \\
FilterSym       & 0.022 & 0.080 & 0.107 & 0.976 & 1.545 & 0.087 & 0.203 & 0.570 & 1.492 & 3.748 \\
PlainSPND        & 0.095 & 0.251 & 0.704 & 2.503 & 6.113 & 0.180 & 0.459 & 1.676 & 4.421 & 11.41 \\
PruneSPND       & 0.031 & 0.069 & 0.142 & 0.594 & 1.349 & 0.074 & 0.171 & 0.558 & 1.494 & 3.289 \\ \hline
\textbf{PreND}           & 0.013 & 0.032 & 0.064 & 0.244 & 0.547 & 0.030 & 0.064 & 0.174 & 0.467 & 1.205 \\
\hline
\end{tabular}
\end{small}

\vspace{1.2em}

\begin{small}
\begin{tabular}{|r|rrrrr|rrrrr|}
\hline
\rowcolor{gray!50!white}
\textbf{Algorithm}
& \multicolumn{5}{c|}{\textbf{Union Size ($\mathbf{d=8}$, URSP)}}
& \multicolumn{5}{c|}{\textbf{Union Size ($\mathbf{d=10}$, URSP)}} \\
\rowcolor{gray!25!white}
& \textbf{10k} & \textbf{20k} & \textbf{50k} & \textbf{100k} & \textbf{200k}
& \textbf{10k} & \textbf{20k} & \textbf{50k} & \textbf{100k} & \textbf{200k} \\ \hline

NonDomDC        & 0.170 & 0.495 & 1.859 & 5.196 & 13.81 & 0.198 & 0.580 & 2.277 & 6.536 & 19.48 \\
FilterX2        & 0.124 & 0.347 & 1.322 & 3.678 & 10.07 & 0.132 & 0.388 & 1.596 & 4.515 & 13.77 \\
FilterSym       & 0.110 & 0.321 & 1.222 & 3.371 & 8.889 & 0.125 & 0.365 & 1.530 & 4.371 & 12.96 \\
PlainSPND        & 0.213 & 0.663 & 2.474 & 7.105 & 21.64 & 0.280 & 0.830 & 3.339 & 9.641 & 28.91 \\
PruneSPND       & 0.077 & 0.280 & 0.937 & 2.655 & 13.28 & 0.106 & 0.30 & 1.225 & 3.729 & 12.95 \\ \hline
\textbf{PreND} & 0.041 & 0.112 & 0.369 & 0.973 & 2.454 & 0.047 & 0.120 & 0.447 & 1.327 & 3.965 \\
\textbf{PreQND} & \textbf{0.037} & 0.098 & 0.339 & 0.859 & 2.202 & \textbf{0.044} & \textbf{0.106} & 0.400 & 1.206 & 3.633 \\
\textbf{PreTND}          & 0.039 & 0.100 & 0.343 & 0.902 & 2.292 & \textbf{0.044} & 0.108 & 0.419 & 1.241 & 3.692 \\
\textbf{SymND}           & 0.045 & 0.097 & 0.331 & 0.903 & 2.492 & 0.048 & 0.125 & 0.386 & 1.051 & 3.471 \\
\textbf{SymQND}          & 0.038 & \textbf{0.087} & \textbf{0.284} & \textbf{0.788} & 2.219 & 0.050 & 0.113 & \textbf{0.342} & \textbf{0.896} & \textbf{2.920} \\
\textbf{SymTND}          & 0.039 & 0.096 & 0.315 & 0.854 & 2.381 & 0.047 & 0.119 & 0.373 & 1.006 & 3.300 \\
\textbf{PlainNDred}      & 0.043 & 0.114 & 0.375 & 0.933 & 2.314 & 0.050 & 0.136 & 0.453 & 1.319 & 3.492 \\
\textbf{PlainQNDred}     & 0.040 & 0.101 & 0.359 & 0.858 & \textbf{2.080} & 0.047 & 0.123 & 0.439 & 1.233 & 3.162 \\
\textbf{PlainTNDred}     & 0.043 & 0.101 & 0.370 & 0.933 & 2.173 & 0.047 & 0.123 & 0.439 & 1.257 & 3.220 \\
\hline
\end{tabular}
\end{small}
\label{tab:union_plateau}
\end{table}

\begin{table}[ht]
\centering
\setlength{\tabcolsep}{4pt}
\caption{Running times (in seconds) for all algorithms on 4, 6, 8, and 10 dimensions for the Pareto sum of two equally sized Uniform-Random Pareto sets with plateaus (URSP).}
\begin{small}
\begin{tabular}{|r|rrrrr|rrrrr|}
\hline
\rowcolor{gray!50!white}
\textbf{Algorithm}
& \multicolumn{5}{c|}{\textbf{Sum Size ($\mathbf{d=4}$, URSP)}}
& \multicolumn{5}{c|}{\textbf{Sum Size ($\mathbf{d=6}$, URSP)}} \\
\rowcolor{gray!25!white}
& \textbf{10k} & \textbf{22k} & \textbf{50k} & \textbf{105k} & \textbf{202k}
& \textbf{10k} & \textbf{22k} & \textbf{50k} & \textbf{105k} & \textbf{202k} \\ \hline
NonDomDC         & 0.108 & 0.293 & 0.749 & 0.407 & 0.542 & 0.104 & 0.296 & 0.757 & 1.515 & 3.194 \\
PlainSPND         & 0.122 & 0.156 & 0.416 & 0.371 & 0.616 & 0.157 & 0.413 & 1.178 & 2.241 & 4.778 \\
BatchedSPND      & 0.059 & 0.153 & 0.248 & 0.283 & 0.560 & 0.151 & 0.416 & 1.262 & 2.407 & 5.654 \\
Doubling(Tree)   & 0.186 & 0.453 & 1.072 & 0.806 & 1.472 & 0.178 & 0.525 & 1.432 & 3.018 & 6.718 \\
Doubling(Filter) & 0.261 & 0.768 & 1.682 & 0.779 & 1.245 & 0.183 & 0.592 & 1.560 & 3.065 & 6.897 \\
LimMem           & 0.148 & 0.382 & 0.909 & 0.370 & 0.604 & 0.177 & 0.482 & 1.197 & 2.135 & 4.372 \\ \hline
\textbf{PreND}         & 0.020 & 0.047 & 0.128 & 0.159 & 0.306 & 0.027 & 0.069 & 0.195 & 0.397 & 0.833 \\
\textbf{PreQND}        & 0.020 & 0.050 & 0.128 & 0.161 & 0.289 & 0.025 & 0.068 & 0.174 & 0.368 & 0.780 \\
\textbf{PreTND}        & 0.021 & 0.051 & 0.130 & 0.161 & 0.297 & 0.026 & 0.066 & 0.178 & 0.366 & 0.792 \\
\textbf{PlainNDred}    & 0.009 & 0.022 & 0.061 & 0.105 & 0.225 & 0.025 & 0.061 & 0.160 & 0.298 & 0.594 \\
\textbf{PlainQNDred}   & \textbf{0.008} & \textbf{0.017} & \textbf{0.045} & \textbf{0.099} & \textbf{0.222} & \textbf{0.022} & 0.060 & \textbf{0.149} & \textbf{0.270} & \textbf{0.546} \\
\textbf{PlainTNDred}   & 0.009 & \textbf{0.017} & 0.046 & 0.104 & 0.225 & 0.024 & \textbf{0.059} & 0.163 & 0.284 & 0.569 \\ \hline
\end{tabular}
\end{small}

\vspace{1.2em}

\begin{small}
\begin{tabular}{|r|rrrrr|rrrrr|}
\hline
\rowcolor{gray!50!white}
\textbf{Algorithm}
& \multicolumn{5}{c|}{\textbf{Sum Size ($\mathbf{d=8}$, URSP)}}
& \multicolumn{5}{c|}{\textbf{Sum Size ($\mathbf{d=10}$, URSP)}} \\
\rowcolor{gray!25!white}
& \textbf{10k} & \textbf{22k} & \textbf{50k} & \textbf{105k} & \textbf{202k}
& \textbf{10k} & \textbf{22k} & \textbf{50k} & \textbf{105k} & \textbf{202k} \\ \hline

NonDomDC         & 0.176 & 0.541 & 1.749 & 5.106 & 12.0 & 0.192 & 0.664 & 2.326 & 6.938 & 18.22 \\
PlainSPND         & 0.207 & 0.747 & 2.343 & 7.218 & 17.69 & 0.268 & 0.875 & 3.192 & 9.751 & 26.77 \\
BatchedSPND      & 0.257 & 1.006 & 3.620 & 13.02 & 35.09 & 0.311 & 1.299 & 5.852 & 21.85 & 83.93 \\
Doubling(Tree)   & 0.224 & 0.802 & 2.558 & 7.594 & 18.51 & 0.273 & 0.940 & 3.312 & 10.15 & 27.01 \\
Doubling(Filter) & 0.272 & 0.939 & 2.912 & 8.479 & 20.57 & 0.296 & 1.033 & 3.384 & 10.44 & 28.98 \\
LimMem           & 0.348 & 1.061 & 3.291 & 9.473 & 21.04 & 0.378 & 1.335 & 4.602 & 13.7 & 35.5 \\ \hline
\textbf{PreND}         & 0.041 & 0.129 & 0.393 & 1.082 & 2.460 & 0.043 & 0.141 & 0.516 & 1.601 & 4.107 \\
\textbf{PreQND}        & 0.037 & 0.117 & 0.356 & \textbf{0.987} & 2.251 & \textbf{0.039} & \textbf{0.131} & \textbf{0.461} & 1.447 & 3.754 \\
\textbf{PreTND}        & 0.038 & 0.118 & 0.371 & 1.023 & 2.318 & 0.040 & 0.133 & 0.476 & 1.502 & 3.849 \\
\textbf{PlainNDred}    & 0.039 & 0.116 & 0.379 & 1.087 & 2.305 & 0.045 & 0.141 & 0.526 & 1.265 & 3.403 \\
\textbf{PlainQNDred}   & \textbf{0.035} & \textbf{0.110} & \textbf{0.338} & 0.998 & \textbf{2.104} & 0.042 & 0.138 & 0.472 & \textbf{1.159} & \textbf{3.119} \\
\textbf{PlainTNDred}   & 0.037 & 0.124 & 0.358 & 1.033 & 2.187 & 0.044 & 0.141 & 0.496 & 1.188 & 3.198 \\ \hline
\end{tabular}
\end{small}
\label{tab:sum_plateau}
\end{table}

\begin{table}[ht]
\centering
\setlength{\tabcolsep}{4pt}
\caption{Running times (in seconds) for all algorithms on 4, 6, 8, and 10 dimensions for the Pareto union of two equally sized Uniform-Random Pareto sets with correlated objectives (URSC).}
\begin{small}
\begin{tabular}{|r|rrrrr|rrrrr|}
\hline
\rowcolor{gray!50!white}
\textbf{Algorithm}
& \multicolumn{5}{c|}{\textbf{Union Size ($\mathbf{d=4}$, URSC)}}
& \multicolumn{5}{c|}{\textbf{Union Size ($\mathbf{d=6}$, URSC)}} \\
\rowcolor{gray!25!white}
& \textbf{10k} & \textbf{20k} & \textbf{50k} & \textbf{100k} & \textbf{200k}
& \textbf{10k} & \textbf{20k} & \textbf{50k} & \textbf{100k} & \textbf{200k} \\ \hline

NonDomDC        & 0.060 & 0.133 & 0.525 & 1.443 & 5.762 & 0.139 & 0.390 & 1.436 & 3.641 & 12.45 \\
FilterX2        & 0.044 & 0.093 & 0.582 & 1.612 & 5.730 & 0.097 & 0.230 & 1.225 & 2.880 & 11.58 \\
FilterSym       & 0.030 & 0.089 & 0.578 & 1.765 & 4.536 & 0.090 & 0.194 & 1.149 & 2.977 & 10.12 \\
PlainSPND        & 0.090 & 0.278 & 0.978 & 2.969 & 7.931 & 0.156 & 0.455 & 1.891 & 5.234 & 15.46 \\
PruneSPND       & 0.017 & 0.042 & 0.256 & 0.640 & 2.343 & 0.028 & 0.065 & 0.288 & 0.898 & 2.829 \\ \hline
\textbf{PreND}           & 0.019 & 0.034 & 0.121 & 0.478 & 0.886 & 0.024 & 0.066 & 0.237 & 0.685 & 1.869 \\
\textbf{PreQND}          & 0.017 & 0.031 & 0.114 & 0.420 & 0.809 & 0.022 & 0.059 & 0.203 & 0.602 & 1.695 \\
\textbf{PreTND}          & 0.018 & 0.033 & 0.115 & 0.446 & 0.851 & 0.024 & 0.062 & 0.216 & 0.638 & 1.756 \\
\textbf{SymND}     & 0.011 & 0.029 & 0.096 & 0.295 & 0.712 & 0.016 & 0.037 & 0.111 & 0.288 & 0.781 \\
\textbf{SymQND}  & \textbf{0.010} & \textbf{0.026} & \textbf{0.087} & \textbf{0.274} & \textbf{0.681} & \textbf{0.014} & \textbf{0.033} & \textbf{0.102} & \textbf{0.267} & \textbf{0.733} \\
\textbf{SymTND}    & 0.012 & 0.031 & 0.098 & 0.302 & 0.737 & 0.016 & 0.038 & 0.114 & 0.306 & 0.806 \\
\textbf{PlainNDred}      & 0.014 & 0.045 & 0.151 & 0.537 & 1.066 & 0.033 & 0.083 & 0.294 & 0.887 & 2.025 \\
\textbf{PlainQNDred}     & 0.013 & 0.040 & 0.147 & 0.503 & 0.991 & 0.029 & 0.074 & 0.278 & 0.834 & 1.896 \\
\textbf{PlainTNDred}     & 0.014 & 0.042 & 0.151 & 0.524 & 1.034 & 0.032 & 0.078 & 0.296 & 0.854 & 1.949 \\
\hline
\end{tabular}
\end{small}

\vspace{1.2em}

\begin{small}
\begin{tabular}{|r|rrrrr|rrrrr|}
\hline
\rowcolor{gray!50!white}
\textbf{Algorithm}
& \multicolumn{5}{c|}{\textbf{Union Size ($\mathbf{d=8}$, URSC)}}
& \multicolumn{5}{c|}{\textbf{Union Size ($\mathbf{d=10}$, URSC)}} \\
\rowcolor{gray!25!white}
& \textbf{10k} & \textbf{20k} & \textbf{50k} & \textbf{100k} & \textbf{200k}
& \textbf{10k} & \textbf{20k} & \textbf{50k} & \textbf{100k} & \textbf{200k} \\ \hline

NonDomDC        & 0.171 & 0.495 & 1.922 & 5.408 & 13.57 & 0.192 & 0.595 & 2.417 & 6.954 & 20.26 \\
FilterX2        & 0.102 & 0.311 & 1.083 & 3.969 & 18.73 & 0.119 & 0.376 & 1.382 & 4.003 & 12.72 \\
FilterSym       & 0.098 & 0.282 & 1.183 & 4.277 & 21.56 & 0.113 & 0.372 & 1.674 & 4.703 & 28.63 \\
PlainSPND        & 0.199 & 0.533 & 1.989 & 6.359 & 20.06 & 0.236 & 0.655 & 2.644 & 7.457 & 22.92 \\
PruneSPND       & 0.032 & 0.070 & 0.243 & 0.740 & 2.306 & 0.027 & 0.099 & 0.241 & 0.667 & 1.731 \\ \hline
\textbf{PreND}           & 0.038 & 0.094 & 0.357 & 0.759 & 2.916 & 0.040 & 0.104 & 0.410 & 1.227 & 4.015 \\
\textbf{PreQND}          & 0.036 & 0.085 & 0.323 & 0.696 & 2.682 & 0.036 & 0.095 & 0.363 & 1.085 & 3.612 \\
\textbf{PreTND}          & 0.036 & 0.091 & 0.340 & 0.713 & 2.747 & 0.037 & 0.096 & 0.390 & 1.135 & 3.734 \\
\textbf{SymND}     & 0.017 & 0.038 & 0.124 & 0.347 & 0.891 & 0.016 & 0.051 & 0.128 & 0.335 & 0.925 \\
\textbf{SymQND}  & \textbf{0.016} & \textbf{0.035} & \textbf{0.111} & \textbf{0.319} & \textbf{0.832} & \textbf{0.015} & \textbf{0.047} & \textbf{0.114} & \textbf{0.307} & \textbf{0.869} \\
\textbf{SymTND}    & 0.017 & 0.039 & 0.128 & 0.357 & 0.914 & 0.017 & 0.052 & 0.128 & 0.344 & 0.945 \\
\textbf{PlainNDred}      & 0.046 & 0.104 & 0.388 & 0.851 & 3.087 & 0.047 & 0.127 & 0.451 & 1.284 & 3.043 \\
\textbf{PlainQNDred}     & 0.041 & 0.099 & 0.358 & 0.779 & 2.836 & 0.042 & 0.121 & 0.413 & 1.203 & 2.755 \\
\textbf{PlainTNDred}     & 0.044 & 0.109 & 0.374 & 0.821 & 2.929 & 0.050 & 0.122 & 0.436 & 1.226 & 2.843 \\
\hline
\end{tabular}
\end{small}
\label{tab:union_correlation}
\end{table}

\begin{table}[ht]
\centering
\setlength{\tabcolsep}{4pt}
\caption{Running times (in seconds) for all algorithms on 4, 6, 8, and 10 dimensions for the Pareto sum of two equally sized Uniform-Random Pareto sets with correlated objectives (URSC).}
\begin{small}
\begin{tabular}{|r|rrrrr|rrrrr|}
\hline
\rowcolor{gray!50!white}
\textbf{Algorithm}
& \multicolumn{5}{c|}{\textbf{Sum Size ($\mathbf{d=4}$, URSC)}}
& \multicolumn{5}{c|}{\textbf{Sum Size ($\mathbf{d=6}$, URSC)}} \\
\rowcolor{gray!25!white}
& \textbf{10k} & \textbf{22k} & \textbf{50k} & \textbf{105k} & \textbf{202k}
& \textbf{10k} & \textbf{22k} & \textbf{50k} & \textbf{105k} & \textbf{202k} \\ \hline

NonDomDC         & 0.093 & 0.328 & 0.976 & 0.274 & 0.557 & 0.058 & 0.139 & 0.442 & 0.965 & 2.343 \\
PlainSPND         & 0.061 & 0.249 & 0.681 & 0.481 & 1.105 & 0.085 & 0.209 & 0.610 & 1.419 & 3.222 \\
BatchedSPND      & 0.039 & 0.196 & 0.463 & 0.467 & 1.138 & 0.068 & 0.172 & 0.624 & 1.526 & 4.482 \\
Doubling(Tree)   & 0.146 & 0.668 & 1.301 & 0.727 & 1.670 & 0.099 & 0.272 & 0.762 & 1.972 & 4.331 \\
Doubling(Filter) & 0.235 & 0.849 & 1.889 & 0.655 & 1.622 & 0.114 & 0.270 & 0.929 & 2.253 & 4.328 \\
LimMem           & 0.148 & 0.538 & 1.060 & 0.352 & 0.767 & 0.086 & 0.212 & 0.674 & 1.462 & 3.105 \\ \hline
\textbf{PreND}         & 0.020 & 0.064 & 0.112 & 0.120 & 0.324 & 0.015 & 0.042 & 0.107 & 0.272 & 0.574 \\
\textbf{PreQND}        & 0.021 & 0.063 & 0.116 & \textbf{0.118} & 0.324 & 0.014 & 0.034 & 0.103 & 0.249 & 0.539 \\
\textbf{PreTND}        & 0.022 & 0.068 & 0.125 & 0.119 & 0.329 & 0.014 & 0.035 & 0.102 & 0.265 & 0.545 \\
\textbf{PlainNDred}    & 0.010 & 0.033 & 0.075 & 0.127 & 0.319 & 0.014 & 0.033 & 0.106 & 0.240 & 0.547 \\
\textbf{PlainQNDred}   & \textbf{0.009} & \textbf{0.026} & \textbf{0.062} & \textbf{0.118} & \textbf{0.318} & \textbf{0.013} & \textbf{0.030} & \textbf{0.096} & \textbf{0.223} & \textbf{0.507} \\
\textbf{PlainTNDred}   & 0.010 & 0.028 & 0.064 & 0.120 & 0.330 & \textbf{0.013}  & 0.032 & 0.105 & 0.229 & 0.528 \\ \hline
\end{tabular}
\end{small}

\vspace{0.4cm}

\begin{small}
\begin{tabular}{|r|rrrrr|rrrrr|}
\hline
\rowcolor{gray!50!white}
\textbf{Algorithm}
& \multicolumn{5}{c|}{\textbf{Sum Size ($\mathbf{d=8}$, URSC)}}
& \multicolumn{5}{c|}{\textbf{Sum Size ($\mathbf{d=10}$, URSC)}} \\
\rowcolor{gray!25!white}
& \textbf{10k} & \textbf{22k} & \textbf{50k} & \textbf{105k} & \textbf{202k}
& \textbf{10k} & \textbf{22k} & \textbf{50k} & \textbf{105k} & \textbf{202k} \\ \hline

NonDomDC         & 0.135 & 0.420 & 1.249 & 3.514 & 6.093 & 0.196 & 0.650 & 2.007 & 6.004 & 15.75 \\
PlainSPND         & 0.163 & 0.555 & 1.659 & 4.235 & 8.684 & 0.241 & 0.836 & 2.822 & 8.476 & 21.78 \\
BatchedSPND      & 0.188 & 0.689 & 2.217 & 7.680 & 15.08 & 0.264 & 1.086 & 4.378 & 16.83 & 57.13 \\
Doubling(Tree)   & 0.181 & 0.590 & 1.773 & 4.543 & 10.12 & 0.251 & 0.865 & 2.920 & 8.490 & 22.19 \\
Doubling(Filter) & 0.225 & 0.731 & 1.989 & 5.565 & 11.15 & 0.287 & 0.984 & 3.147 & 9.316 & 24.79 \\
LimMem           & 0.229 & 0.727 & 2.060 & 5.947 & 9.455 & 0.389 & 1.254 & 3.822 & 11.12 & 28.96 \\ \hline
\textbf{PreND}         & 0.035 & 0.091 & 0.282 & 0.740 & 1.551 & 0.043 & 0.149 & 0.489 & 1.355 & 3.694 \\
\textbf{PreQND}        & 0.032 & \textbf{0.083} & 0.257 & 0.681 & 1.433 & \textbf{0.039} & 0.136 & \textbf{0.433} & 1.236 & 3.348 \\
\textbf{PreTND}        & 0.033 & 0.086 & 0.271 & 0.697 & 1.456 & 0.043 & 0.137 & 0.461 & 1.272 & 3.468 \\
\textbf{PlainNDred}    & 0.032 & 0.102 & 0.279 & 0.650 & 1.531 & 0.047 & 0.149 & 0.480 & 1.143 & 2.889 \\
\textbf{PlainQNDred}   & \textbf{0.029} & 0.086 & \textbf{0.248} & \textbf{0.616} & \textbf{1.419} & 0.049 & \textbf{0.135} & 0.456 & \textbf{1.062} & \textbf{2.668} \\
\textbf{PlainTNDred}   & 0.031 & 0.095 & 0.258 & 0.629 & 1.445 & 0.045 & 0.141 & 0.464 & 1.084 & 2.716 \\ \hline
\end{tabular}
\end{small}
\label{tab:sum_correlation}
\end{table}

\begin{table}[ht]
\centering
\setlength{\tabcolsep}{3.7pt}
\caption{Running times (in seconds) for all algorithms on 5, 6, 8, and 10 dimensions for the Pareto union of two equally sized Uniform-Random Pareto sets with plateaus and correlated objectives (URSPC).}
\begin{small}
\begin{tabular}{|r|rrrrr|rrrrr|}
\hline
\rowcolor{gray!50!white}
\textbf{Algorithm}
& \multicolumn{5}{c|}{\textbf{Union Size ($\mathbf{d=5}$, URSPC)}}
& \multicolumn{5}{c|}{\textbf{Union Size ($\mathbf{d=6}$, URSPC)}} \\
\rowcolor{gray!25!white}
& \textbf{10k} & \textbf{20k} & \textbf{50k} & \textbf{100k} & \textbf{200k}
& \textbf{10k} & \textbf{20k} & \textbf{50k} & \textbf{100k} & \textbf{200k} \\ \hline

NonDomDC        & 0.127 & 0.255 & 0.919 & 2.456 & 8.459 & 0.354 & 0.572 & 1.315 & 3.746 & 10.679 \\
FilterX2        & 0.079 & 0.190 & 0.682 & 2.559 & 7.496 & 0.226 & 0.401 & 0.932 & 3.212 & 11.150 \\
FilterSym       & 0.073 & 0.158 & 0.605 & 2.553 & 6.138 & 0.202 & 0.345 & 1.176 & 3.181 & 9.878 \\
PlainSPND        & 0.135 & 0.373 & 1.458 & 4.054 & 11.952 & 0.224 & 0.574 & 1.888 & 5.604 & 15.384 \\
PruneSPND       & 0.028 & 0.064 & 0.248 & 0.634 & 2.451 & 0.050 & 0.111 & 0.355 & 1.033 & 2.665 \\ \hline
\textbf{PreND}           & 0.021 & 0.050 & 0.172 & 0.518 & 4.864 & 0.047 & 0.097 & 0.290 & 0.687 & 1.800 \\
\textbf{PreQND}          & 0.019 & 0.046 & 0.156 & 0.492 & 4.774 & 0.047 & 0.097 & 0.265 & 0.621 & 1.621 \\
\textbf{PreTND}          & 0.020 & 0.049 & 0.164 & 0.489 & 4.703 & 0.049 & 0.099 & 0.278 & 0.636 & 1.689 \\
\textbf{SymND}     & 0.015 & 0.035 & 0.100 & 0.257 & 0.883 & 0.042 & 0.065 & 0.116 & 0.311 & 0.767 \\
\textbf{SymQND}    & \textbf{0.013} & \textbf{0.032} & \textbf{0.094} & \textbf{0.238} & \textbf{0.839} & \textbf{0.041} & \textbf{0.062} & \textbf{0.106} & \textbf{0.291} & \textbf{0.721} \\
\textbf{SymTND}    & 0.015 & 0.036 & 0.104 & 0.261 & 0.893 & 0.042 & 0.065 & 0.120 & 0.322 & 0.791 \\
\textbf{PlainNDred}      & 0.018 & 0.046 & 0.173 & 0.549 & 5.257 &\textbf{0.041} & 0.091 & 0.333 & 0.701 & 1.633 \\
\textbf{PlainQNDred}     & 0.017 & 0.044 & 0.172 & 0.517 & 4.903 & 0.039 & 0.090 & 0.315 & 0.661 & 1.519 \\
\textbf{PlainTNDred}     & 0.018 & 0.045 & 0.170 & 0.523 & 4.886 & \textbf{0.041} & 0.094 & 0.320 & 0.690 & 1.561 \\
\hline
\end{tabular}
\end{small}

\vspace{1.2em}

\begin{small}
\begin{tabular}{|r|rrrrr|rrrrr|}
\hline
\rowcolor{gray!50!white}
\textbf{Algorithm}
& \multicolumn{5}{c|}{\textbf{Union Size ($\mathbf{d=8}$, URSPC)}}
& \multicolumn{5}{c|}{\textbf{Union Size ($\mathbf{d=10}$, URSPC)}} \\
\rowcolor{gray!25!white}
& \textbf{10k} & \textbf{20k} & \textbf{50k} & \textbf{100k} & \textbf{200k}
& \textbf{10k} & \textbf{20k} & \textbf{50k} & \textbf{100k} & \textbf{200k} \\ \hline

NonDomDC        & 0.310 & 0.491 & 2.142 & 4.598 & 16.310 & 0.485 & 0.526 & 2.222 & 6.403 & 21.604 \\
FilterX2        & 0.198 & 0.320 & 1.291 & 3.603 & 14.778 & 0.310 & 0.346 & 1.449 & 3.941 & 14.564 \\
FilterSym       & 0.180 & 0.301 & 1.209 & 3.233 & 13.379 & 0.248 & 0.414 & 1.949 & 5.900 & 22.568 \\
PlainSPND        & 0.269 & 0.567 & 2.105 & 6.743 & 19.866 & 0.362 & 0.663 & 2.760 & 7.900 & 24.350 \\
PruneSPND       & 0.042 & 0.080 & 0.244 & 0.808 & 2.477 & 0.045 & 0.073 & 0.252 & 0.951 & 1.956 \\ \hline
\textbf{PreND}           & 0.061 & 0.101 & 0.253 & 0.863 & 3.076 & 0.062 & 0.109 & 0.373 & 1.131 & 3.805 \\
\textbf{PreQND}          & 0.057 & 0.099 & 0.232 & 0.797 & 2.818 & 0.060 & 0.100 & 0.333 & 1.067 & 3.481 \\
\textbf{PreTND}          & 0.058 & 0.097 & 0.237 & 0.829 & 2.896 & 0.063 & 0.103 & 0.349 & 1.075 & 3.623 \\
\textbf{SymND}     & 0.034 & 0.041 & 0.112 & 0.336 & 0.903 & 0.046 & 0.045 & 0.127 & 0.389 & 0.966 \\
\textbf{SymQND}    & \textbf{0.032} & \textbf{0.038} & \textbf{0.105} & \textbf{0.311} & \textbf{0.844} & \textbf{0.044} & \textbf{0.044} & \textbf{0.119} & \textbf{0.362} & \textbf{0.899} \\
\textbf{SymTND}    & 0.033 & 0.042 & 0.117 & 0.345 & 0.911 & 0.045 & 0.046 & 0.129 & 0.393 & 0.987 \\
\textbf{PlainNDred}      & 0.051 & 0.114 & 0.304 & 0.911 & 3.102 & 0.053 & 0.120 & 0.419 & 1.196 & 3.634 \\
\textbf{PlainQNDred}     & 0.047 & 0.105 & 0.284 & 0.849 & 2.933 & 0.047 & 0.115 & 0.405 & 1.075 & 3.234 \\
\textbf{PlainTNDred}     & 0.058 & 0.119 & 0.294 & 0.878 & 2.909 & 0.050 & 0.124 & 0.406 & 1.096 & 3.330 \\
\hline
\end{tabular}
\end{small}
\label{tab:union_plateau_correlation}
\end{table}

\begin{table}[ht]
\centering
\setlength{\tabcolsep}{3.7pt}
\caption{Running times (in seconds) for all algorithms on 5, 6, 8, and 10 dimensions for the Pareto sum of two equally sized Uniform-Random Pareto sets with plateaus and correlated objectives (URSPC).}
\begin{small}
\begin{tabular}{|r|rrrrr|rrrrr|}
\hline
\rowcolor{gray!50!white}
\textbf{Algorithm}
& \multicolumn{5}{c|}{\textbf{Sum Size ($\mathbf{d=5}$, URSPC)}}
& \multicolumn{5}{c|}{\textbf{Sum Size ($\mathbf{d=6}$, URSPC)}} \\
\rowcolor{gray!25!white}
& \textbf{10k} & \textbf{22k} & \textbf{50k} & \textbf{105k} & \textbf{202k}
& \textbf{10k} & \textbf{22k} & \textbf{50k} & \textbf{105k} & \textbf{202k} \\ \hline

NonDomDC        & 0.268 & 0.454 & 0.806 & 0.486 & 1.094 & 0.063 & 0.143 & 0.415 & 0.886 & 1.647 \\
PlainSPND        & 0.546 & 0.463 & 1.414 & 0.672 & 1.731 & 0.092 & 0.216 & 0.576 & 1.125 & 2.299 \\
BatchedSPND     & 0.302 & 0.247 & 0.747 & 0.676 & 1.957 & 0.084 & 0.178 & 0.590 & 1.187 & 2.624 \\
Doubling(Tree)  & 0.531 & 0.745 & 1.534 & 1.023 & 2.684 & 0.111 & 0.275 & 0.747 & 1.907 & 3.431 \\
Doubling(Filter)& 0.461 & 0.837 & 1.452 & 1.087 & 2.736 & 0.130 & 0.252 & 0.816 & 2.300 & 3.482 \\
LimMem          & 0.320 & 0.601 & 0.830 & 0.611 & 1.462 & 0.095 & 0.209 & 0.609 & 1.614 & 2.177 \\ \hline
\textbf{PreND}           & 0.035 & 0.055 & 0.128 & 0.170 & 0.377 & 0.019 & 0.038 & 0.101 & 0.303 & 0.451 \\
\textbf{PreQND}          & 0.033 & 0.057 & 0.130 & 0.166 & 0.370 & 0.018 & 0.036 & 0.095 & 0.294 & 0.421 \\
\textbf{PreTND}          & 0.034 & 0.059 & 0.131 & 0.170 & 0.369 & 0.018 & 0.037 & 0.097 & 0.264 & 0.432 \\
\textbf{PlainNDred}      & 0.022 & 0.031 & 0.087 & 0.144 & 0.382 & 0.018 & 0.033 & 0.091 & 0.222 & 0.331 \\
\textbf{PlainQNDred}  & \textbf{0.020} & \textbf{0.024} & \textbf{0.076} & \textbf{0.142} & \textbf{0.368} & \textbf{0.016} & \textbf{0.030} & \textbf{0.083} & \textbf{0.177} & \textbf{0.318} \\
\textbf{PlainTNDred}     & 0.021 & 0.025 & 0.077 & 0.143 & 0.376 & 0.017 & 0.031 & 0.087 & 0.199 & 0.324 \\
\hline
\end{tabular}
\end{small}

\vspace{1.2em}

\begin{small}
\begin{tabular}{|r|rrrrr|rrrrr|}
\hline
\rowcolor{gray!50!white}
\textbf{Algorithm}
& \multicolumn{5}{c|}{\textbf{Sum Size ($\mathbf{d=8}$, URSPC)}}
& \multicolumn{5}{c|}{\textbf{Sum Size ($\mathbf{d=10}$, URSPC)}} \\
\rowcolor{gray!25!white}
& \textbf{10k} & \textbf{22k} & \textbf{50k} & \textbf{105k} & \textbf{202k}
& \textbf{10k} & \textbf{22k} & \textbf{50k} & \textbf{105k} & \textbf{202k} \\ \hline

NonDomDC        & 0.127 & 0.371 & 1.133 & 2.895 & 6.498 & 0.201 & 0.665 & 2.011 & 5.434 & 14.00 \\
PlainSPND        & 0.151 & 0.504 & 1.485 & 4.069 & 9.233 & 0.263 & 0.847 & 2.860 & 8.728 & 20.81 \\
BatchedSPND     & 0.169 & 0.560 & 1.999 & 6.336 & 15.70 & 0.292 & 1.107 & 4.440 & 16.287 & 52.88 \\
Doubling(Tree)  & 0.169 & 0.556 & 1.673 & 4.973 & 12.90 & 0.263 & 0.894 & 2.958 & 8.354 & 21.51 \\
Doubling(Filter)& 0.219 & 0.669 & 1.836 & 5.725 & 12.03 & 0.299 & 1.098 & 3.484 & 9.449 & 23.07 \\
LimMem          & 0.214 & 0.621 & 1.853 & 4.785 & 10.11 & 0.388 & 1.288 & 3.824 & 10.05 & 25.23 \\ \hline
\textbf{PreND}           & 0.031 & 0.084 & 0.251 & 0.778 & 1.523 & 0.044 & 0.145 & 0.455 & 1.336 & 3.537 \\
\textbf{PreQND}          & 0.029 & 0.079 & 0.243 & 0.731 & 1.436 & \textbf{0.041} & \textbf{0.135} & \textbf{0.427} & 1.239 & 3.282 \\
\textbf{PreTND}          & 0.030 & 0.081 & 0.243 & 0.762 & 1.465 & 0.043 & 0.145 & 0.442 & 1.289 & 3.379 \\
\textbf{PlainNDred}      & 0.029 & 0.085 & 0.230 & 0.632 & 1.247 & 0.048 & 0.147 & 0.454 & 1.260 & 3.053 \\
\textbf{PlainQNDred}     & \textbf{0.027} & \textbf{0.078} & \textbf{0.210} & \textbf{0.600} & \textbf{1.180} & 0.043 & 0.139 & 0.429 & \textbf{1.178} & \textbf{2.818} \\
\textbf{PlainTNDred}     & 0.028 & 0.082 & 0.217 & 0.612 & 1.195 & 0.045 & 0.146 & 0.449 & 1.210 & 2.887 \\
\hline
\end{tabular}
\end{small}
\label{tab:sum_plateau_correlation}
\end{table}

\clearpage

\begin{table}[ht]
\centering
\setlength{\tabcolsep}{3pt}
\renewcommand{\arraystretch}{1.2}
\caption{Min, Avg, and Max speedups for each algorithm for the \emph{Pareto Sum} operation for URS, URSP, URSC, and URSPC data sets (same with Table \ref{tab:sum_speedup_summary}, but also including average speedup).}
\begin{small}
\begin{tabular}{|r|ccc|ccc|ccc|ccc|}
\hline
\rowcolor{gray!50!white}
\textbf{Algorithm}
& \multicolumn{3}{c|}{\textbf{URS}}
& \multicolumn{3}{c|}{\textbf{URSP}}
& \multicolumn{3}{c|}{\textbf{URSC}}
& \multicolumn{3}{c|}{\textbf{URSPC}} \\
\rowcolor{gray!25!white}
& Min & Avg & Max & Min & Avg & Max & Min & Avg & Max & Min & Avg & Max \\
\hline
PreND       & 1.8 & 3.86 & 5.1 & 1.8 & 3.82 & 4.9 & 1.7 & 3.77 & 4.7 & 2.9 & 4.20 & 7.7 \\
PreQND      & 1.8 & 4.18 & 5.6 & 1.8 & 4.11 & 5.3 & 1.7 & 4.06 & 5.2 & 2.9 & 4.42 & 8.2 \\
PreTND      & 1.8 & 4.09 & 5.5 & 1.8 & 4.02 & 5.2 & 1.7 & 3.92 & 5.0 & 2.9 & 4.30 & 7.8 \\
PlainNDred  & 2.9 & 4.75 & 6.3 & 2.4 & 4.74 & 7.0 & 1.7 & 4.32 & 6.2 & 2.9 & 5.09 & 12.2 \\
PlainQNDred & 2.9 & 5.13 & 6.9 & 2.4 & 5.27 & 9.2 & 1.7 & 4.77 & 7.5 & 3.0 & 5.62 & 13.2 \\
PlainTNDred & 3.0 & 5.01 & 6.7 & 2.4 & 5.03 & 8.8 & 1.7 & 4.59 & 7.2 & 2.9 & 5.42 & 12.7 \\
\hline
\end{tabular}
\end{small}
\label{tab:sum_urs_speedup_summary}
\end{table}

\begin{table}[ht]
\centering
\setlength{\tabcolsep}{3pt}
\renewcommand{\arraystretch}{1.2}
\caption{Min, Avg, and Max speedups for each algorithm for the \emph{Pareto Union} operation for URS, URSP, URSC, and URSPC data sets (same with Table \ref{tab:union_speedup_summary}, but also including average speedup).}
\begin{small}
\begin{tabular}{|r|ccc|ccc|ccc|ccc|}
\hline
\rowcolor{gray!50!white}
\textbf{Algorithm}
& \multicolumn{3}{c|}{\textbf{URS}}
& \multicolumn{3}{c|}{\textbf{URSP}}
& \multicolumn{3}{c|}{\textbf{URSC}}
& \multicolumn{3}{c|}{\textbf{URSPC}} \\
\rowcolor{gray!25!white}
& Min & Avg & Max & Min & Avg & Max & Min & Avg & Max & Min & Avg & Max \\
\hline
PreND       & 1.5 & 3.2 & 5.2 & 1.7 & 2.6 & 3.6 & 0.4 & 1.1 & 2.6 & 0.5 & 1.0 & 1.5 \\
PreQND      & 1.6 & 3.6 & 5.8 & 1.8 & 2.9 & 4.0 & 0.5 & 1.2 & 2.9 & 0.5 & 1.1 & 1.7 \\
PreTND      & 1.6 & 3.5 & 5.7 & 1.8 & 2.8 & 3.9 & 0.5 & 1.1 & 2.8 & 0.5 & 1.0 & 1.6 \\
SymND       & 1.7 & 3.4 & 4.8 & 1.3 & 2.6 & 3.7 & 1.4 & 2.2 & 3.6 & 1.0 & 2.2 & 3.5 \\
SymQND      & 1.7 & 3.8 & 5.4 & 1.4 & 2.9 & 4.4 & 1.6 & 2.4 & 3.9 & 1.0 & 2.3 & 3.7 \\
SymTND      & 1.6 & 3.6 & 5.2 & 1.3 & 2.7 & 3.9 & 1.4 & 2.1 & 3.5 & 1.0 & 2.2 & 3.4 \\
PlainNDred  & 1.7 & 3.6 & 6.3 & 1.8 & 2.8 & 3.8 & 0.5 & 0.9 & 2.2 & 0.5 & 1.0 & 1.6 \\
PlainQNDred & 1.9 & 3.8 & 6.8 & 1.9 & 3.0 & 4.3 & 0.6 & 1.0 & 2.4 & 0.5 & 1.1 & 1.8 \\
PlainTNDred & 1.8 & 3.7 & 6.7 & 1.8 & 2.9 & 4.1 & 0.5 & 1.0 & 2.3 & 0.5 & 1.0 & 1.7 \\
\hline
\end{tabular}
\end{small}
\label{tab:union_urs_speedup_summary}
\end{table}









\end{document}